\documentclass[fleqn,usenatbib]{mnras}

\usepackage{newtxtext,newtxmath}
\usepackage{longtable,lscape}

\usepackage[T1]{fontenc}

\DeclareRobustCommand{\VAN}[3]{#2}
\let\VANthebibliography\thebibliography
\def\thebibliography{\DeclareRobustCommand{\VAN}[3]{##3}\VANthebibliography}

\usepackage{graphicx}	
\usepackage{amsmath}	
\usepackage{booktabs}
\usepackage[figuresright]{rotating}
\usepackage{textcase}
\usepackage{multicol}
\usepackage{subfig}
\usepackage{comment}

\usepackage{array}
\usepackage{caption}
\usepackage{threeparttable}
\usepackage{scalerel}
\usepackage{tikz}
\usetikzlibrary{svg.path}

\definecolor{orcidlogocol}{HTML}{A6CE39}
\tikzset{
  orcidlogo/.pic={
    \fill[orcidlogocol] svg{M256,128c0,70.7-57.3,128-128,128C57.3,256,0,198.7,0,128C0,57.3,57.3,0,128,0C198.7,0,256,57.3,256,128z};
    \fill[white] svg{M86.3,186.2H70.9V79.1h15.4v48.4V186.2z}
                 svg{M108.9,79.1h41.6c39.6,0,57,28.3,57,53.6c0,27.5-21.5,53.6-56.8,53.6h-41.8V79.1z M124.3,172.4h24.5c34.9,0,42.9-26.5,42.9-39.7c0-21.5-13.7-39.7-43.7-39.7h-23.7V172.4z}
                 svg{M88.7,56.8c0,5.5-4.5,10.1-10.1,10.1c-5.6,0-10.1-4.6-10.1-10.1c0-5.6,4.5-10.1,10.1-10.1C84.2,46.7,88.7,51.3,88.7,56.8z};
  }
}

\newcommand\orcidicon[1]{\href{https://orcid.org/#1}{\mbox{\scalerel*{
\begin{tikzpicture}[yscale=-1,transform shape]
\pic{orcidlogo};
\end{tikzpicture}
}{|}}}}

\newcommand{\FT}[1]{}

\newcommand{\Mgii}{Mg\,\textsc{ii}}
\newcommand{\Civ}{C\,\textsc{iv}}

\newcommand{\hii}{H\,\textsc{ii}}

    \title[UV/X-ray and radius-luminosity relations]{Quasar UV/X-ray relation luminosity distances are shorter than reverberation-measured radius-luminosity relation luminosity distances}

\author[Khadka et al.]{
Narayan Khadka,$^{\orcidicon{0000-0001-5512-2716}{1}}$\thanks{E-mail: nkhadka@bellarmine.edu}
Michal Zaja\v{c}ek,$^{\orcidicon{0000-0001-6450-1187}{2}}$\thanks{E-mail: zajacek@mail.muni.cz}
Raj Prince,$^{\orcidicon{0000-0002-1173-7310}{3}}$\thanks{E-mail: raj@cft.edu.pl}
Swayamtrupta Panda,$^{\orcidicon{0000-0002-5854-7426}{4}}$\thanks{CNPq Fellow,\ E-mail: spanda@lna.br}
Bo\.{z}ena Czerny,$^{\orcidicon{0000-0001-5848-4333}{3}}$\thanks{E-mail: bcz@cft.edu.pl}
\newauthor \hspace{0.1mm}
Mary Loli Mart\'inez-Aldama,$^{\orcidicon{0000-0002-7843-7689}{5,6}}$\thanks{E-mail: mary.martinez@uv.cl}
Vikram Kumar Jaiswal,$^{\orcidicon{0000-0003-0547-4654}{3}}$\thanks{Email: vkj005@gmail.com}
Bharat Ratra$^{\orcidicon{0000-0002-7307-0726}{7}}$\thanks{E-mail: ratra@phys.ksu.edu}
\\
$^{1}$Department of Physics, Bellarmine University, 2001 Newburg Rd, Louisville, KY 40205, USA\\
$^{2}$Department of Theoretical Physics and Astrophysics, Faculty of Science, Masaryk University, Kotl\'a\v{r}sk\'a 2, 611 37 Brno, Czech Republic\\
$^{3}$Center for Theoretical Physics, Polish Academy of Sciences, Al.\ Lotnik\'{o}w 32/46, 02-668 Warsaw, Poland\\
$^{4}$Laborat\'orio Nacional de Astrof\'isica - MCTIC, R. dos Estados Unidos, 154 - Na\c{c}\~oes, Itajub\'a - MG, 37504-364, Brazil\\
$^{5}$Instituto de Física y Astronomía, Facultad de Ciencias, Universidad de Valparaíso, Gran Bretaña 1111, Playa Ancha, Valparaíso, Chile\\
$^{6}$Departamento de Astronomía, Universidad de Chile, Camino del Observatorio 1515, Casilla 36-D, Correo Central, Santiago, Chile\\
$^{7}$Department of Physics, Kansas State University, 116 Cardwell Hall, Manhattan, KS 66506, USA\\}

\date{Accepted XXX. Received YYY; in original form ZZZ}

\pubyear{2022}

\begin{document}
\label{firstpage}
\pagerange{\pageref{firstpage}--\pageref{lastpage}}
\maketitle

\begin{abstract}
We use measurements of 59/58 quasars (QSOs), over a redshift range $0.0041\leq z \leq 1.686$, to do a comparative study of the radius--luminosity ($R-L$) and X-ray$-$UV luminosity ($L_X-L_{UV}$) relations and the implication of these relations for cosmological parameter estimation. By simultaneously determining $R-L$ or $L_X-L_{UV}$ relation parameters and cosmological parameters in six different cosmological models, we find that both $R-L$ and $L_X-L_{UV}$ relations are standardizable but provide only weak cosmological parameter constraints, with $L_X-L_{UV}$ relation data favoring larger current non-relativistic matter density parameter $\Omega_{m0}$ values than $R-L$ relation data and most other available data. We derive $L_X-L_{UV}$ and $R-L$ luminosity distances for each of the sources in the six cosmological models and find that $L_X-L_{UV}$ relation luminosity distances are shorter than $R-L$ relation luminosity distances as well as standard flat $\Lambda$CDM model luminosity distances. This explains why $L_X-L_{UV}$ relation QSO data favor larger $\Omega_{m0}$ values than do $R-L$ relation QSO data or most other cosmological measurements. While our sample size is small and only spans a small $z$ range, these results indicate that more work is needed to determine whether the $L_X-L_{UV}$ relation can be used as a cosmological probe.
\end{abstract}

\begin{keywords}
\textit{(cosmology:)} cosmological parameters -- \textit{(cosmology:)} observations -- \textit{(cosmology:)} dark energy -- \textit{(galaxies:) quasars: emission lines}
\end{keywords}



\section{Introduction}

If general relativity provides an accurate description of gravity on cosmological scales, dark energy is responsible for the observed current accelerated cosmological expansion and contributes $\sim 70\%$ of the current cosmological energy budget. In the standard spatially-flat $\Lambda$CDM cosmological model \citep{Peebles1984} dark energy is the cosmological constant $\Lambda$ and non-relativistic cold dark matter (CDM) contributes $\sim 25\%$ of the current cosmological energy budget with non-relativistic baryonic matter contributing $\sim 5\%$. While the standard model is reasonably consistent with low redshift $z \lesssim 2.3$ observations \citep{Scolnicetal2018, Yuetal2018, eBOSSCollaboration2021} and $z \sim 1100$ measurements \citep{PlanckCollaboration2020}, it might not be able to accommodate some data \citep{DiValentinoetal2021b, PerivolaropoulosSkara2021, Abdallaetal2022}. 

A new reliable cosmological probe, especially one in the largely unexplored part of redshift space between the highest $z \sim 2.3$ baryon acoustic oscillation (BAO) measurements and cosmic microwave background anisotropy data at $z \sim 1100$, might help clarify whether the standard flat $\Lambda$CDM model needs to be improved on.     

The last decade has seen the initial development of a number of such probes, including \hii\ starburst galaxy apparent magnitude observations which reach to $z\sim2.5$ \citep{ManiaRatra2012, Chavezetal2014, GonzalezMoranetal2021, Caoetal2020, Caoetal2021a, Caoetal_2021c, Johnsonetal2022, Mehrabietal2022}, quasar (hereafter QSO) angular size observations which reach to $z\sim2.7$ \citep{Caoetal2017, Ryanetal2019, Caoetal2020, Caoetal2021b, Zhengetal2021, Lianetal2021}, \Mgii\ and \Civ\ reverberation mapped QSO observations which reach to $z\sim3.4$ \citep{Michal2021, khadka2021, Khadkaetal2022, Caoetal2022}, and gamma-ray burst (GRB) observations which reach to $z\sim8.2$ \citep{Wang_2016, Wangetal2022, Dirirsa2019, Demianskietal_2021, KhadkaRatra2020c, Khadkaetal2021a, Huetal2021, OrlandoMarco2021, CaoKhadkaRatra2022, CaoDainottiRatra2022a, CaoDainottiRatra2022b, Dainottietal2022b, Liuetal2022, Jiaetal2022, Liangetal2022, Kumaretal2022}. 

Another potentially promising probe makes use of QSO X-ray and UV flux measurements which reach to $z \sim 7.5$ \citep{RisalitiLusso2015, RisalitiLusso2019, KhadkaRatra2020a, KhadkaRatra2020b, KhadkaRatra2021a, KhadkaRatra2022, Lussoetal2020, Rezaeietal2022, Luongoetal2021, HuWang2022, Colgainetal2022, dainotti2022, Petrosianetal2022, 2022MNRAS.517.1901L, 2022arXiv221014432W, Pourojaghietal2022} and is the main subject of our paper. With the progress of knowledge of such QSO properties and an increase in the number of these sources, such QSO data have been used to constrain cosmological model parameters. Strong constraints, and tension with the standard flat $\Lambda$CDM model, have been claimed \citep{RisalitiLusso2019,Lussoetal2020} from a method based on an assumed non-linear relation between the QSO UV and X-ray luminosities, $L_{UV}$ and $L_X$ \citep{Tananbaumetal1979, Zamoranietal1981, AvniTananbaum1986, Steffenetal2006, Justetal2007, Greenetal2009, Youngetal2010, Lussoetal2010, Grupeetal2010, Vagnettietal2010}. We note that the analyses of \citet{RisalitiLusso2019} and \citet{Lussoetal2020} were approximate and based on incorrect assumptions \citep{KhadkaRatra2020a, KhadkaRatra2020b, KhadkaRatra2021a, KhadkaRatra2022, Banerjeeetal2021, Petrosianetal2022}. The correct technique for analyses of $L_X-L_{UV}$ relation QSO data was developed in \citet{KhadkaRatra2020a}. Here one must use these QSO data to simultaneously determine the $L_X-L_{UV}$ relation parameters and the cosmological model parameters. In this case, one must also study a number of different cosmological models to determine whether $L_X-L_{UV}$ relation parameter values are independent of the assumed cosmological model, and if they are, then these QSOs are standardizable and the circularity problem is circumvented. Unfortunately, the most recent \citet{Lussoetal2020} QSO compilation is not standardizble \citep{KhadkaRatra2021a, KhadkaRatra2022}, because the $L_X-L_{UV}$ relation parameters depend on the assumed cosmological model and on redshift \citep{KhadkaRatra2021a, KhadkaRatra2022}. See \citet{dainotti2022}\footnote{\citet{dainotti2022} applied a statistical treatment of QSO selection bias and also treated the redshift evolution of their X-ray and UV luminosities by introducing so-called de-evolved luminosities which resulted in de-evolved $L_{X}-L_{UV}$ relation parameters consistent with the parameters of the original \citet{Lussoetal2020} $L_{X}-L_{UV}$ relation as well as in tighter cosmological constraints consistent with the standard flat $\Lambda$CDM model \citep{2022arXiv221110785L}. However, \citet{2022arXiv221110785L} study their corrected (but non-calibrated) QSO data alone in only the flat $\Lambda$CDM model and also do not divide them into redshift bins and so have not checked whether their de-evolved $L_{X}-L_{UV}$ relation parameters are cosmological model and redshift independent, i.e, they have not determined whether their de-evolved $L_{X}-L_{UV}$ QSOs are standardizable.}, \citet{2022MNRAS.517.1901L}, and \citet{2022arXiv221014432W} for more recent discussions of the redshift evolution, but note that out of these studies, only \citet{2022MNRAS.517.1901L} do analyses of redshift-space subsets of these QSO data \citep[as did][]{KhadkaRatra2021a, KhadkaRatra2022}. \citet{KhadkaRatra2022} found that the largest of the seven QSO sub-samples in the \citet{Lussoetal2020} compilation, the SDSS-4XMM, i.e. the one that contains about 2/3 of the total QSOs, has an $L_{X}-L_{UV}$ relation that depends on the cosmological model as well as on redshift and is the main, but possibly not only, cause of the problem with the \citet{Lussoetal2020} compilation.

Recently, there has been considerable progress in the development of another QSO probe that is based on the correlation between the rest-frame time-delay of the broad-line response with respect to the variable continuum and the monochromatic continuum luminosity, known as the radius-luminosity ($R-L$) relation. The application of this relation was proposed over a decade ago \citep{2011ApJ...740L..49W,2011A&A...535A..73H,2013ApJ...767..149B,czerny2013} but it was successfully implemented only recently by using \Mgii\ and \Civ\ time delays \citep{Michal2021, khadka2021, Khadkaetal2022, Caoetal2022}; see also \citet{2021bhns.confE...1K} and \citet{Czernyetal2022} for recent reviews. We note that using currently available H$\beta$ measurements is still problematic \citep{2022MNRAS.513.1985K}. Although H$\beta$ $R-L$ relation parameters are independent of the assumed cosmological model, the constraints are weak, favour decelerated expansion, and are in 2$\sigma$ tension with better established probes. Corrections related to the Eddington ratio introduced in the fixed flat $\Lambda$CDM model \citep{Mary2019,Mary2020,Panda_2022FrASS...950409P, 2022arXiv221015041P} do not yield a significant improvement when cosmological parameters are set free \citep{2022MNRAS.513.1985K}. Future reassessment of H$\beta$ time delays, including a careful removal of outliers \citep{2021AcPPA.139..389C}, appears necessary as an attempt to resolve this H$\beta$ QSO problem. In contrast to H$\beta$ results, cosmological constraints based on the \Mgii\ and \Civ\ $R-L$ relations are consistent with the standard flat $\Lambda$CDM model.

Given the problems with the \citet{Lussoetal2020} $L_X-L_{UV}$ QSO compilation, X-ray detected \Mgii\ reverberation-measured QSOs provide a unique opportunity to determine whether the QSO $L_X-L_{UV}$ relation can be used as a cosmological probe. X-ray detected \Mgii\ QSOs can be used to derive $L_X-L_{UV}$ relation and cosmological model parameter values and these can be compared to the corresponding $R-L$ relation and cosmological model parameter values. Such a sample provides a unique opportunity to probe potential systematic effects of these two relations that are independent of each other. The only correlation present is between corresponding UV flux densities at 2500 \AA\ (for the $L_X-L_{UV}$ relation) and 3000 \AA\ (for the $R-L$ relation). However, the measurements of the \Mgii\ line-emission time-delay and the X-ray flux density at 2 keV are independent of each other.

\citet{khadka2021} and  \citet{Khadkaetal2022} have shown that a larger compilation of \Mgii\ $R-L$ QSOs are standardizable and so can be used as a cosmological probe, and we find that the smaller sample of 58 X-ray detected \Mgii\ $R-L$ QSOs, over $0.0041\leq z \leq 1.686$, that we study here also share these attributes. We find that the corresponding $L_X-L_{UV}$ relation for these QSOs are also standardizable, which is encouraging but possibly a consequence of the much smaller sample size and smaller redshift range compared to those of the \citet{Lussoetal2020} $L_X-L_{UV}$ QSO compilation. However, we go on to derive $L_X-L_{UV}$ and $R-L$ luminosity distances for each of the 58 sources in six different cosmological models and find that $L_X-L_{UV}$ relation luminosity distances are shorter than $R-L$ relation luminosity distances as well as standard flat $\Lambda$CDM model luminosity distances. This result explains why $L_X-L_{UV}$ relation QSO data favor larger current non-relativistic matter density parameter $\Omega_{m0}$ values than do $R-L$ relation QSO data or most other cosmological measurements. While our sample size is small and these QSOs span only $0.0041\leq z \leq 1.686$, our results indicate that more work is needed before we can determine whether the $L_X-L_{UV}$ relation can be used as a cosmological probe.

Our paper is structured as follows. In Sec.~\ref{sec:models} we introduce the cosmological models and $R-L$ and $L_X-L_{UV}$ relations and their parameters. The data we use are described in Sec.~\ref{sec:data}. The method we use to infer parameter values and uncertainties is outlined in Sec.~\ref{sec:methods}. In Sec.~\ref{sec:results} we present the main results obtained using the two independent methods. We conclude in Sec.~\ref{sec:conclusions}.

\section{Cosmological models and parameters}
\label{sec:models}

In this study we constrain cosmological model parameters, $R-L$ relation parameters, and $L_X-L_{UV}$ relation parameters in three pairs of dark-energy general-relativistic cosmological models with flat and non-flat spatial geometries,\footnote{For discussions of constraints on spatial curvature see \citet{Ranaetal2017}, \citet{Oobaetal2018a, Oobaetal2018b}, \citet{ParkRatra2019a, ParkRatra2019b}, \citet{DESCollaboration2019}, \citet{EfstathiouGratton2020}, \citet{DiValentinoetal2021a},  \citet{KiDSCollaboration2021}, \citet{ArjonaNesseris2021}, \citet{Dhawanetal2021}, \citet{Renzietal2021}, \citet{Gengetal2022}, \citet{WeiMelia2022}, \citet{MukherjeeBanerjee2022}, \citet{Glanvilleetal2022}, \citet{Wuetal2022}, \citet{deCruzPerez2022}, \citet{DahiyaJain2022}, and references therein.} so in total six cosmological models, by using QSO measurements. This allows us to compare two sets of cosmological constraints, those derived using the $R-L$ relation and those derived using the $L_X-L_{UV}$ relation. On the other hand, since we constrain correlation parameters for both these correlation relations using the same set of sources, these results can indicate which correlation relation better holds for the QSOs we consider. 

Observational data used in this paper are QSO time-delays and 3000 \AA, 2500 \AA, and 2 keV flux densities. The $R-L$ and $L_X-L_{UV}$ relations involve luminosity so flux needs to be converted to luminosity. To do this we need the luminosity distance $D_L(z,\mathbf{p})$ for each source. Given a cosmological model, the luminosity distance (in cm) can be computed as a function of redshift ($z$) and cosmological parameters ($\mathbf{p}$),
\begin{equation}
  \label{eq:DL}
\resizebox{0.475\textwidth}{!}{%
    $D_L(z,\mathbf{p}) = 
    \begin{cases}
    \frac{c(1+z)}{H_0\sqrt{\Omega_{k0}}}\sinh\left[\frac{H_0\sqrt{\Omega_{k0}}}{c}
    D_C(z,\mathbf{p})\right] & \text{if}\ \Omega_{k0} > 0, \\
    \vspace{1mm}
    (1+z)D_C(z,\mathbf{p}) & \text{if}\ \Omega_{k0} = 0,\\
    \vspace{1mm}
    \frac{c(1+z)}{H_0\sqrt{|\Omega_{k0}|}}\sin\left[\frac{H_0\sqrt{|\Omega_{k0}|}}{c}
    D_C(z,\mathbf{p})\right] & \text{if}\ \Omega_{k0} < 0.
    \end{cases}$%
    }
\end{equation}
Here $c$ is the speed of light, $H_0$ is the Hubble constant, $\Omega_{k0}$ is the current value of the spatial curvature energy density parameter, and $D_C(z,\mathbf{p})$ is the comoving distance. This is computed as a function of $z$ and $\mathbf{p}$ for a given cosmological model from
\begin{equation}
\label{eq:gz}
   D_C(z,\mathbf{p}) = c\int^z_0 \frac{dz'}{H(z',\mathbf{p})},
\end{equation}
where $H(z,\mathbf{p})$ is the Hubble parameter, and is given below for the six cosmological models we use in this paper. The luminosity distance can be used to compute the luminosity $L(z,\mathbf{p})$, in units of ${\rm erg\,s^{-1}}$ (or the luminosity per frequency in units of ${\rm erg\,s^{-1}Hz^{-1}}$) from the flux density $F$, in units of ${\rm erg\,s^{-1}cm^{-2}}$ (or the flux density per frequency in units of ${\rm erg\,s^{-1}cm^{-2}Hz^{-1}}$), through
\begin{equation}
    L(z,\mathbf{p})=4\pi D_L(z,\mathbf{p})^2 F.
    \label{eq_L}
\end{equation}

Observations indicate that for \Mgii\ QSOs the reverberation measured time delay $\tau$ and $L_{3000}(z,\mathbf{p})$, the monochromatic luminosity at 3000 \AA, obey an $R-L$ relation \citep{2019ApJ...880...46C, Czernyetal2022, Michal2020, Michal2021, Homayouni2020, Mary2020, Zhefu2021, Yuetal2022, khadka2021, Khadkaetal2022, Caoetal2022}
\begin{equation}
\label{eq:corr}
   \log \left({\frac{\tau} {\rm day}}\right) = \beta + \gamma \log\left[{\frac{L_{3000}(z,\mathbf{p})}{10^{44}\,{\rm erg\,s^{-1}}}}\right]\,,
\end{equation}
where $\log$ = $\log_{10}$ and the intercept $\beta$ and the slope $\gamma$ are free parameters to be determined from data concurrently with the cosmological-model parameters $\mathbf{p}$ that influence the $R-L$ relation through $L_{3000}(z,\mathbf{p})$,  see Sec.~\ref{sec:methods}. Using instead the measured flux density $F_{3000}$ and the luminosity distance, eq.~\eqref{eq:corr} can be expressed as
\begin{align}
\label{eq:R-L_DL}
    \log{\left(\frac{\tau}{\rm day}\right)} = & \beta +\gamma\log(4\pi) \\ & + \gamma \left[\log\left(\frac{F_{3000}}{{\rm erg\,s^{-1}cm^{-2}}}\right)-44\right] + 2\gamma \log(D_L), \notag
\end{align}

Observations also indicate that QSO X-ray (2 keV) and UV (2500 \AA) luminosities, $L_X$ and $L_{UV}$ (expressed per frequency), are correlated \citep{Tananbaumetal1979, Zamoranietal1981, AvniTananbaum1986, Steffenetal2006, Justetal2007, Greenetal2009, Youngetal2010, Lussoetal2010, Grupeetal2010, Vagnettietal2010} through the $L_X-L_{UV}$ relation. This relation is
\begin{equation}
\label{eq:xuv}
    \log\left(\frac{L_{X}}{{\rm erg\,s^{-1}Hz^{-1}}} \right) = \beta + \gamma \log\left(\frac{L_{UV}}{10^{29}\,{\rm erg\,s^{-1}Hz^{-1}}}\right),
\end{equation}
where the slope $\gamma$ and the intercept $\beta$ are free parameters [the difference between these free parameters and those of eq.\ (\ref{eq:corr}) should be clear from the context] to be determined from the data concurrently with the cosmological-model parameters, as described in Sec.~\ref{sec:methods}. Luminosities and flux densities are related through the luminosity distance, so eq.\ (\ref{eq:xuv}) can be rewritten as
\begin{align}
\label{eq:xuvflux}
    \log{\left(\frac{F_{X}}{\rm erg\,s^{-1} cm^{-2} Hz^{-1}}\right)} & = \beta +(\gamma - 1)\log(4\pi) + 2(\gamma - 1)\log(D_L) \notag \\ & + \gamma \left[\log\left(\frac{F_{UV}}{{\rm erg\,s^{-1}cm^{-2} Hz^{-1}}}\right)-29\right],
\end{align}
where $F_{UV}$ and $F_X$ are the quasar UV and X-ray flux densities (per frequency), and $L_{UV}=4\pi D_L^2 F_{UV}$ is scaled to $10^{29}\,{\rm erg\,s^{-1} cm^{-2} Hz^{-1}}$, which tightens the constraints for the intercept $\beta$.

For the computation of luminosity distance, the fundamental quantity needed is the Hubble parameter $H(z)$ which is computed using the assumed cosmological model. In what follows we give the functional form of $H(z)$ for each cosmological model we use.

In the $\Lambda$CDM model the Hubble parameter is
\begin{equation}
\label{eq:friedLCDM}
    H(z,\mathbf{p}) = H_0\sqrt{\Omega_{m0}(1+z)^3 + \Omega_{k0}(1+z)^2 + \Omega_{\Lambda}}\,.
\end{equation}
Here $\Omega_{\Lambda}$ is the cosmological constant density parameter and the three energy density parameters obey the current energy budget equation $\Omega_{m0}$ + $\Omega_{k0}$ + $\Omega_{\Lambda}$ = 1.  In the spatially non-flat $\Lambda$CDM model we choose $\Omega_{m0}$, $\Omega_{k0}$, and $H_0$ to be the free parameters. For the spatially-flat $\Lambda$CDM model we choose the same set of free parameters but now set $\Omega_{k0} = 0$ as required for flat spatial hypersurfaces. 

In the XCDM dynamical dark energy parametrization the Hubble parameter is
\begin{equation}
\label{eq:XCDM}
    H(z,\mathbf{p}) = H_0\sqrt{\Omega_{m0}(1+z)^3 + \Omega_{k0}(1+z)^2 + \Omega_{X0}(1+z)^{3(1+\omega_X)}},
\end{equation}
where $\Omega_{X0}$ is the current value of the $X$-fluid dark energy density parameter and, together with $\Omega_{m0}$ and $\Omega_{k0}$, it obeys the current energy budget equation $\Omega_{m0}$ + $\Omega_{k0}$ + $\Omega_{X0}$ = 1. The equation of state parameter of the $X$-fluid $\omega_X = P_X/\rho_X$, where $P_X$ and $\rho_X$ are the pressure and energy density of the $X$-fluid. In the spatially non-flat XCDM parametrization we choose $\Omega_{m0}$, $\Omega_{k0}$, $\omega_X$, and $H_0$ to be the free parameters. For the spatially-flat XCDM parametrization we choose the same set of free parameters but now set $\Omega_{k0} = 0$ as required for flat spatial hypersurfaces. When $\omega_X = -1$ the XCDM parametrization reduces to the $\Lambda$CDM model. 

In the $\phi$CDM model the dynamical dark energy is a scalar field $\phi$ \citep{PeeblesRatra1988, RatraPeebles1988, Pavlovetal2013}.\footnote{For discussions of constraints on the $\phi$CDM model see \citet{Zhaietal2017}, \citet{Oobaetal2018c, Oobaetal2019}, \citet{ParkRatra2018, ParkRatra2019c, ParkRatra2020}, \citet{SolaPercaulaetal2019}, \citet{Singhetal2019}, \citet{UrenaLopezRoy2020}, \citet{SinhaBanerjee2021}, \citet{Xuetal2021}, \citet{deCruzetal2021}, \citet{Jesusetal2021}, \citet{Adiletal2022}, and references therein.} Here we assume that the scalar field potential energy density $V(\phi)$ is an inverse power law of $\phi$ and this potential energy density, defined next, determines the scalar field dark energy density parameter $\Omega_{\phi}(z, \alpha)$. The functional form of $V(\phi)$ we use is
\begin{equation}
\label{eq:phiCDMV}
    V(\phi) = \frac{1}{2}\kappa m_{p}^2 \phi^{-\alpha},
\end{equation}
where $m_{p}$ and $\alpha$ are the Planck mass and a positive parameter respectively, and $\kappa$ is a constant whose value is determined by using the shooting method to guarantee that the current energy budget equation $\Omega_{m0} + \Omega_{k0} + \Omega_{\phi}(z = 0, \alpha) = 1$ holds.

With this potential energy density, coupled differential equations, i.e.\ the scalar field equation of motion and the Friedmann equation, govern the dynamics of $\phi$ and the cosmological scale factor $a$. For a spatially homogeneous scalar field these two coupled equations of motion are 
\begin{align}
\label{eq:field}
   & \ddot{\phi}  + 3\frac{\dot{a}}{a}\dot\phi - \frac{1}{2}\alpha \kappa m_{p}^2 \phi^{-\alpha - 1} = 0, \\
\label{eq:friedpCDM}
   & \left(\frac{\dot{a}}{a}\right)^2 = \frac{8 \pi}{3 m_{p}^2}\left(\rho_m + \rho_{\phi}\right) - \frac{k}{a^2}.
\end{align}
Here an overdot indicates a derivative with respect to time, $k$ is positive, zero, and negative for closed, flat, and open spatial geometry (corresponding to $\Omega_{k0} < 0, =0, {\rm and} >0$), $\rho_m$ is the non-relativistic matter energy density, and the scalar field energy density is given by
\begin{equation}
    \rho_{\phi} = \frac{m^2_p}{32\pi}\left[\dot{\phi}^2 + \kappa m^2_p \phi^{-\alpha}\right].
\end{equation}
The numerical solution of the coupled differential equations (\ref{eq:field}) and (\ref{eq:friedpCDM}) is used to compute $\rho_{\phi}$ and
\begin{equation}
    \Omega_{\phi}(z, \alpha) = \frac{8 \pi \rho_{\phi}}{3 m^2_p H^2_0}.
\end{equation}

The Hubble parameter in the $\phi$CDM model is
\begin{equation}
    H(z,\mathbf{p}) = H_0\sqrt{\Omega_{m0}(1+z)^3 + \Omega_{k0}(1+z)^2 + \Omega_{\phi}\left(z, \alpha\right)}.
\end{equation}
In the spatially non-flat $\phi$CDM model we choose $\Omega_{m0}$, $\Omega_{k0}$, $\alpha$, and $H_0$ to be the free parameters. For the spatially-flat $\phi$CDM model we choose the same set of free parameters but now set $\Omega_{k0} = 0$ as required for flat spatial hypersurfaces. When $\alpha = 0$ the $\phi$CDM model reduces to the $\Lambda$CDM model.

QSO data cannot constrain $H_0$ because there is a degeneracy between the intercept ($\beta$) of the correlation relations and $H_0$, so we set $H_0$ to 70 km s$^{-1}$ Mpc$^{-1}$ in all QSO data-only analyses.

\section{Data description}
\label{sec:data}

 By cross-matching the previously studied 78 reverberation-mapped \Mgii\ QSOs \citep{khadka2021} with the XMM-Newton X-ray source catalog (4XMM\-DR11), we found that 58 of the 78 sources were also detected in the X-ray domain. The X-ray fluxes in the 4XMM\-DR11 catalog are measured at various energy bands and listed in units of erg s$^{-1}$ cm$^{-2}$. For our study, we used spline interpolation to determine the X-ray fluxes at 2 keV energy, i.e. $F_{\rm 2keV}$. In our analysis, we use 2 keV flux densities per unit frequency $F_{X}\equiv F_{\rm 2keV}/\nu_{\rm 2keV}$ (in ${\rm erg\,s^{-1}\,cm^{-2}\,Hz^{-1}}$). For the \Mgii\ QSOs, we had flux densities at 3000 \AA, $F_{3000}$ in ${\rm erg\,s^{-1}cm^{-2}}$, available from the continuum flux determination in the surroundings of the broad \Mgii\ line. To obtain UV flux densities at 2500 \AA\ we used the continuum slope $\alpha_{\nu}=-0.45\pm 0.01$ of the mean quasar spectrum from \citet{2001AJ....122..549V}, specifically $F_{UV}=F_{3000,\nu}(2500/3000)^{-(\alpha_{\nu}+1)}$, where $F_{3000,\nu} = F_{3000}/\nu_{3000}$ is the continuum flux density per unit frequency at 3000 \AA\ in ${\rm erg\,s^{-1}cm^{-2}Hz^{-1}}$. The continuum slope can take different values depending on sample characteristics such as the redshift range, the radio classification, or the sample size \citep[][]{shull_2012}. Since our sample was selected based only on reverberation mapping measurements and the X-ray detection, the $\alpha_\nu$ value reported by  \citet{2001AJ....122..549V} is more appropriate for our analysis, which, in turn, was obtained from a sample without any particular characteristics. The continuum slope used in our analysis is in a good agreement with the median $\alpha_\nu$ of the values reported in Table~3 of \citet{shull_2012}, which guarantees a correct approximation of $\alpha_{\nu}$.
  
 As a check on our sample and on the X-ray (2 keV) and UV (2500 \AA) flux data points, we computed the $\alpha_{OX}$ parameters \citep{Tananbaumetal1979}
\begin{equation}
\alpha_{OX} = -0.384 \log(F_{UV} /F_{X})\,,
\label{eq_alpha_OX}
\end{equation}
which follows from the power-law approximation of the spectral energy distribution in the form $F\propto \nu^{+\alpha_{OX}}$.
We found that $\alpha_{OX}$ values for all sources in our sample are within the limits provided in \citet{1994AJ....108..374B} and \citet{2021ApJ...908...53W} (also see Table~\ref{tab_xray_uv_data}), so none of the sources appears to be heavily obscured. More specifically, we have 58 pairs of monochromatic X-ray and UV flux densities per unit frequency ($F_X$, $F_{UV}$) measurements and 59 pairs of time-delay and monochromatic 3000 \AA\ flux density ($\tau$, $F_{3000}$) measurements since the \Mgii\ line-emission time delay of NGC4151 was measured twice, in 1988 and 1991 \citep{2006ApJ...647..901M}, see Table~\ref{tab_xray_uv_data}. 

\begin{figure}
    \centering
    \includegraphics[scale=0.5]{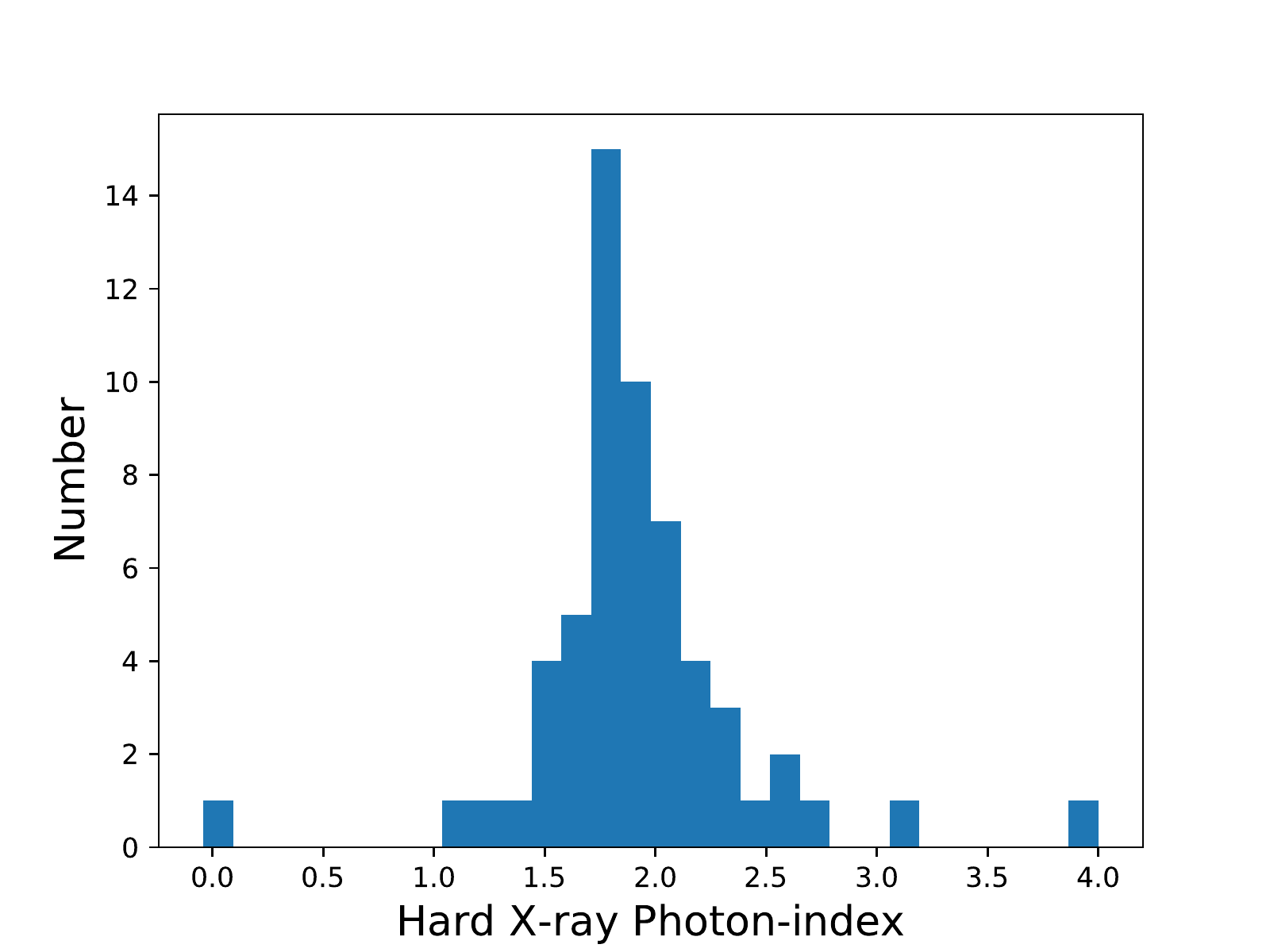}
    \includegraphics[scale=0.5]{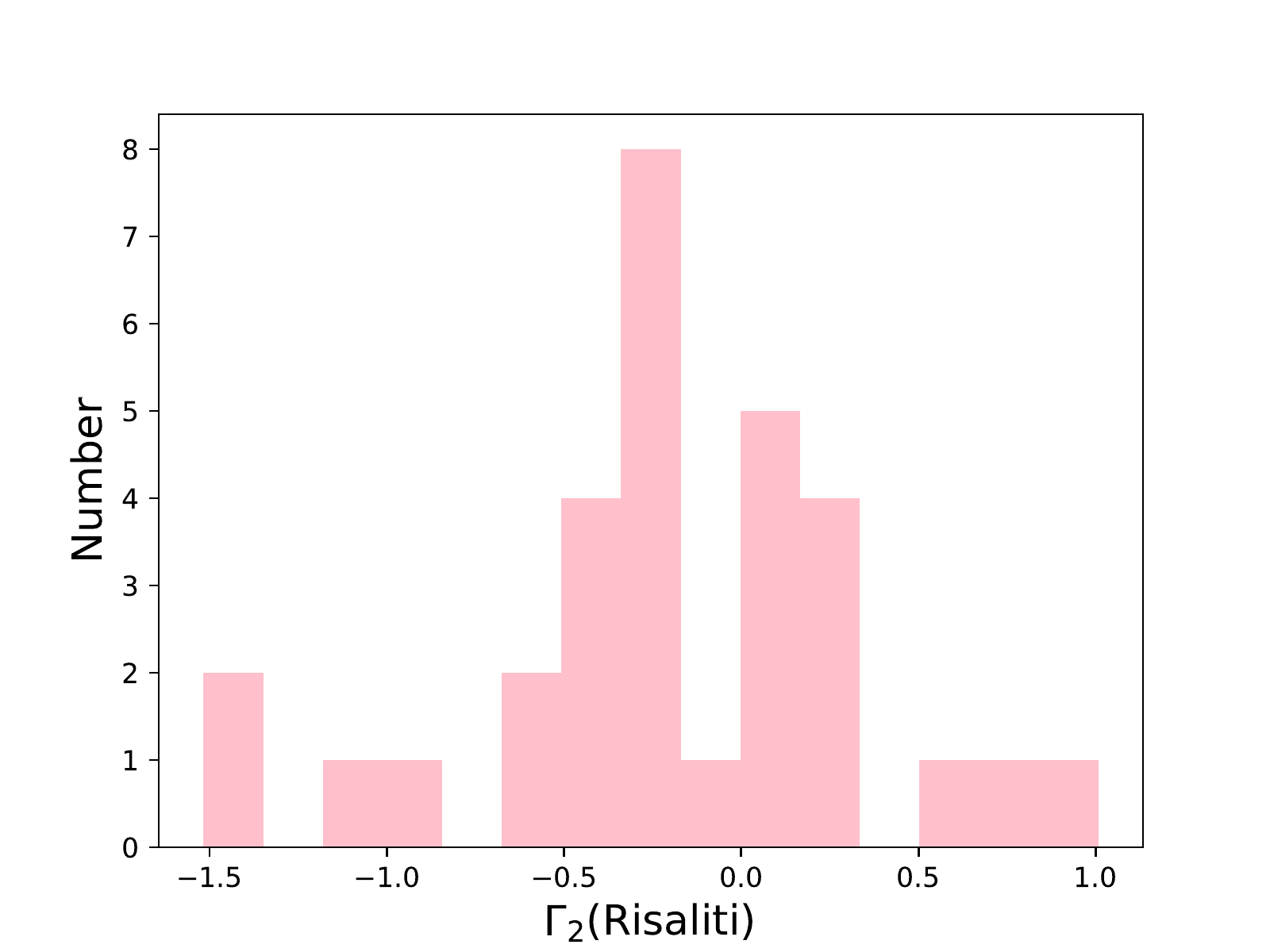}
    \caption{Top:Hard X-ray photon index derived from the fluxes collected from the XMM catalog in different X-ray bands. Bottom: UV spectral index estimated for 31 sources which satisfy the criteria of \citet{Lussoetal2020}, i.e.\ that the UV spectral index should lie between $-0.7$ and 1.5.}
    \label{fig:Xray_slope}
\end{figure}

\citet{Lussoetal2020} recommended a pre-selection of the sources based on criteria concerning the UV, optical/IR and X-ray slopes. We did not apply such criteria since our starting sample is already very small in comparison with their initial sample and further sample reduction is not a viable option. But we performed additional tests, apart from $\alpha_{ox}$, following the recommendations of \citet{Lussoetal2020}. For all 58 objects, having the X-ray fluxes in several X-ray bands from 4XMMDR11 we fitted a broken power law to the data, fixing the frequency break at 1 keV rest frame, and obtained best fits for the slopes. The histogram of the hard X-ray slopes is shown in Fig.~\ref{fig:Xray_slope}. Out of 58 sources, 41 satisfy the criterion that the photon index should be between 1.7 and 2.8. In the whole sample only 3 objects are very strong outliers, two with slopes extremely soft and one (NGC 4151) with a very hard slope indicating strong absorption.

We also collected the GALEX far-UV magnitudes using the GALEX EUV quasar colors catalogue of SDSS QSOs DR14 \citep{VandenBerk_etal_2020}, but data were available only for 31 sources out of 58. Since the GALEX magnitude is a broad-band index, very sensitive to the spectral shape, we performed the estimate of the far-UV slope by using the 3000 \AA~rest frame flux from Table~\ref{tab_xray_uv_data}, assuming an arbitrary slope for a power-law extending from 3000 \AA~rest frame to far-UV,  correcting the spectral shape for Galactic extinction by using the value $A_V$\footnote{The foreground Galactic extinction were collected for each source from the \href{http://ned.ipac.caltech.edu/}{NASA/IPAC Extragalactic Database}.} and the extinction curve from \citet{cardelli1989}, and calculated the predicted GALEX flux by folding the resulting spectrum with the profile of the GALEX far-UV filter.\footnote{The GALEX far-UV transmission curve was obtained from the \href{http://svo2.cab.inta-csic.es/svo/theory/fps3/index.php?id=GALEX/GALEX.FUV\&\&mode=browse\&gname=GALEX\&gname2=GALEX\#filter}{Spanish Virtual Observatory's Filter Profile Service}.} We thus obtained the best fit slope by matching the predicted and the measured magnitudes. \citet{Lussoetal2020} recommended using only sources with spectral index between $-0.7$ and 1.5 in the far-UV. A distribution of UV spectral indices of 31 sources is shown in the bottom panel of Fig.~\ref{fig:Xray_slope}. Among the 31 sources for which we have the data, only 4 do not satisfy this criterion. According to \citet{Lussoetal2020} we should combine the two criteria. This would leave us with only 21 objects.

Finally, \citet{Lussoetal2020} recommend removing the Eddington bias by actually removing all outliers from the expected X-ray flux. This is done assuming a standard relation and comparing the predicted 2 keV flux with the minimum which can be detected observationally. In a single short XMM exposure they estimate the minimum flux at $4.6 \times 10^{-32}$ erg s$^{-1}$ cm$^{-2}$ Hz$^{-1}$. Our fluxes are frequently lower but the reported fluxes are almost always the results of the multiple exposures, so the errors quoted in Table~\ref{tab_xray_uv_data} are small and we do not believe we have problems with marginal detections. However, the reduction of the sample to 21 objects would undermine the effort of the comparison of the two methods. So we use the whole sample, and the issue can be addressed in the future when many more AGN will have reverberation measured time delays \citep[see e.g.,][and references therein]{2023arXiv230108975C}.

\begin{figure}
    \centering
    \includegraphics[width=\columnwidth]{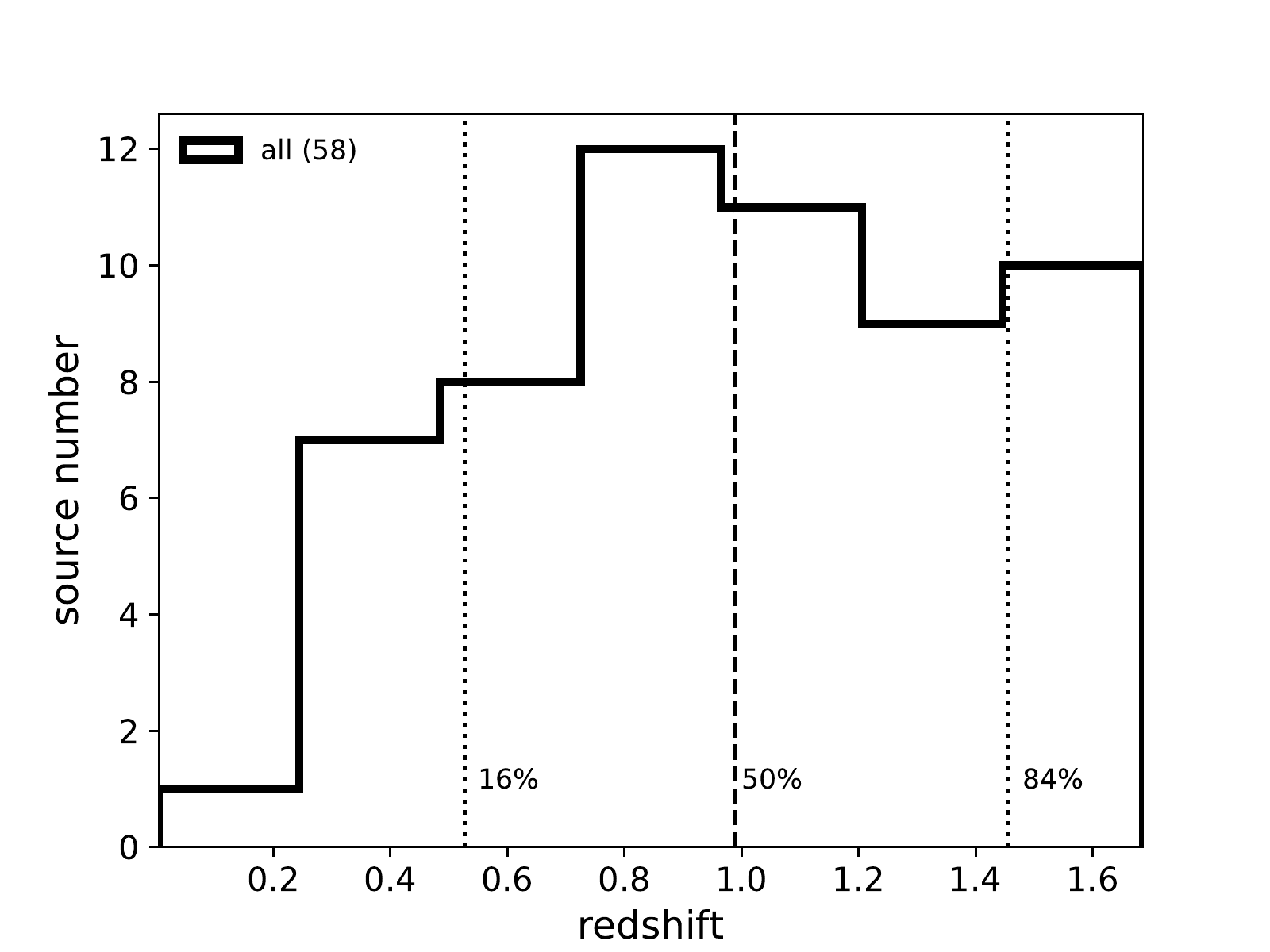}
    \caption{Redshift distribution of the QSO sample (58 sources). The vertical lines mark the percentiles as indicated (16\%, 50\%, 84\%). The bin size is $\Delta z\simeq 0.24$ (7 bins according to Doane's rule).}
    \label{fig_redshift_dist}
\end{figure}

The redshift range of the X-ray/UV and ($\tau$, $F_{3000}$) sample is $0.0041\leq z \leq 1.686$, with median value of $0.990$. Except for NGC4151, peculiar velocities can be neglected. For NGC4151, the original $z=0.0033$ was corrected to $z=0.0041$ after accounting for the peculiar velocity effect. The 16\%-percentile (25\%-percentile) value of redshift is $0.527$ ( 0.716), while the $84\%$-percentile (75\%-percentile) value is $1.454$ (1.270). The redshift distribution is displayed in Fig.~\ref{fig_redshift_dist} with bin size $\Delta z\simeq 0.24$ (7 bins) determined using Doane's rule, which takes into account the distribution skewness.

\begin{figure}
    \centering
    \includegraphics[width=\columnwidth]{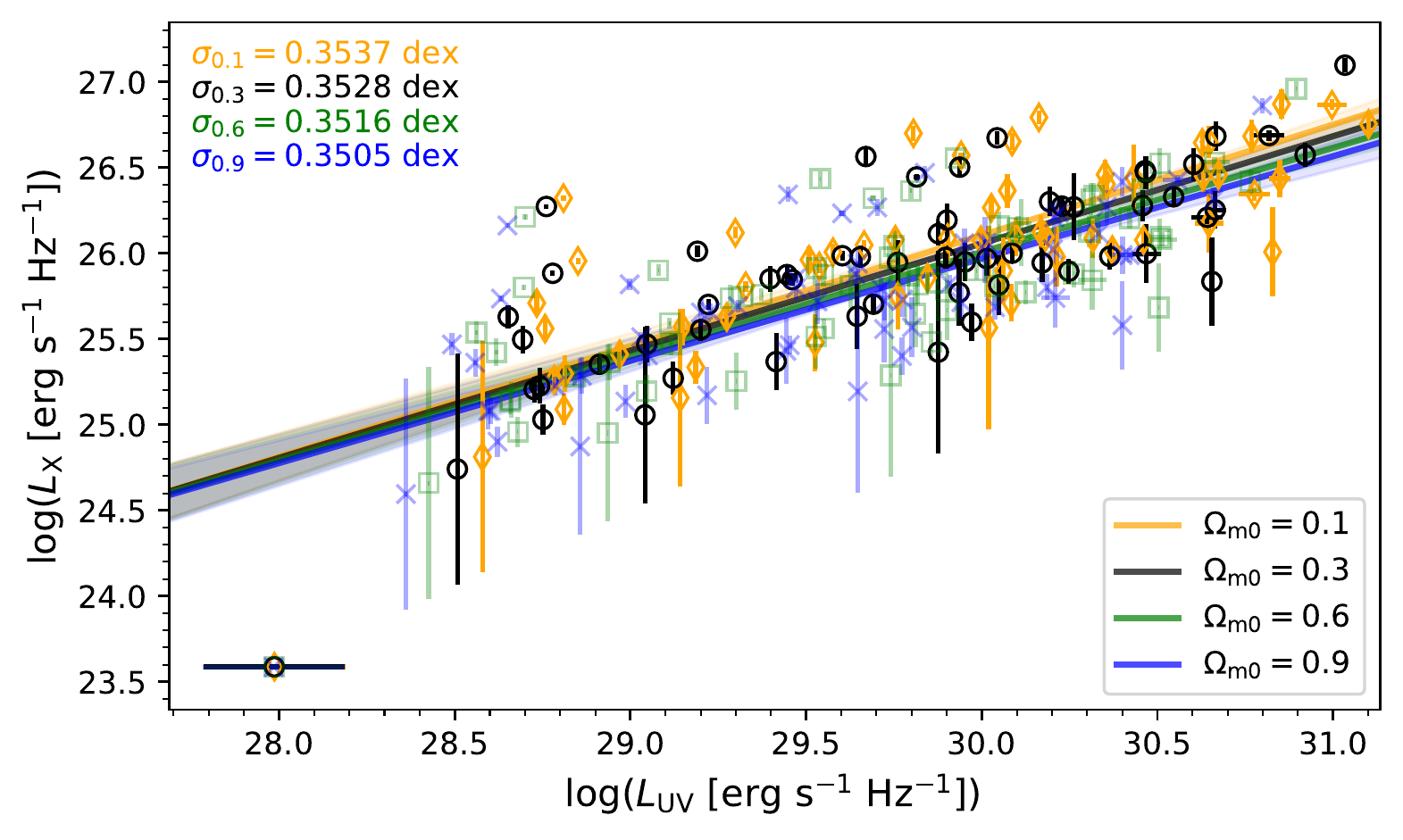}   \caption{The correlation between 2 keV X-ray and 2500\,\AA\ UV luminosities (in ${\rm erg\,s^{-1}Hz^{-1}}$ and UV luminosities scaled to $10^{29}\,{\rm erg\,s^{-1}Hz^{-1}}$) for 58 sources. These are computed for a fixed flat $\Lambda$CDM cosmological model ($H_0=70\,{\rm km\,s^{-1}\,Mpc^{-1}}$) for four different values of $\Omega_{m0}$ (0.1, 0.3, 0.6, 0.9). The black solid line is the best-fit linear relation for $\Omega_{m0}=0.3$. The gray shaded area indicates $1\sigma$ uncertainties of the fit. The intrinsic RMS scatter of data points around the best-fit relation is $\sigma_{\rm ext}=0.35$ dex. The best-fit relations for $\Omega_{\rm m0}=0.1$, $\Omega_{m0}=0.6$, and $\Omega_{m0}=0.9$ are shown in orange, green, and blue, respectively.}
    \label{fig_LxLuv_flatLCDM}
\end{figure}

For the cross-matched UV/X-ray sample, we determine the correlation between X-ray and UV monochromatic luminosities at 2 keV and 2500 \AA, respectively, in the fixed flat $\Lambda$CDM model (with $H_0=70\,{\rm km\,s^{-1}\,Mpc^{-1}}$ and $\Omega_{m0}=0.3$). The correlation is positive and significant, with Pearson correlation coefficient $r=0.78$ ($p=6.54\times 10^{-13}$) and Spearman rank-order correlation coefficient $\rho=0.74$ ($p=2.90\times 10^{-11}$). When we determine the correlation coefficients from the smaller to the larger $\Omega_{m0}$, we obtain $r=0.80$ ($p=3.82\times 10^{-14}$) and $\rho=0.78$ ($p=8.31\times 10^{-13}$), $r=0.76$ ($p=5.45\times 10^{-12}$) and $\rho=0.72$ ($p=2.63\times 10^{-10}$), $r=0.75$ ($p=1.91\times 10^{-11}$) and $\rho=0.70$ ($p=7.36\times 10^{-10}$) for $\Omega_{\rm m0}=0.1$, $0.6$, and $0.9$, respectively. Hence, the correlation slightly decreases for larger values of $\Omega_{m0}$. We fit a linear function to the $L_X-L_{UV}$ distribution for 58 sources, neglecting $L_{\rm X}$ and $L_{\rm UV}$ uncertainties. Looking for the solution with the smallest RMS scatter we obtain, 
\begin{align}
\log L_{X} &= (0.65 \pm 0.06)\log L_{29} + (25.46 \pm 0.07),\  \Omega_{m0}=0.1,\notag\\
\log L_{X} &= (0.63 \pm 0.07)\log L_{29} + (25.43 \pm 0.07),\  \Omega_{m0}=0.3,\notag\\
\log L_{X} &= (0.61 \pm 0.07)\log L_{29} + (25.40 \pm 0.07),\  \Omega_{m0}=0.6,\notag\\
\log L_{X} &= (0.60 \pm 0.07)\log L_{29} + (25.37 \pm 0.06),\ \Omega_{m0}=0.9\,,
\label{eq_Lx_Luv}
\end{align}
where $L_{29}\equiv L_{UV}/10^{29}\,{\rm erg\,s^{-1}\,Hz^{-1}}$. The global RMS scatter around the best-fit relation is $\sigma_{\rm ext} \simeq 0.35$ dex. The X-ray/UV correlation for 58 data points and the linear fit with 1$\sigma$ uncertainties in eq.~\eqref{eq_Lx_Luv} are shown in Fig.~\ref{fig_LxLuv_flatLCDM}. The best-fit coefficients in eq.~\eqref{eq_Lx_Luv} are overall consistent within $1\sigma$ uncertainties with the best-fit relation of \citet{2016ApJ...819..154L}, their Fig.\ 3, with $\log{L_{\rm X}}=0.642^{+0.015}_{-0.005}\log{L_{\rm UV}}+6.965^{+0.461}_{-0.465}$ (for $L_{29}$ normalization, their mean value of the intercept would be $\beta=25.583$), with their dispersion being $\sigma_{\rm ext} = 0.24$. High redshift quasars give the same slope, $0.60 \pm 0.02$ as lower redshift sources \citep{sacchi2022}. These results are also consistent with the more accurate results we derive from a joint determination of the $L_X - L_{UV}$ relation parameters and the cosmological model parameters, discussed in Sec.~\ref{sec:results} below. In Fig.~\ref{fig_LxLuv_flatLCDM} we compare the eq.~\eqref{eq_Lx_Luv} $L_{X}-L_{UV}$ relations for four different values of $\Omega_{m0}$ (0.1, 0.3, 0.6, 0.9) using orange, black, green, and blue lines and points, respectively. The best-fit coefficients are overall consistent within $1\sigma$ uncertainties, with a mild indication of slope and intercept decrease as well as intrinsic RMS scatter decrease for higher values of $\Omega_{m0}$.

\begin{figure}
    \centering
    \includegraphics[width=\columnwidth]{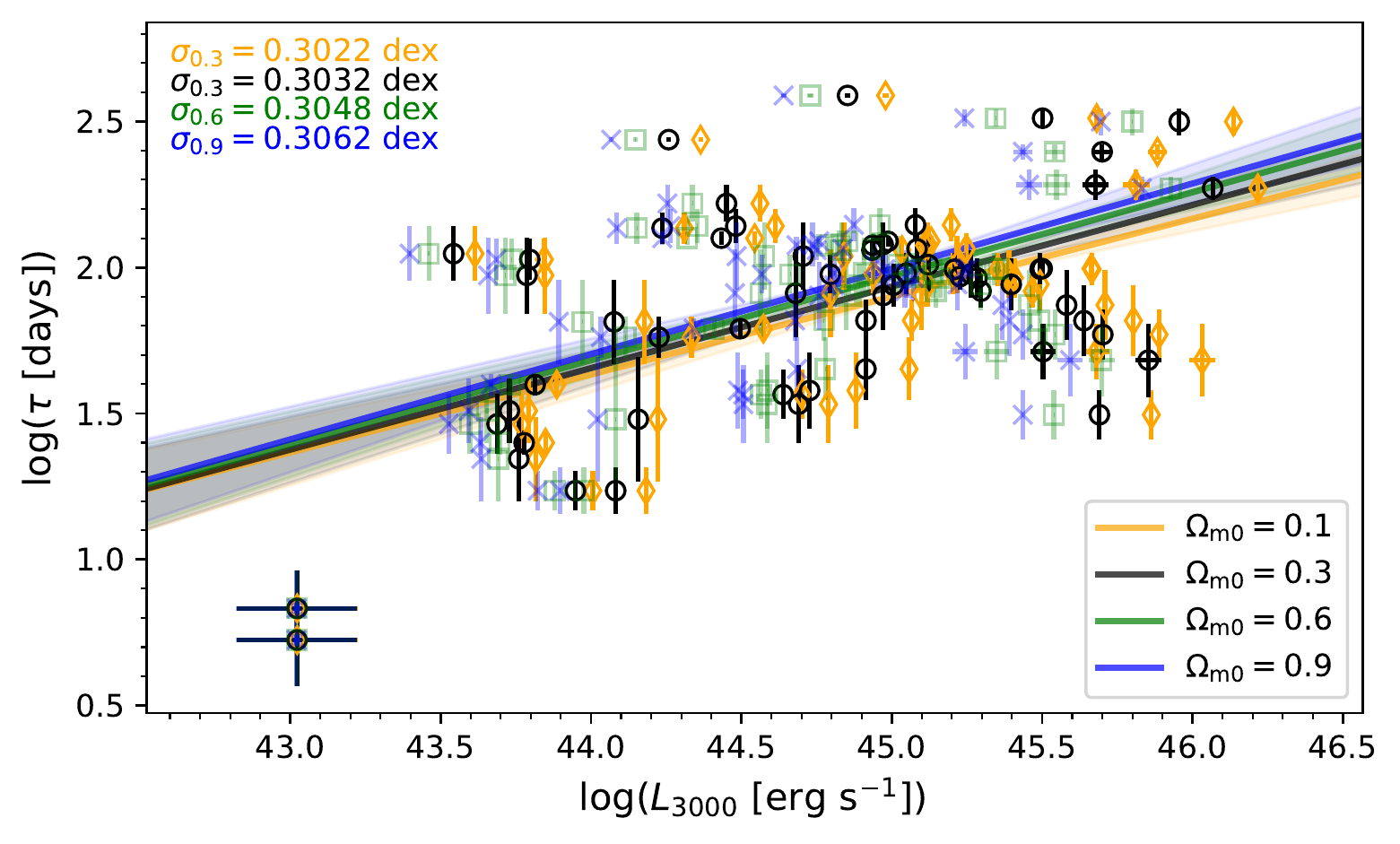}
    \caption{The correlation between rest-frame \Mgii\ time-delay (in days) and monochromatic luminosity at 3000 \AA\ (in ${\rm erg\,s^{-1}}$ and scaled to $10^{44}\,{\rm erg\,s^{-1}}$) for 59 sources. These are computed for a fixed flat $\Lambda$CDM model with $H_0=70\,{\rm km\,s^{-1}\,Mpc^{-3}}$ and four values of $\Omega_{m0}$ (0.1, 0.3,0.6,0.9). The best-fit relation for $\Omega_{m0}=0.3$ is depicted by the black solid line, with the $1\sigma$ uncertainty range shown by the shaded gray region. The 59 data points have an intrinsic RMS scatter of $\sim 0.30$ dex around the best-fit relation. For comparison, we also show $\tau - L_{3000}$ correlations and their best-fit relations for $\Omega_{\rm m0}=0.1$, $\Omega_{m0}=0.6$, and $\Omega_{m0}=0.9$ in orange, green, and blue, respectively.}
    \label{fig_RL_flatLCDM}
\end{figure}

The second correlation expected for the cross-matched data set is the relation between the UV monochromatic luminosity at 3000 \AA\, and the rest-frame \Mgii\ broad-line time delay, the $R-L$ relation. This correlation is again positive and significant, with Pearson correlation coefficient $r=0.56$ ($p=4.42\times 10^{-6}$) and Spearman rank-order coefficient $\rho=0.39$ ($p=0.0024$) for flat $\Lambda$CDM cosmology with $H_0=70\,{\rm km\,s^{-1}\,Mpc^{-1}}$ and $\Omega_{m0}=0.3$. When going from the smaller to the higher $\Omega_{m0}$, one obtains $r=0.56$ ($p=3.63\times 10^{-6}$) and $\rho=0.39$ ($p=0.0024$), $r=0.55$ ($p=6.05\times 10^{-6}$) and $\rho=0.39$ ($p=0.0022$), and $r=0.55$ ($p=7.93\times 10^{-6}$) and $\rho=0.39$ ($p=0.0024$) for $\Omega_{m0}=0.1$, $0.6$, and $0.9$, respectively. In comparison with the X-ray/UV luminosity relation, the $R-L$ relation is generally weaker and less significant. The correlation and its significance is, however, not significantly affected by different values of $\Omega_{m0}$. When we infer the monochromatic luminosities at 3000 \AA\, using the fixed flat $\Lambda$CDM model with the same fixed parameters as before ($\Omega_{m0}=0.1, 0.3, 0.6, 0.9$ and $H_0=70\,{\rm km\,s^{-1}\,Mpc^{-1}}$), we obtain the best-fit $R-L$ relations for the 59 sources (neglecting time-delay and luminosity uncertainties),
\begin{align}
 \log{\tau}&=(0.27 \pm 0.05)\log{L_{44}}+(1.63 \pm 0.06),\ \Omega_{m0}=0.1, \notag\\
   \log{\tau}&=(0.28 \pm 0.06)\log{L_{44}}+(1.66 \pm 0.06),\ \Omega_{m0}=0.3, \notag\\
    \log{\tau}&=(0.29 \pm 0.06)\log{L_{44}}+(1.68 \pm 0.06),\ \Omega_{m0}=0.6, \notag\\
     \log{\tau}&=(0.29 \pm 0.06)\log{L_{44}}+(1.70 \pm 0.05),\ \Omega_{m0}=0.9, 
    \label{eq_RL}
\end{align}
where $L_{44}\equiv L_{3000}/10^{44}\,{\rm erg\,s^{-1}}$ is the monochromatic luminosity at 3000 \AA, $L_{3000}$, scaled to $10^{44}\,{\rm erg\,s^{-1}}$. The relations in eq.~\eqref{eq_RL} with 1$\sigma$ uncertainties are depicted in Fig.~\ref{fig_RL_flatLCDM} alongside 59 data points, which have $\sigma_{\rm ext} = 0.30$ dex global RMS scatter around the best-fit relation. The best-fit coefficients in eq.~\eqref{eq_RL} derived based on 59 \Mgii\ QSOs are, within uncertainties, consistent with the previously inferred slopes and intercepts for slightly larger \Mgii\ data sets \citep{Mary2020, khadka2021, Khadkaetal2022, 2022Prince, Caoetal2022}. These results are also consistent with the more accurate results we derive from a joint determination of the $R - L$ relation parameters and the cosmological model parameters, discussed in Sec.~\ref{sec:results} below. In Fig.~\ref{fig_RL_flatLCDM} we compare eq.~\eqref{eq_RL} $R-L$ relations for different values of $\Omega_{m0}$ (0.1, 0.3, 0.6, 0.9). While the best-fit coefficients are overall consistent within 1$\sigma$ uncertainties, we observe a mild indication of slope and intercept increasing as well as intrinsic RMS scatter increasing for higher values of $\Omega_{m0}$, i.e. the opposite trend in comparison with $L_{X} - L_{UV}$ relation for the slope and the intrinsic scatter.

Here we also use 11 BAO data points given in Table~1 of \cite{KhadkaRatra2021a} and 31 $H(z)$ observations data points give in Table~2 of \cite{Ryanetal2018}. We use the better-established joint BAO+$H(z)$ data set cosmological parameter constraints for comparison with the cosmological parameters constraints determined here using \Mgii\ $R-L$ data and \Mgii\ $L_X-L_{UV}$ data.

\section{Methods}
\label{sec:methods}

We use 58 (59) pairs of QSO measurements to constrain cosmological parameters and $L_X-L_{UV}$ ($R-L$) relation parameters. 

In the case of the $R-L$ relation, for a given model, we predict rest-frame time-delays of QSOs at known redshift $z_i$ using eq.\ (\ref{eq:R-L_DL}) and these predicted time-delays are compared with the corresponding observed time-delays using the likelihood function LF \citep{Dago2005}
\begin{equation}
\label{eq:chi2}
    \ln({\rm LF}) = -\frac{1}{2}\sum^{N}_{i = 1} \left[\frac{[\log(\tau^{\rm obs}_{i}) - \log(\tau^{\rm th}_{i})]^2}{s^2_i} + \ln(2\pi s^2_i)\right].
\end{equation}
Here $\ln$ = $\log_e$, $\tau^{\rm th}_{i}(\mathbf{p})$ and $\tau^{\rm obs}_{i}$ are the predicted and observed time-delays at redshift $z_i$, and $s^2_i = \sigma^2_{\log{\tau_{\rm obs},i}} + \gamma^2 \sigma^2_{\log{F_{3000},i}} + \sigma_{\rm ext}^2$, where $\sigma_{\log{\tau_{\rm obs},i}}$ and $\sigma_{\log{F_{3000},i}}$ are the measurement error on the observed time-delay and the measured flux ($F_{3000}$) respectively, and $\sigma_{\rm ext}$ is the global intrinsic dispersion of the $R-L$ relation.

\begin{table}
	\centering
	\caption{Summary of the non-zero flat prior parameter ranges.}
	\label{tab:prior}
	\begin{threeparttable}
	\begin{tabular}{l|c}
	\hline
	Parameter & Prior range \\
	\hline
	$\Omega_bh^2$ & $[0, 1]$ \\
	$\Omega_ch^2$ & $[0, 1]$ \\
    $\Omega_{m0}$ & $[0, 1]$ \\
    $\Omega_{k0}$ & $[-2, 2]$ \\
    $\omega_{X}$ & $[-5, 0.33]$ \\
    $\alpha$ & $[0, 10]$ \\
    $\sigma_{\rm ext}$ & $[0, 5]$ \\
    $\beta$ & \begin{tabular}{c}$[0, 10]$ for the $R-L$ relation \\and $[0,50]$ for the $L_X-L_{UV}$ relation.\end{tabular}\\
    $\gamma$ & $[0, 5]$ \\
	\hline
	\end{tabular}
	\begin{tablenotes}[flushleft]
\item Here $\Omega_b$ and $\Omega_c$ are the current values of the non-relativistic baryonic and CDM density parameters and $h$ is $H_0$ in units of 100 km s$^{-1}$ Mpc$^{-1}$. We use $\Omega_b$ and $\Omega_c$ as free parameters in the BAO data analyses, instead of $\Omega_{m0} = \Omega_b + \Omega_c$.
\end{tablenotes}
    \end{threeparttable}
\end{table}

Similarly, in the case of the $L_X-L_{UV}$ relation, in a given model, we can predict X-ray fluxes of QSOs at known redshift $z_i$ using eq. (\ref{eq:xuvflux}) and these fluxes are compared with the corresponding observed X-ray fluxes using the likelihood function LF \citep{Dago2005}
\begin{equation}
\label{eq:chi22}
    \ln({\rm LF}) = -\frac{1}{2}\sum^{N}_{i = 1} \left[\frac{[\log(F^{\rm obs}_{X,i}) - \log(F^{\rm th}_{X,i})]^2}{s^2_i} + \ln(2\pi s^2_i)\right],
\end{equation}
with $s^2_i = \sigma^2_{\log{F_{UV}}} + \gamma^2 \sigma^2_{\log{F_{UV},i}} +  \sigma_{\rm ext}^2$, where $\sigma^2_{\log{F_{UV}}}$ and $\sigma^2_{\log{F_{UV},i}}$ are the measurement error on the observed X-ray and UV fluxes respectively, and $\sigma_{\rm ext}$ is the global intrinsic dispersion of the $L_X - L_{UV}$ relation. $F^{\rm th}_{X,i}(\mathbf{p})$ is the predicted flux at redshift $z_i$.

We maximize the likelihood functions given in eqs. (\ref{eq:chi2}) and (\ref{eq:chi22}) by performing Markov chain Monte Carlo (MCMC) sampling implemented in the \textsc{MontePython} code \citep{Brinckmann2019}. We use flat priors on each free parameter involved in the computation and the priors we use are given in Table \ref{tab:prior}. The convergence of each chain corresponding to free parameters is confirmed by satisfying the Gelman-Rubin convergence criterion, $R-1 < 0.05$. Each chain obtained from the MCMC sampling is analysed using the \textsc{Python} package \textsc{Getdist} \citep{Lewis_2019}.

We compare the performance of the $R-L$ and the $L_X-L_{UV}$ relations by computing the Akaike and the Bayesian information criterion ($AIC$ and $BIC$) values for each analysis. The $AIC$ and the $BIC$ values are given by
\begin{align}
\label{eq:AIC}
    AIC =& \chi^2_{\rm min} + 2d,\\
\label{eq:BIC}
    BIC =& \chi^2_{\rm min} + d\ln{N}\, ,
\end{align}
where $\chi^2_{\rm min} = -2 \ln({\rm LF}_{\rm max})$. Here $N$ is the number of data points, $d$ is the number of free parameters, and the number of degrees of freedom is $dof = N - d$.

\begin{sidewaystable*}
\centering
\resizebox*{\columnwidth}{0.75\columnwidth}{%
\begin{threeparttable}
\caption{Unmarginalized best-fit parameters from all data sets.}\label{tab:um_BFP}
\begin{tabular}{lcccccccccccc}
\toprule
Model & Data set & $\Omega_{m0}$ & $\Omega_{k0}$ & $\omega_{X}$ & $\alpha$  & $\sigma_{\rm ext}$ & $\beta$\tnote{a} & $\gamma$\tnote{b} & $dof$ & $-$2$\ln(LF_{\rm max})$ & $AIC$ & $BIC$\\
\midrule
Flat $\Lambda$CDM & $H(z)$ + BAO & 0.298 & -- & -- & -- & -- & -- & -- & 39 & 23.66 & 29.66 & 34.87 \\
& $R-L$ QSOs  & 0.053 & -- & -- &--& 0.288 & 1.642 & 0.259 & 55 & 25.78 & 33.78 & 42.09\\
& $L_X-L_{UV}$ QSOs  & 0.995 & -- & -- & -- & 0.329 & 25.393 & 0.599 & 54 & 44.10 & 52.10 & 60.34\\

\midrule
Non-flat $\Lambda$CDM & $H(z)$ + BAO  & 0.294 & 0.031 & -- & -- & -- & -- & -- & 38 & 23.60 & 31.60 & 38.55 \\
& $R-L$ QSOs  & 0.315 & $-$0.974 & -- & -- & 0.284 & 1.590 & 0.336 & 54 & 22.80 & 32.80 & 43.19\\
& $L_X-L_{UV}$ QSOs  & 0.986 & 1.987 & -- & -- & 0.329 & 25.355 & 0.587 & 53 & 42.96 & 52.96 & 63.26\\

\midrule
Flat XCDM & $H(z)$ + BAO  & 0.280 & -- & $-$0.691 & -- & -- & -- & -- & 38 & 19.66 & 27.66 & 34.61\\
& $R-L$ QSOs  & 0.002 & -- & $-$4.997 & -- & 0.278 & 1.343 & 0.215  & 54 & 19.76 & 29.76 & 40.15\\
& $L_X-L_{UV}$ QSOs  & 0.103 & -- & 0.139 & -- & 0.329 & 25.385 & 0.580  & 53 & 43.96 & 53.96 & 64.26\\

\midrule
Non-flat XCDM & $H(z)$ + BAO  & 0.291 & $-$0.147 & $-$0.641 & -- & -- & -- & --  & 37 & 18.34 & 28.34 & 37.03\\
& $R-L$ QSOs  & 0.052 & $-$0.101 & $-$3.055 & -- & 0.281 & 1.402 & 0.281  & 53 & 21.44 & 33.44 & 45.91\\
& $L_X-L_{UV}$ QSOs  & 0.952 & 1.952 & $-$4.840 & -- & 0.329 & 25.290 & 0.591  & 52 & 41.48 & 53.48 & 65.84\\

\midrule
Flat $\phi$CDM & $H(z)$ + BAO  & 0.265 & -- & -- & 1.445 & -- & -- & --  & 38 & 19.56 & 27.56 & 34.51\\
& $R-L$ QSOs & 0.066 & -- & -- & 0.003 & 0.292 & 1.640 & 0.264 & 54 & 25.8 & 35.8 & 46.19\\
& $L_X-L_{UV}$ QSOs & 0.997 & -- & -- & 9.105 & 0.331 &  25.399 & 0.596 & 53 & 44.08 & 54.08 & 64.38\\

\midrule
Non-flat $\phi$CDM & $H(z)$ + BAO & 0.261 & $-$0.155 & -- & 2.042 & -- & -- & -- & 37 & 18.16 & 28.16 & 36.85 \\
& $R-L$ QSOs  & 0.126 & $-$0.124 & -- & 0.083 & 0.293 & 1.631 & 0.280 & 53 & 25.8 & 37.8 & 50.27\\
& $L_X-L_{UV}$ QSOs  & 0.970 & $-$0.919 & -- & 9.321 & 0.330 & 25.372 & 0.567 & 52 & 43.94 & 55.94 & 68.30\\
\bottomrule
\end{tabular}
\begin{tablenotes}[flushleft]
\item [a] For $R-L$ data $\beta$ values are for the $R-L$ relation and for $L_X-L_{UV}$ data $\beta$ values are for the $L_X-L_{UV}$ relation.
\item [b] For $R-L$ data $\gamma$ values are for the $R-L$ relation and for $L_X-L_{UV}$ data $\gamma$ values are for the $L_X-L_{UV}$ relation.
\end{tablenotes}
\end{threeparttable}
}
\end{sidewaystable*}

\begin{sidewaystable*}
\centering
\resizebox*{\columnwidth}{0.75\columnwidth}{%
\begin{threeparttable}
\caption{One-dimensional marginalized posterior mean values and uncertainties ($\pm 1\sigma$ error bars or $2\sigma$ limits) of the parameters for all models for all data sets.}\label{tab:1d_BFP}
\begin{tabular}{lccccccccc}
\toprule
Model & Data set & $\Omega_{m0}$ & $\Omega_{k0}$ & $\omega_{X}$ & $\alpha$  & $\sigma_{\rm ext}$ & $\beta$\tnote{a} & $\gamma$\tnote{b} \\
\midrule
Flat $\Lambda$CDM & $H(z)$ + BAO & $0.299^{+0.015}_{-0.017}$ & -- & -- & -- & -- & -- & --  \\
& $R-L$ QSOs  & $< 0.614$ (1$\sigma$) & -- & -- &--& $0.305^{+0.028}_{-0.037}$ & $1.681\pm 0.066$ & $0.276\pm 0.061$ \\
& $L_X-L_{UV}$ QSOs  & $> 0.407$ (1$\sigma$) & -- & -- & -- & $0.350^{+0.033}_{-0.044}$ & $25.436\pm 0.077$ & $0.616\pm 0.074$\\

\midrule
Non-flat $\Lambda$CDM & $H(z)$ + BAO  & $0.292\pm0.023$ & $-0.014\pm0.075$ & -- & -- & -- & -- & -- & --  \\
& $R-L$ QSOs  & $0.54^{+0.370}_{-0.220}$ & $-0.200^{+0.780}_{-1.400}$ & -- & -- & $0.304^{+0.028}_{-0.037}$ & $1.677^{+0.074}_{-0.067}$ & $0.290\pm 0.065$ \\
& $L_X-L_{UV}$ QSOs  & $> 0.412$ (1$\sigma$) & $> -0.998$ (2$\sigma$) & -- & -- & $0.349^{+0.032}_{-0.043}$ & $25.417\pm 0.076$ & $0.609\pm 0.073$ \\

\midrule
Flat XCDM & $H(z)$ + BAO  & $0.282\pm0.021$ & -- & $-0.744^{+0.149}_{-0.097}$ & -- & -- & -- & -- \\
& $R-L$ QSOs  & $<0.513$ (1$\sigma$) & -- & $<-0.390$ (2$\sigma$) & -- & $0.302^{+0.028}_{-0.037}$ & $1.630^{+0.120}_{-0.074}$ & $0.275\pm 0.060$  \\
& $L_X-L_{UV}$ QSOs  & $> 0.475$ (1$\sigma$) & -- & $-2.100^{+2.200}_{-1.100}$ & -- & $0.351^{+0.033}_{-0.044}$ & $25.438^{+0.073}_{-0.087}$ & $0.614\pm 0.075$  \\

\midrule
Non-flat XCDM & $H(z)$ + BAO  & $0.293\pm0.027$ & $-0.120\pm0.130$ & $-0.693^{+0.130}_{-0.077}$ & -- & -- & -- & --  \\
& $R-L$ QSOs  & $0.510\pm 0.270$ & $-0.070^{+0.640}_{-0.970}$ & $-2.400^{+1.900}_{-1.400}$ & -- & $0.303^{+0.028}_{-0.037}$ & $1.650^{+0.110}_{-0.073}$ & $0.289\pm 0.065$  \\
& $L_X-L_{UV}$ QSOs  & $> 0.427$ (1$\sigma$) & $> 0.355$ ($(2\sigma)$) & $-2.300^{+2.300}_{-1.100}$ & -- & $0.348^{+0.033}_{-0.044}$ & $25.394^{+0.075}_{-0.085}$ & $0.608\pm 0.075$  \\

\midrule
Flat $\phi$CDM & $H(z)$ + BAO  & $0.266\pm0.023$ & -- & -- & $1.530^{+0.620}_{-0.850}$ & -- & -- & --  \\
& $R-L$ QSOs & -- & -- & -- & -- & $0.305^{+0.027}_{-0.037}$ & $1.701\pm 0.060$ & $0.280\pm 0.061$ \\
& $L_X-L_{UV}$ QSOs & $0.530^{+0.420}_{-0.200}$ & -- & -- & -- & $0.349^{+0.032}_{-0.043}$ &  $25.419\pm 0.070$ & $0.609\pm 0.073$\\

\midrule
Non-flat $\phi$CDM & $H(z)$ + BAO & $0.271^{+0.024}_{-0.028}$ & $-0.080\pm 0.100$ & -- & $1.660^{+0.670}_{-0.830}$ & -- & -- & --  \\
& $R-L$ QSOs  & $0.460^{+0.180}_{-0.430}$ & $0.060\pm 0.370$ & -- & -- & $0.305^{+0.029}_{-0.037}$ & $1.702\pm 0.060$ & $0.280\pm 0.061$\\
& $L_X-L_{UV}$ QSOs  & -- & $0.030\pm 0.370$ & -- & $5.100^{+2.700}_{-2.700}$ & $0.348^{+0.034}_{-0.044}$ & $25.418\pm 0.070$ & $0.610\pm 0.073$ \\
\bottomrule
\end{tabular}
\begin{tablenotes}[flushleft]
\item [a] For $R-L$ data $\beta$ values are for the $R-L$ relation and for $L_X-L_{UV}$ data $\beta$ values are for the $L_X-L_{UV}$ relation.
\item [b] For $R-L$ data $\gamma$ values are for the $R-L$ relation and for $L_X-L_{UV}$ data $\gamma$ values are for the $L_X-L_{UV}$ relation.
\end{tablenotes}
\end{threeparttable}
}
\end{sidewaystable*}

\begin{figure*}
\begin{multicols}{2}    
    \includegraphics[width=\linewidth]{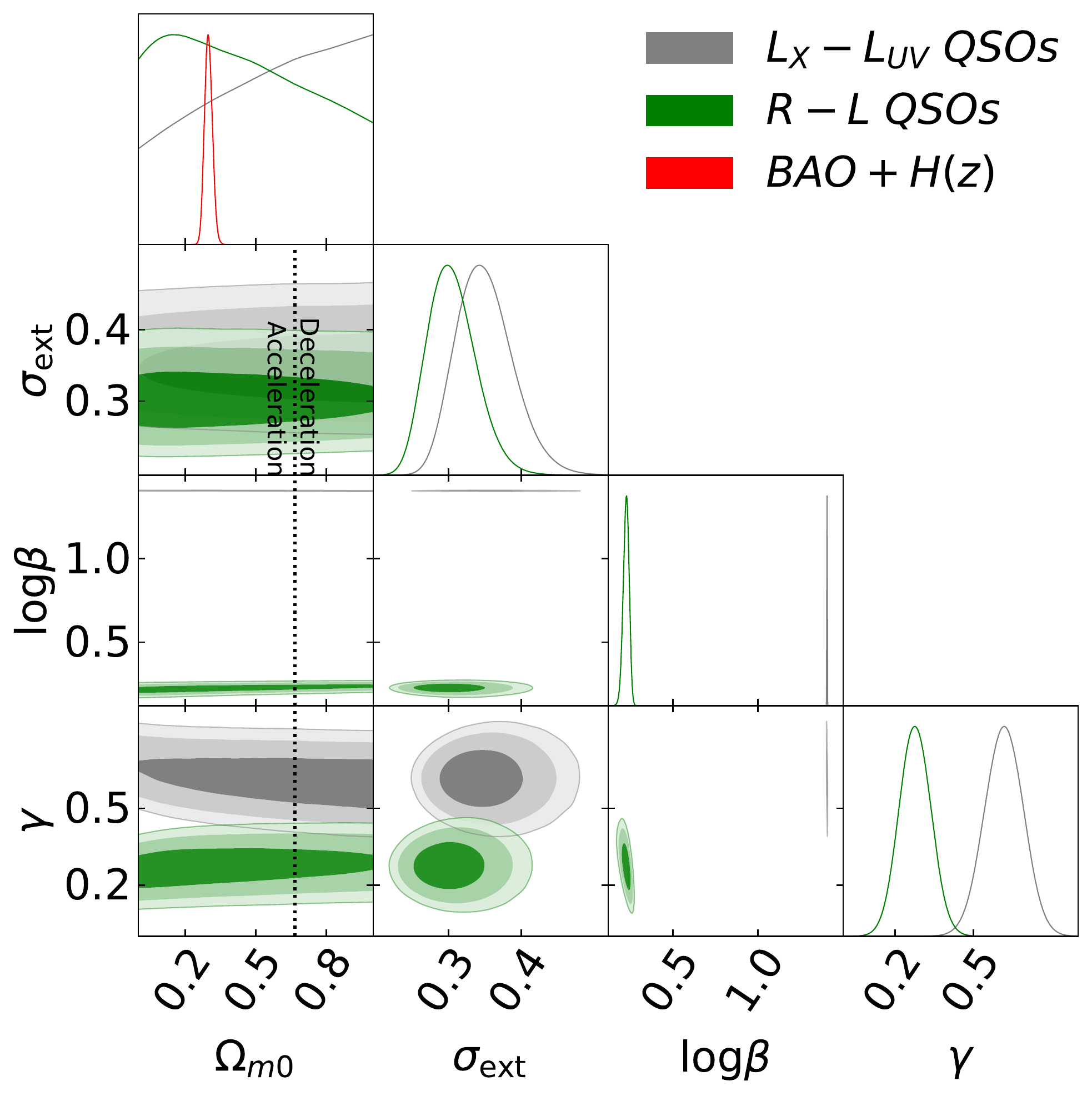}\par
    \includegraphics[width=\linewidth]{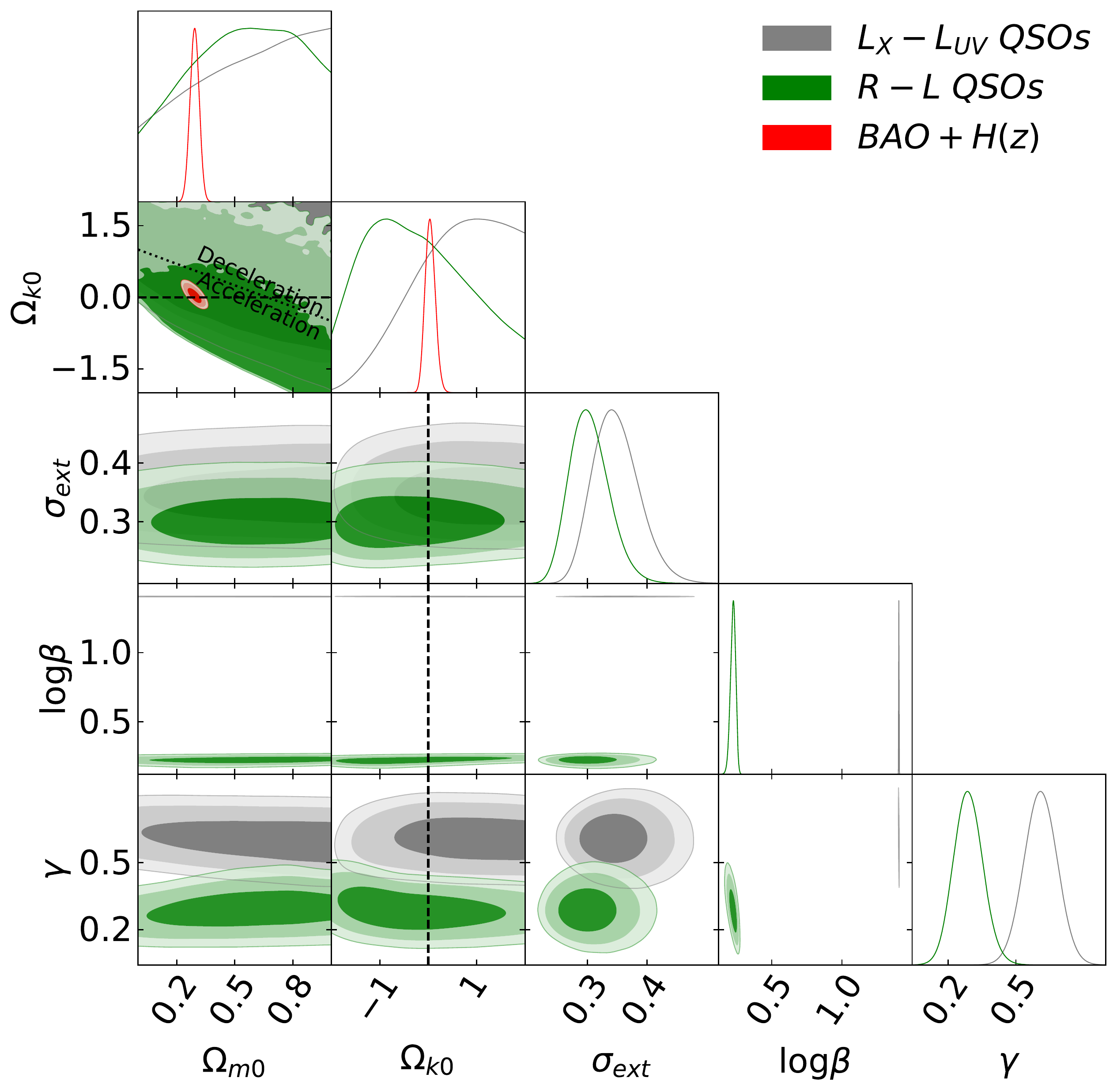}\par
\end{multicols}
\caption{One-dimensional likelihood distributions and two-dimensional contours at 1$\sigma$, 2$\sigma$, and 3$\sigma$ confidence levels using $L_X-L_{UV}$ QSO (gray), $R-L$ QSO (green),  and BAO + $H(z)$ (red) data for all free parameters. Left panel shows the flat $\Lambda$CDM model. The black dotted vertical lines are the zero acceleration lines with currently accelerated cosmological expansion occurring to the left of the lines. Right panel shows the non-flat $\Lambda$CDM model. The black dotted sloping line in the $\Omega_{k0}-\Omega_{m0}$ subpanel is the zero acceleration line with currently accelerated cosmological expansion occurring to the lower left of the line. The black dashed horizontal or vertical line in the $\Omega_{k0}$ subpanels correspond to $\Omega_{k0} = 0$.}
\label{fig:LCDM_triangle}
\end{figure*}

\begin{figure*}
\begin{multicols}{2}    
    \includegraphics[width=\linewidth]{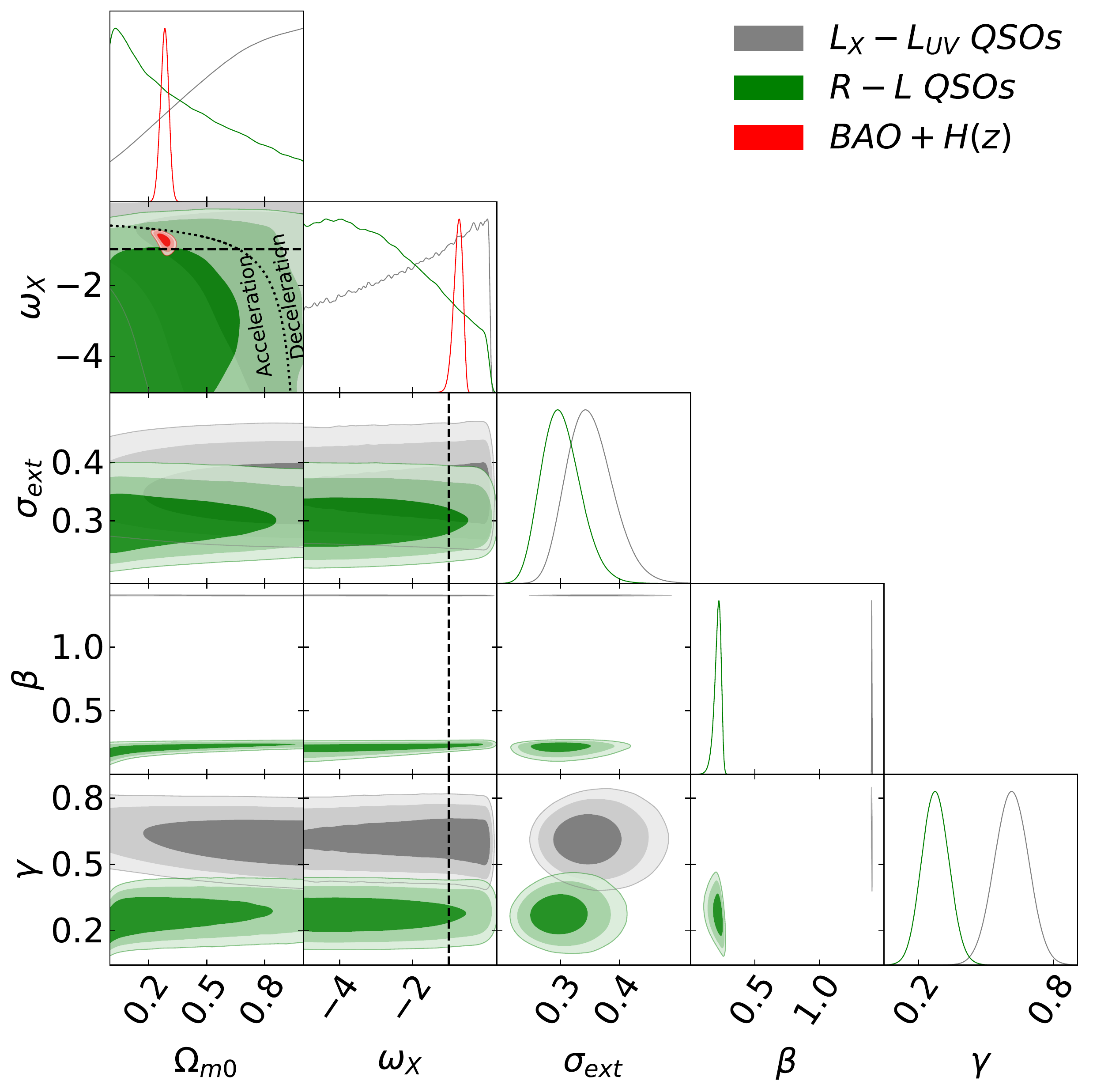}\par
    \includegraphics[width=\linewidth]{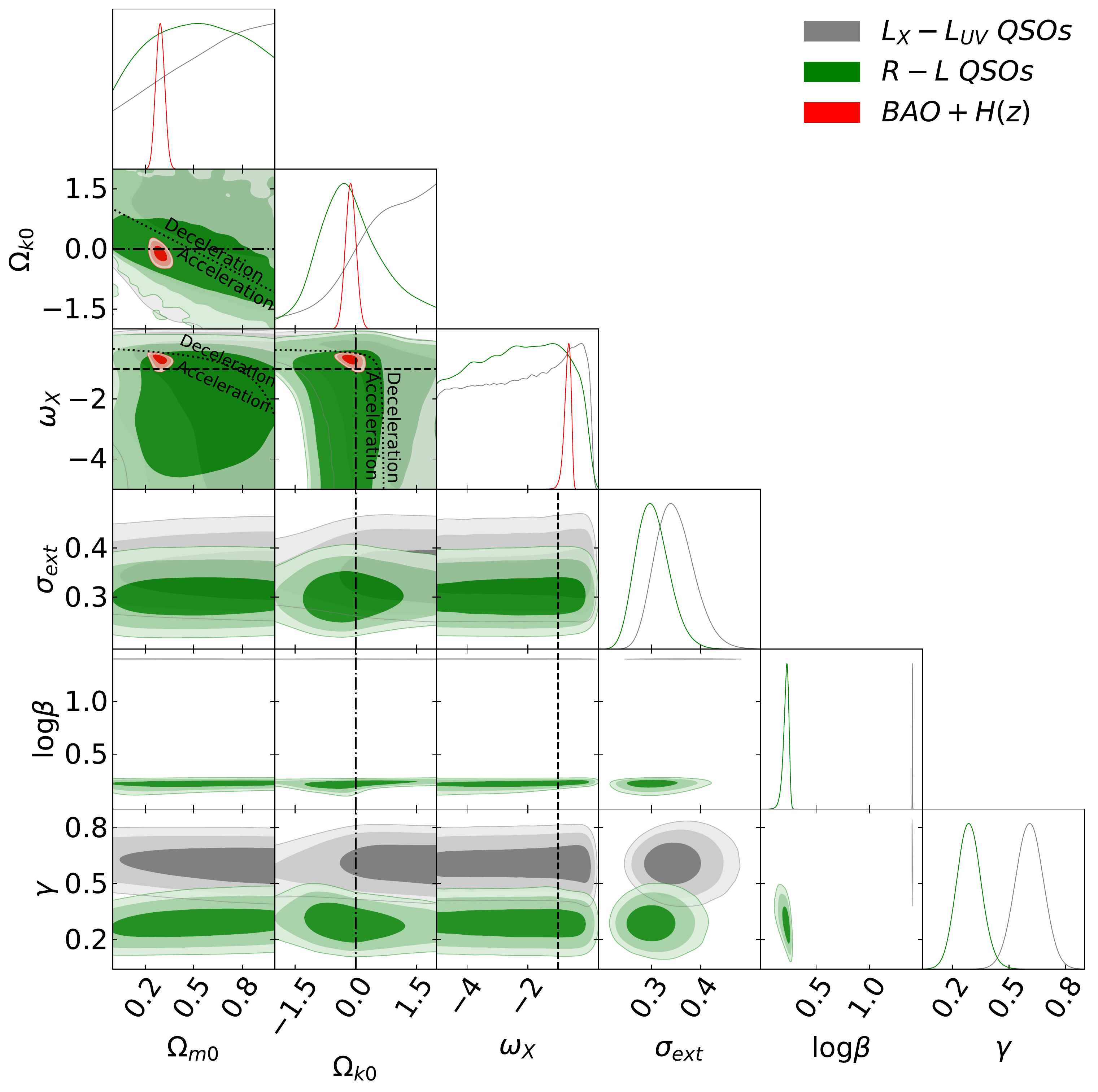}\par
\end{multicols}
\caption{One-dimensional likelihood distributions and two-dimensional contours at 1$\sigma$, 2$\sigma$, and 3$\sigma$ confidence levels using $L_X-L_{UV}$ QSO (gray), $R-L$ QSO (green),  and BAO + $H(z)$ (red) data for all free parameters. Left panel shows the flat XCDM parametrization. The black dotted curved line in the $\omega_X-\Omega_{m0}$ subpanel is the zero acceleration line with currently accelerated cosmological expansion occurring below the line and the black dashed straight lines correspond to the $\omega_X = -1$ $\Lambda$CDM model. Right panel shows the non-flat XCDM parametrization. The black dotted lines in the $\Omega_{k0}-\Omega_{m0}$, $\omega_X-\Omega_{m0}$, and $\omega_X-\Omega_{k0}$ subpanels are the zero acceleration lines with currently accelerated cosmological expansion occurring below the lines. Each of the three lines is computed with the third parameter set to the BAO + $H(z)$ data best-fit value. The black dashed straight lines correspond to the $\omega_x = -1$ $\Lambda$CDM model. The black dotted-dashed straight lines correspond to $\Omega_{k0} = 0$.}
\label{fig:XCDM_triangle}
\end{figure*}

\begin{figure*}
\begin{multicols}{2}    
    \includegraphics[width=\linewidth]{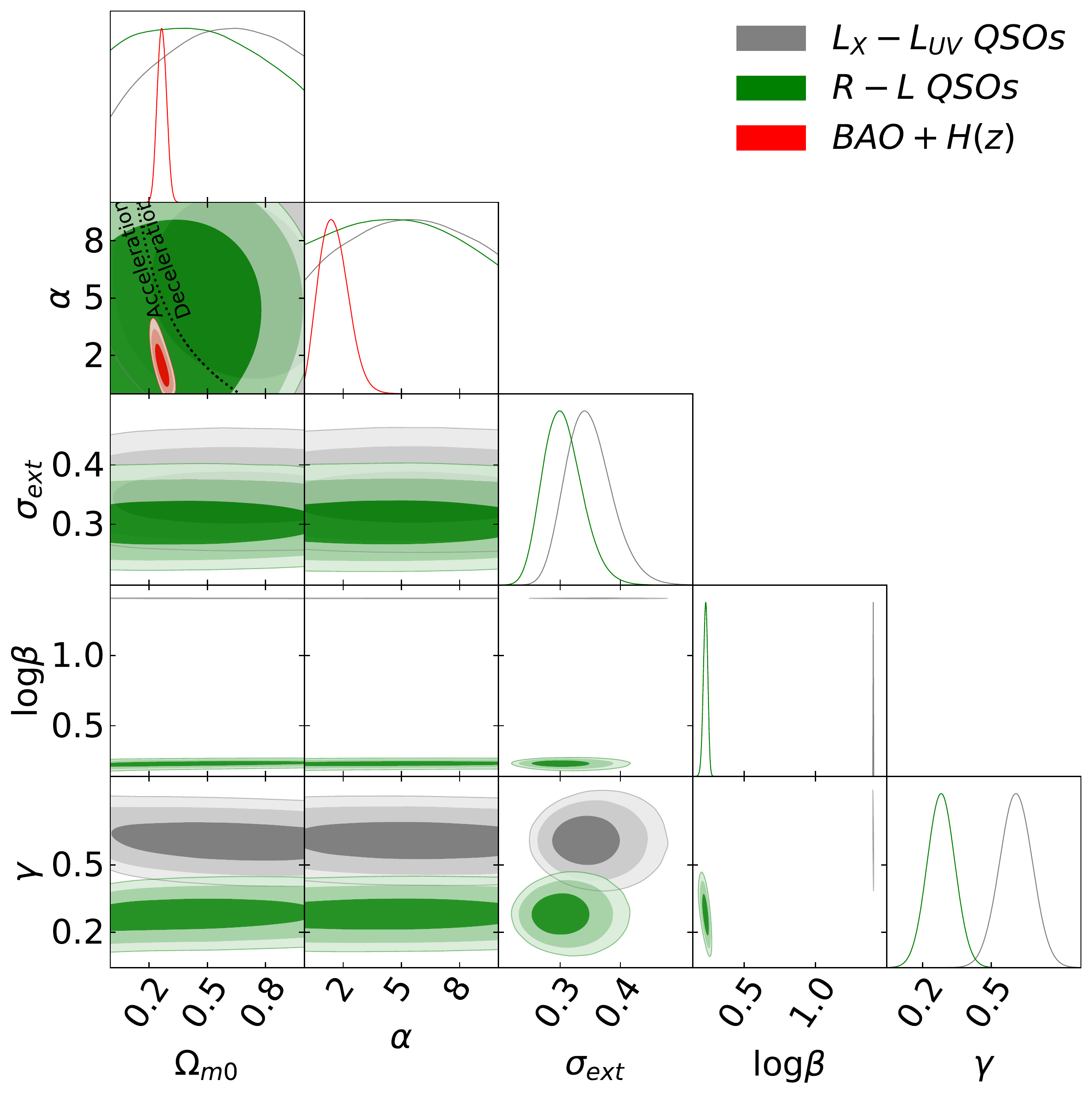}\par
    \includegraphics[width=\linewidth]{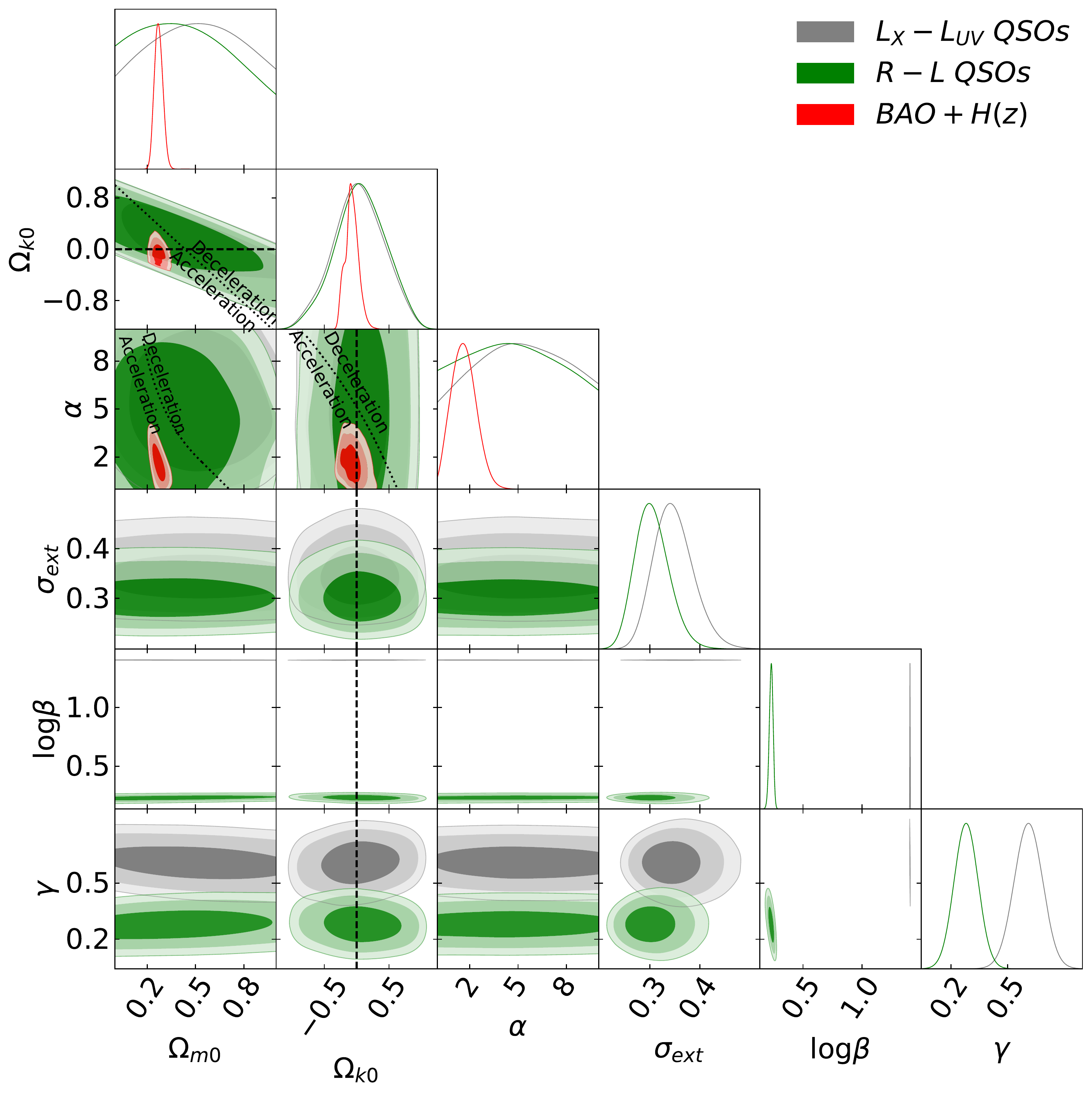}\par
\end{multicols}
\caption{One-dimensional likelihood distributions and two-dimensional contours at 1$\sigma$, 2$\sigma$, and 3$\sigma$ confidence levels using $L_X-L_{UV}$ QSO (gray), $R-L$ QSO (green),  and BAO + $H(z)$ (red) data for all free parameters. The $\alpha = 0$ axes correspond to the $\Lambda$CDM model. Left panel shows the flat $\phi$CDM model. The black dotted curved line in the $\alpha - \Omega_{m0}$ subpanel is the zero acceleration line with currently accelerated cosmological expansion occurring to the left of the line. Right panel shows the non-flat $\phi$CDM model. The black dotted lines in the $\Omega_{k0}-\Omega_{m0}$, $\alpha-\Omega_{m0}$, and $\alpha-\Omega_{k0}$ subpanels are the zero acceleration lines with currently accelerated cosmological expansion occurring below the lines. Each of the three lines is computed with the third parameter set to the BAO + $H(z)$ data best-fit value. The black dashed straight lines correspond to $\Omega_{k0} = 0$.}
\label{fig:phiCDM_triangle}
\end{figure*}

\section{Results}
\label{sec:results}

We have compiled a QSO sample that allows us to obtain cosmological constraints from two independent methods:
\begin{itemize}
\item[(i)] by using the \Mgii\ emission-line time delay with respect to the UV continuum and the resulting $R-L$ relation, and
\item[(ii)] by using the X-ray luminosity -- UV luminosity, $L_X - L_{UV}$, relation.
\end{itemize}
In this section, we present our results and then discuss them in two ways. First, we analyse the cosmological parameter and $R-L$ relation or $L_X - L_{UV}$ relation parameter constraints. We then compare the luminosity distance measurements for each source from these two independent methods.

Our cosmological parameter, $R-L$ relation parameter, and $L_X-L_{UV}$ relation parameter constraint results are listed in Tables \ref{tab:um_BFP} and \ref{tab:1d_BFP}. Unmarginalized best-fit parameters are listed in Table \ref{tab:um_BFP} and one-dimensional marginalized posterior mean values and corresponding uncertainties are listed in Table \ref{tab:1d_BFP}. The one-dimensional likelihoods and two-dimensional confidence contours are plotted in Figs.\ \ref{fig:LCDM_triangle}--\ref{fig:phiCDM_triangle}. $R-L$ QSO data, $L_X-L_{UV}$ QSO data, and BAO + $H(z)$ data results are shown in green, gray, and red. Here we have used the better-established BAO + $H(z)$ data results from \cite{KhadkaRatra2021a} to compare with results obtained from $R-L$ QSO data and $L_X-L_{UV}$ QSO data. For a detailed discussion of the BAO + $H(z)$ results see \cite{KhadkaRatra2021a}.

\subsection{Parameter constraint results from the two methods}

In this paper we use 59 (58) pairs of QSO measurements to simultaneously constrain the cosmological parameters and the $R-L$ ($L_X-L_{UV}$) relation parameters in six different cosmological models. 

For the $R-L$ relation, in all six models, the value of intercept $\beta$ ranges from $1.630^{+0.120}_{-0.074}$ to $1.702\pm0.060$ and the value of slope $\gamma$ ranges from $0.275\pm0.060$  to $0.290\pm0.065$. For the $L_X-L_{UV}$ relation, in all six models, the value of intercept $\beta$ ranges from $25.417\pm0.076$ to $25.438^{+0.073}_{-0.087}$ and the value of slope $\gamma$ ranges from $0.608\pm0.075$  to $0.616\pm0.074$. These $\gamma$ and $\beta$ values are almost completely model independent, indicating that these $R-L$ relation and $L_X-L_{UV}$ relation QSOs are standardizable. While the slope $\gamma$ for the $R-L$ relation is determined to within $\sim 22$\% at 1$\sigma$, for the $L_X-L_{UV}$ relation $\gamma$ is better determined to within $\sim 12$\% at 1$\sigma$. The intercept $\beta$ for the $R-L$ relation is determined to within $\sim 4-6$\% at 1$\sigma$ and for the $L_X-L_{UV}$ relation $\beta$ is also better determined to within $\sim 0.3$\% at 1$\sigma$. 
 
For these QSOs the value of the intrinsic scatter $\sigma_{\rm ext}$ for $R-L$ and $L_X-L_{UV}$ relations are $0.302^{+0.028}_{-0.037}$--$0.305^{+0.029}_{-0.037}$ and $0.348^{+0.033}_{-0.044}$--$0.351^{+0.033}_{-0.044}$, respectively. The slightly larger value of $\sigma_{\rm ext}$ makes the $L_X-L_{UV}$ relation slightly less reliable than the $R-L$ relation. (An increase in the number of sources can decrease the value of $\sigma_{\rm ext}$ but this comparison is only for those sources for which both $R-L$ and $L_X-L_{UV}$ relations are applicable.)

These QSOs provide very weak constraints on cosmological parameters from both the $R-L$ and the $L_X-L_{UV}$ relation. From Table \ref{tab:1d_BFP}, using the $R-L$ relation, in all six cosmological models the value of $\Omega_{m0}$ ranges from $0.460^{+0.180}_{-0.430}$ to $< 0.614$ (1$\sigma$). The minimum and maximum values correspond to the flat $\Lambda$CDM model and the non-flat $\phi$CDM model, respectively. From Table~\ref{tab:1d_BFP}, using the $L_X-L_{UV}$ relation, in all six cosmological models the value of $\Omega_{m0}$ ranges from $> 0.407$ (1$\sigma$) to $0.530^{+0.430}_{-0.200}$. The minimum and maximum values correspond to the flat $\Lambda$CDM model and the flat $\phi$CDM model, respectively. In both cases constraints on $\Omega_{m0}$ are very weak. From the plots shown in Figs.\ \ref{fig:LCDM_triangle}--\ref{fig:phiCDM_triangle} we can see that the $\Omega_{m0}$ distributions corresponding to the $R-L$ relation have a higher probability tendency towards smaller $\Omega_{m0}$ values but $\Omega_{m0}$ distributions corresponding to the $L_X-L_{UV}$ relation show a higher probability tendency towards larger $\Omega_{m0}$ values. This tendency of the $L_X-L_{UV}$ relation is consistent with results presented in \cite{KhadkaRatra2021a}.

From Table \ref{tab:1d_BFP}, for $R-L$ data, in all three non-flat models the value of $\Omega_{k0}$ ranges from $-0.200^{+0.780}_{-1.400}$ to $0.060\pm0.370$. For $L_X-L_{UV}$ data, in all three non-flat models the value of $\Omega_{k0}$ ranges from $> -0.998$ to $0.030\pm0.370$. 

From Table \ref{tab:1d_BFP}, for $R-L$ data, for the flat and non-flat XCDM parametrization the values of the equation of state parameter $\omega_X$ are $< -0.390$ (2$\sigma$) and $-2.400^{+1.900}_{-1.400}$, respectively. For $L_X-L_{UV}$ data, for the flat and non-flat XCDM parametrization the values of $\omega_X$ are $-2.100^{+2.200}_{-1.100}$ and $-2.300^{+2.300}_{-1.100}$, respectively. In both cases, these data provide weak constraints on $\omega_X$ and mostly the posterior value depends on the prior range. From Table~\ref{tab:1d_BFP}, for $L_X-L_{UV}$ data, for the non-flat $\phi$CDM model the value of the positive parameter $\alpha$ is $5.100\pm2.700$. These data are unable to constrain $\alpha$ in all other cases.

We have listed $AIC$ and $BIC$ values for each case in Table \ref{tab:um_BFP}. From this table, for a given cosmological model, we see that the $AIC$ and $BIC$ values are always lower for the $R-L$ relation compared to the values for the $L_X-L_{UV}$ relation. This indicates that the $R-L$ relation better fits $R-L$ data than the $L_X-L_{UV}$ relation fits $L_X-L_{UV}$. This is consistent with the indications from the $\sigma_{\rm ext}$ values discussed above.

\begin{figure*}  
\includegraphics[width=\columnwidth, height=3cm]{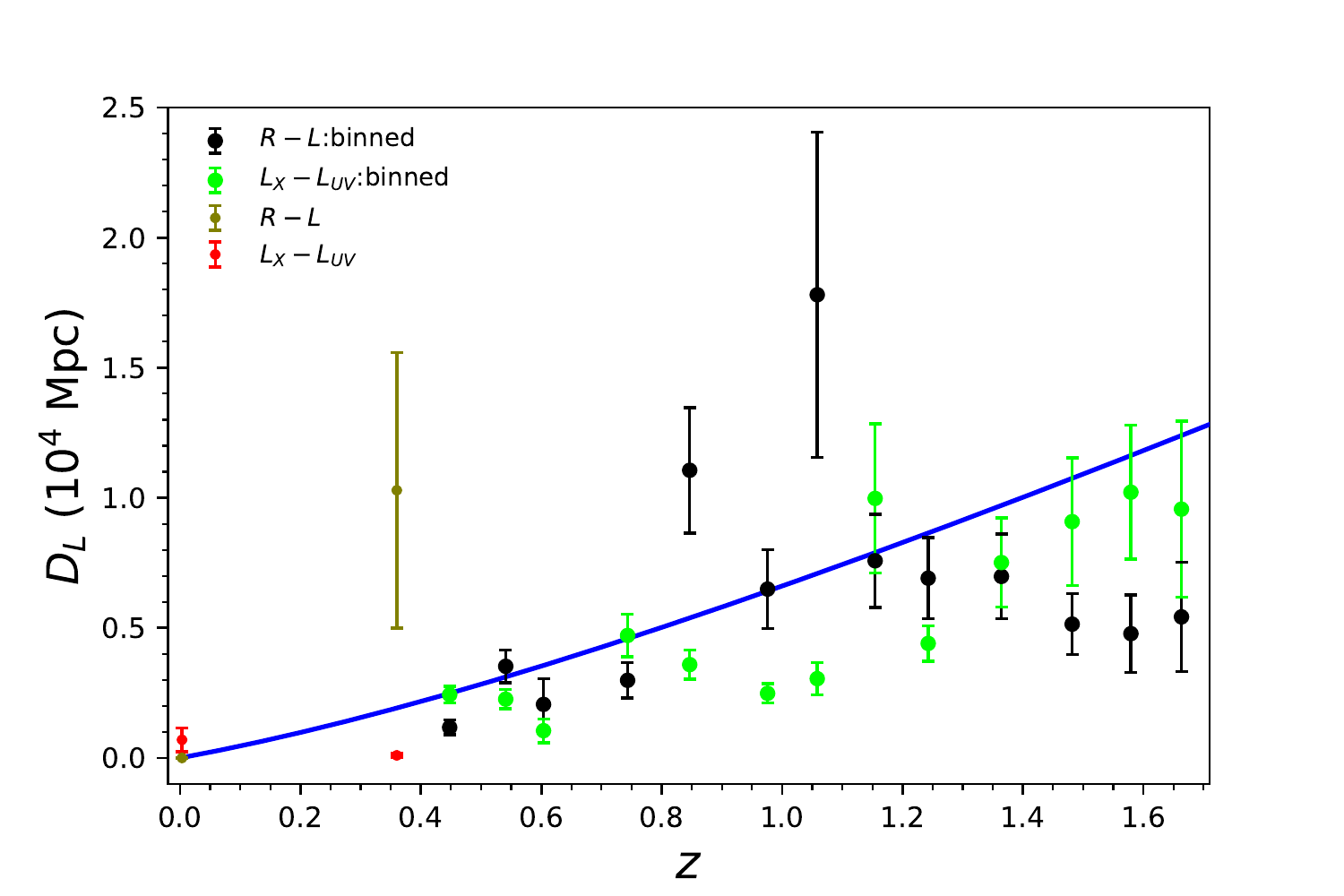}
\includegraphics[width=\columnwidth, height=3cm]{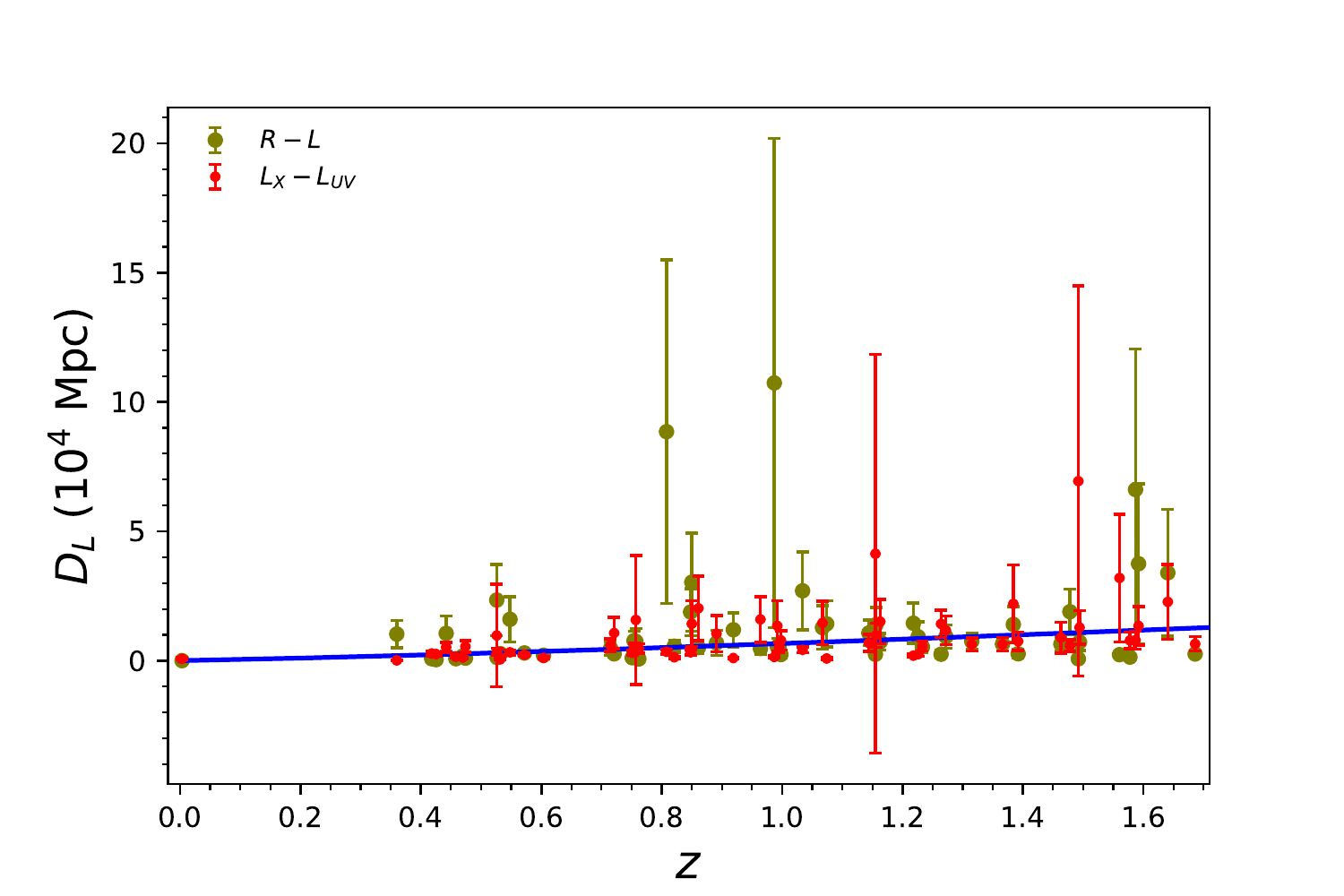}
\includegraphics[width=\columnwidth,height=3.5cm]{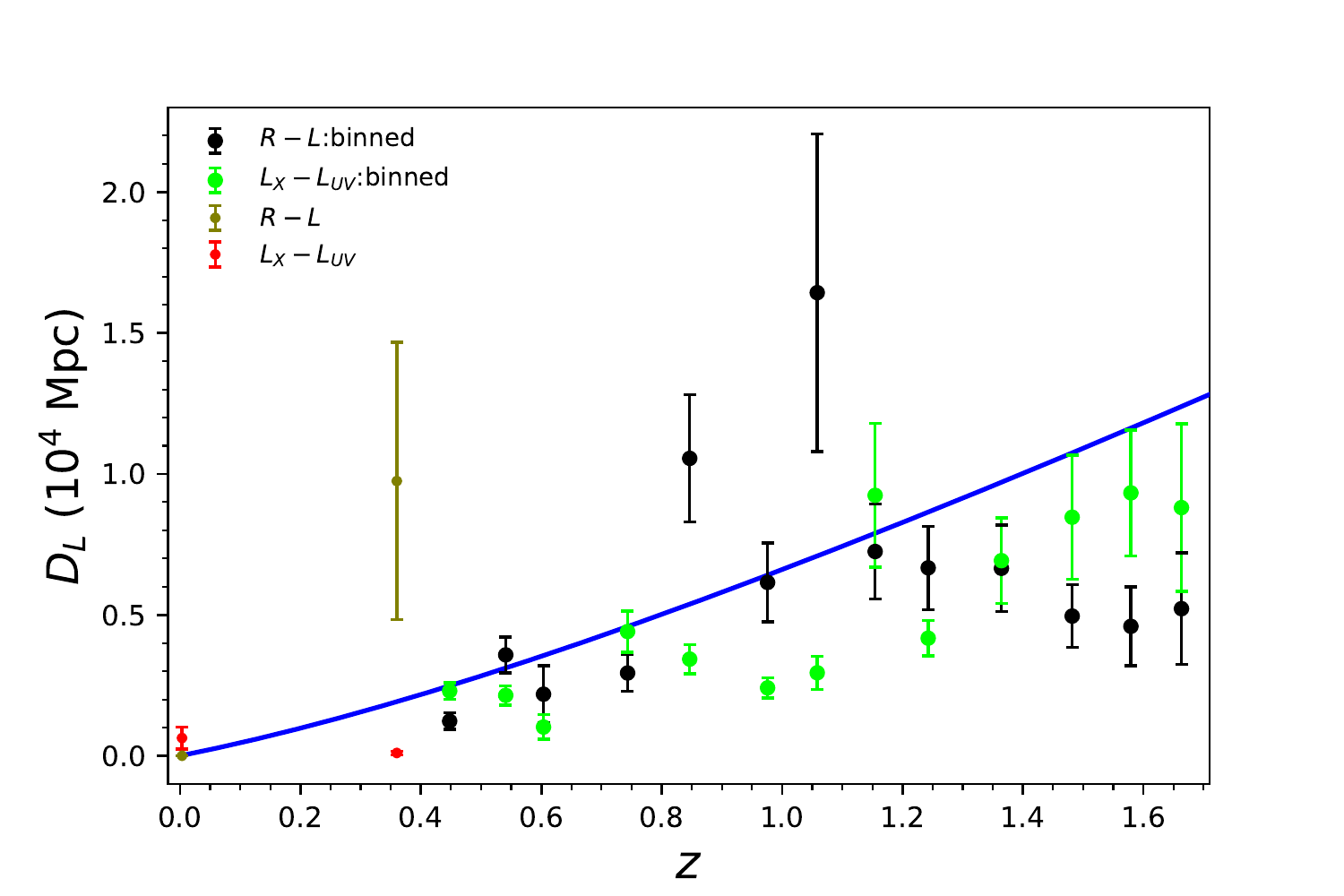}
\includegraphics[width=\columnwidth,height=3.5cm]{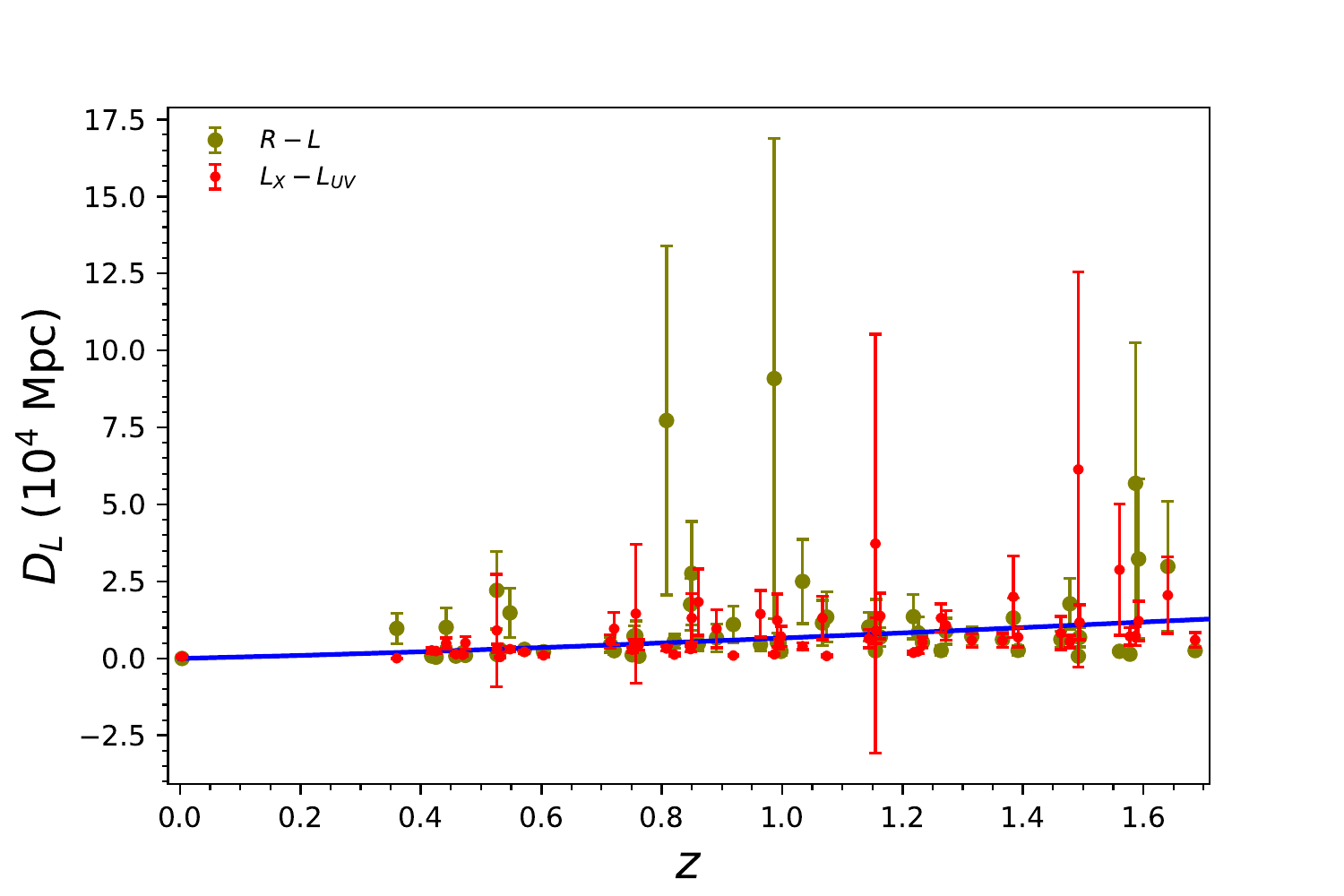}
\includegraphics[width=\columnwidth,height=3.5cm]{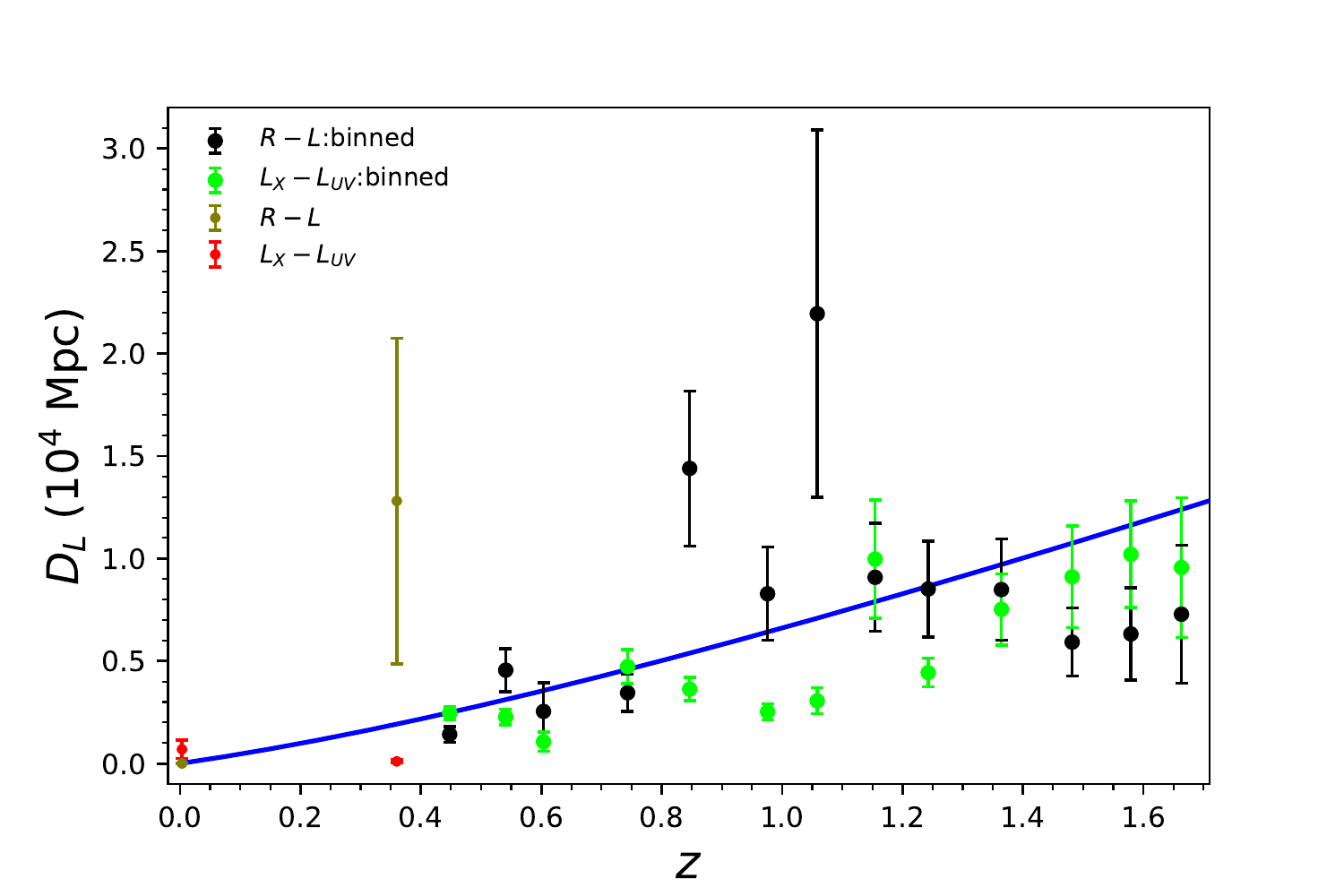}
\includegraphics[width=\columnwidth,height=3.5cm]{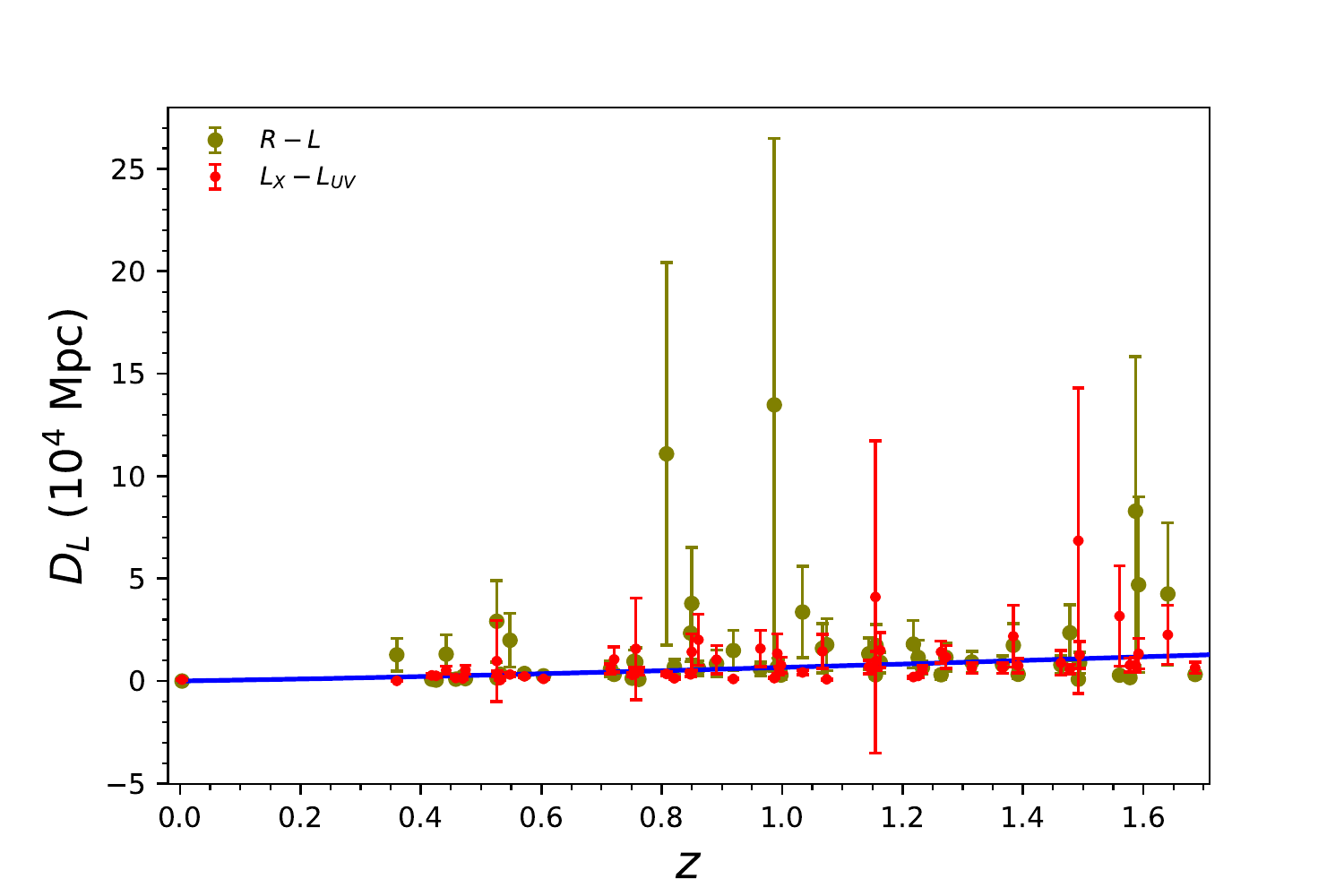}
\includegraphics[width=\columnwidth,height=3.5cm]{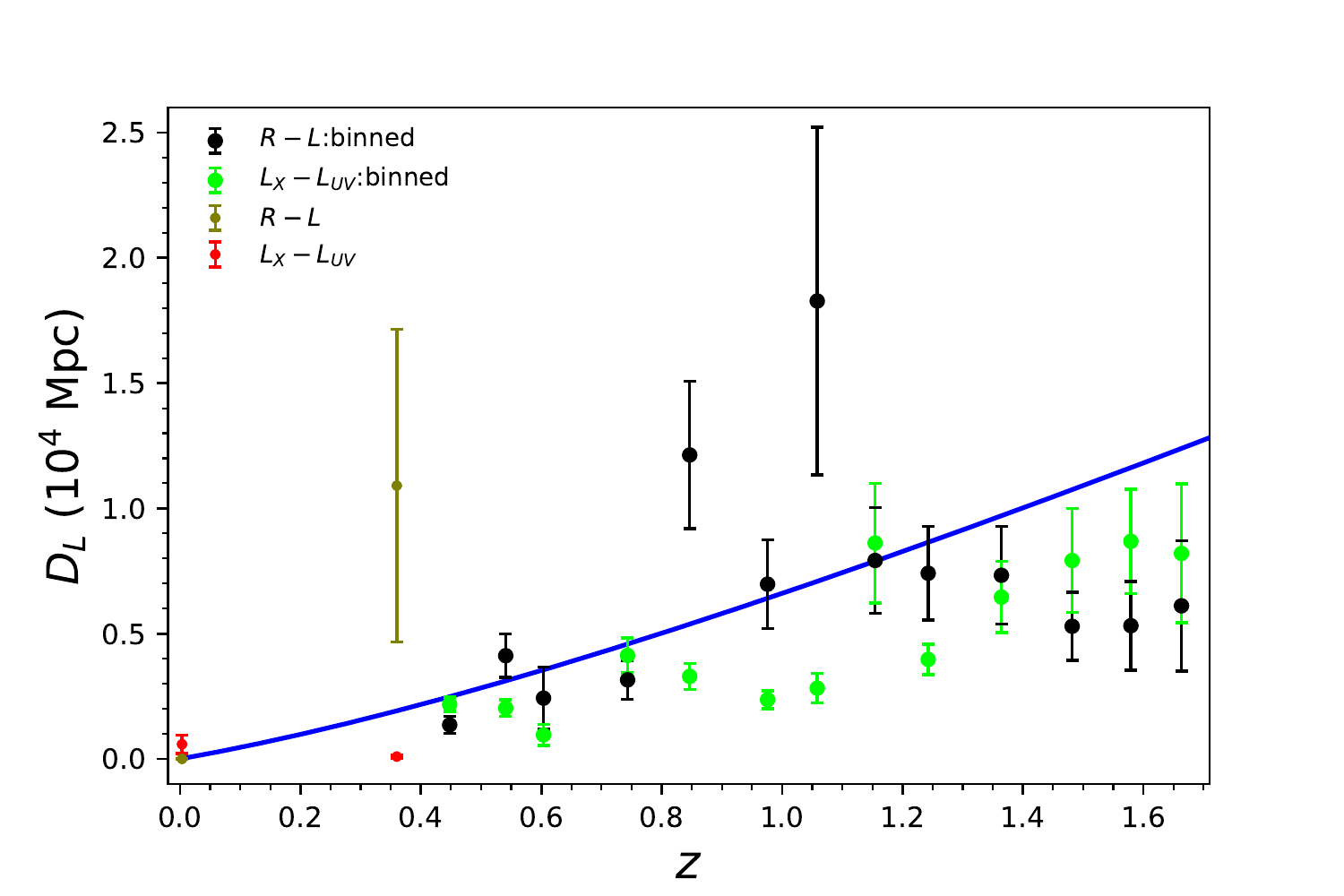}
\includegraphics[width=\columnwidth,height=3.5cm]{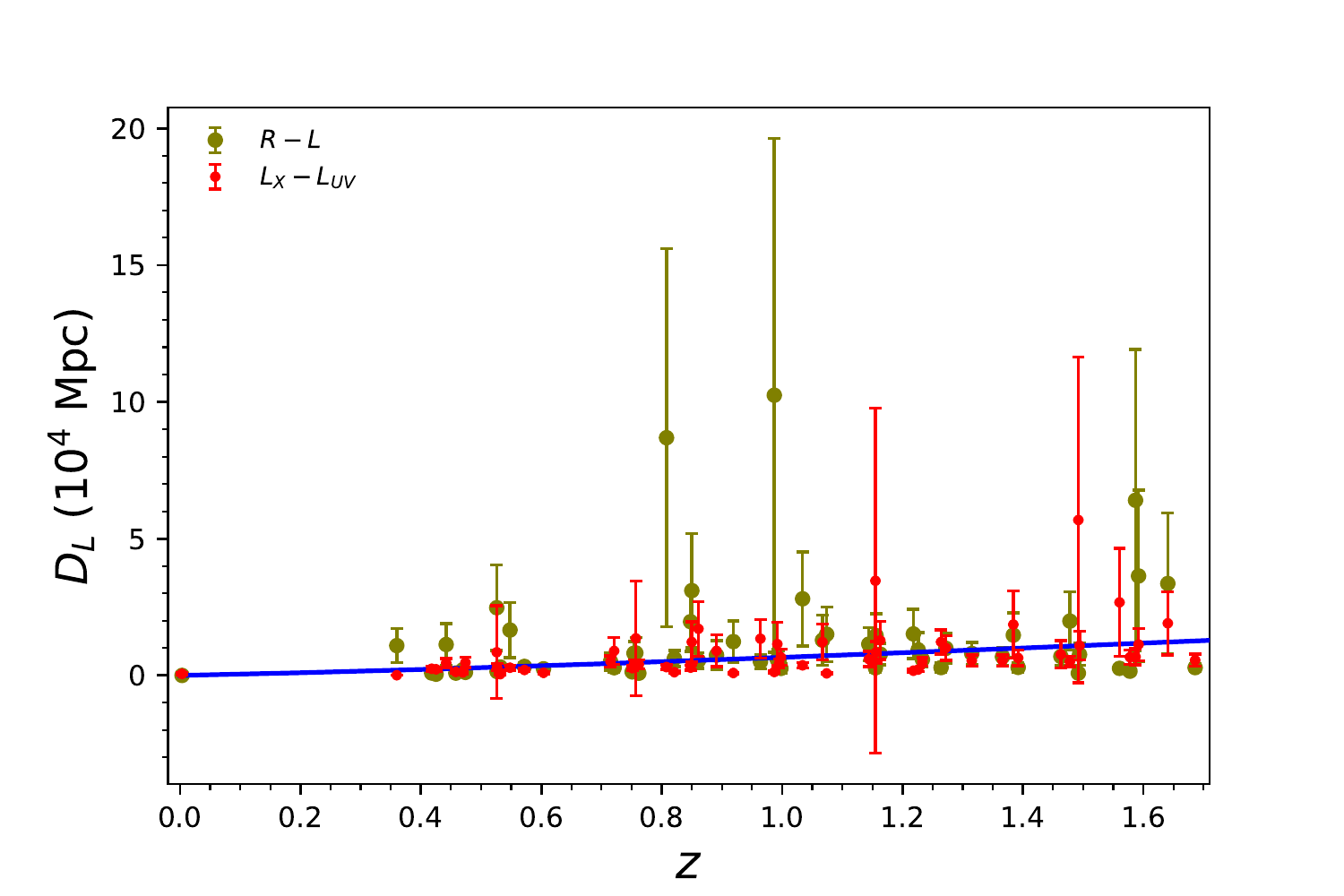}
\includegraphics[width=\columnwidth,height=3.5cm]{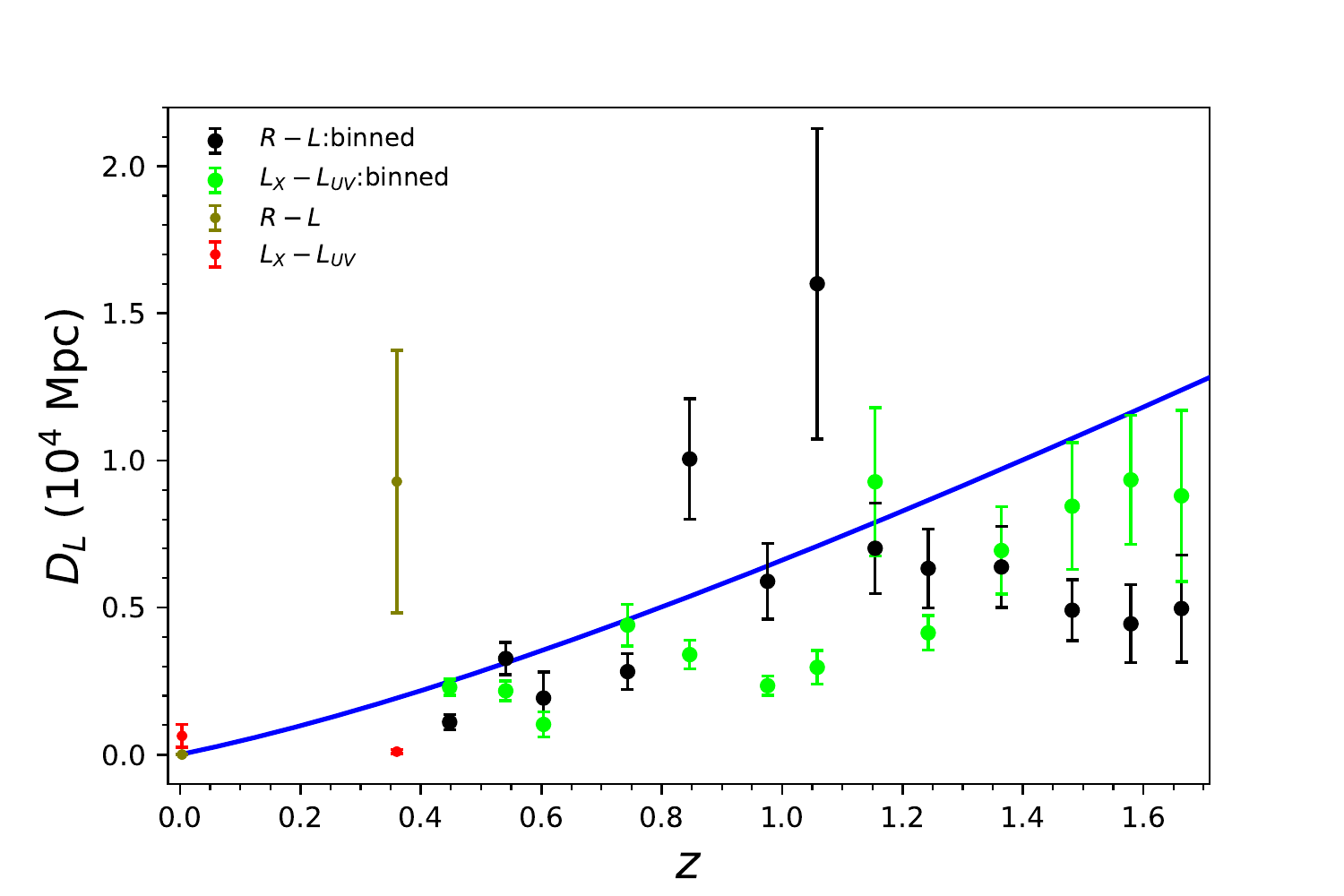}
\includegraphics[width=\columnwidth,height=3.5cm]{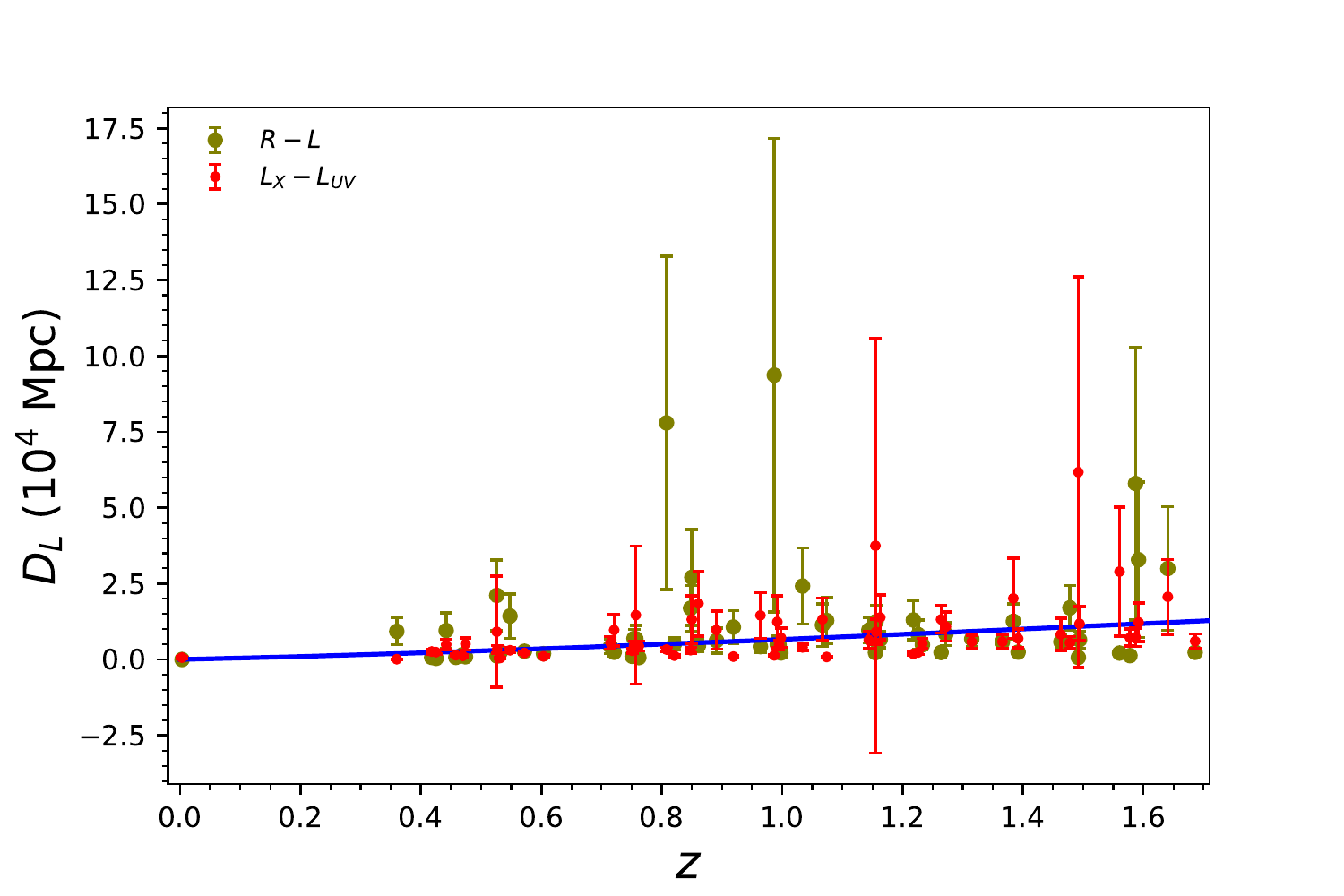}
\includegraphics[width=\columnwidth,height=3.5cm]{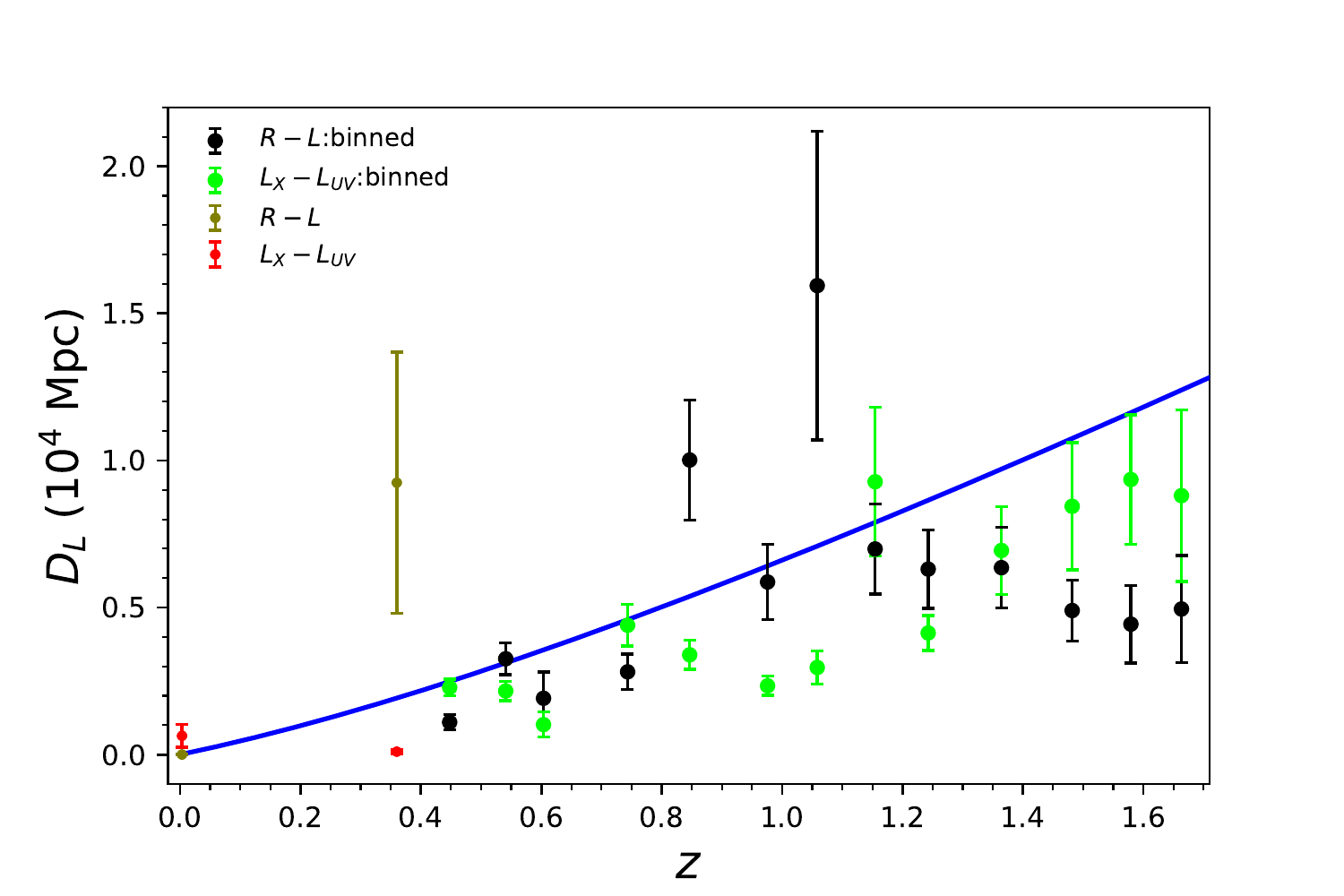}
\includegraphics[width=\columnwidth,height=3.5cm]{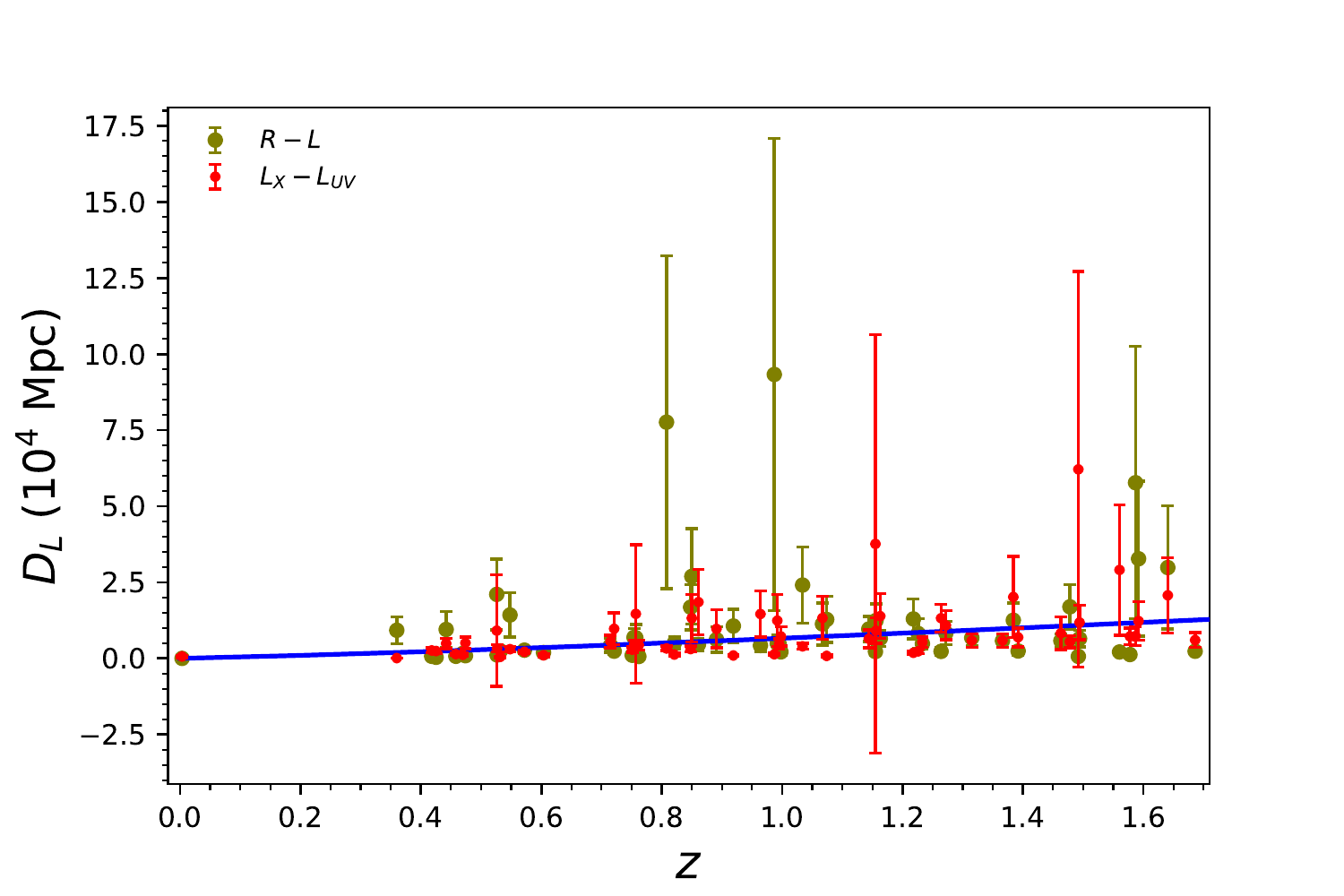}
\caption{Comparison of individual (right column) or (largely) binned (left column) $R-L$ and $L_X-L_{UV}$ luminosity distances for each source in, from top row to bottom row, the flat and non-flat $\Lambda$CDM, flat and non-flat XCDM, and flat and non-flat $\phi$CDM models. Black (green) points show the weighted means and uncertainties of the $R-L$ ($L_X-L_{UV}$) luminosity distances in narrow redshift bins. These are plotted at the average redshift of points in the bin. Olive (red) points show the luminosity distances and uncertainties for each source obtained using the $R-L$ ($L_X-L_{UV}$) relation. The blue solid line in each subpanel is the prediction from the flat $\Lambda$CDM model with $\Omega_{m0} = 0.3$. 
}
\label{fig:luminosity_distance}
\end{figure*}

\subsection{Luminosity distances for individual sources from the two methods}
\label{subsec_DL}

\begin{table*}
    \centering
    \caption{Statistical properties of the 58 point $(\log{D_{L,R-L}}-\log{D_{L,L_X-L_{UV}}})/(\sigma_{\log{D_{L,R-L}}}^2+\sigma_{\log{D_{L,L_X-L_{UV}}}}^2)^{1/2}$ distributions for six cosmological models (see Fig.~\ref{fig_histograms_DL_RL_xuv}). From left to right the columns list models, median values, 16\% and 84\% percentiles, mean values, standard deviations, skewness, kurtosis, and two-sample Kolmogorov-Smirnov (KS) test statistic (comparing the cosmological model distribution to the flat $\Lambda$CDM model distribution) with the corresponding $p$-value.}    
    \begin{tabular}{c|c|c|c|c|c|c|c|c}
    \hline
    \hline
    Model & Median  & 16\%  & 84\%  & Mean  & Standard deviation  & Skewness  & Kurtosis & KS test  \\
    \hline     
    Flat $\Lambda$CDM    & -0.138  & -1.775 & 2.719  & 0.116  & 2.206  & 0.327  & -0.207 & 0.00, $p=1.00$  \\
   Non-flat $\Lambda$CDM & -0.089  & -1.686 &  2.822  & 0.171  & 2.199  &  0.325  &  -0.189 & 0.05, $p=1.00$  \\
     Flat XCDM  & 0.176  &  -1.379 & 2.656  & 0.329  & 2.031  & 0.204  & -0.029 & 0.17, $p=0.36$  \\      
   Non-flat XCDM  & 0.183  & -1.422  & 2.804  & 0.362  & 2.067 & 0.227 &  -0.100 & 0.16, $p=0.49$  \\     
    Flat $\phi$CDM  &  -0.193  & -1.894  & 2.779  & 0.095  & 2.286  & 0.358   & -0.241 & 0.05, $p=1.00$  \\ 
    Non-flat $\phi$CDM & -0.200  & -1.901   & 2.777  & 0.091  & 2.286  & 0.360  & -0.243 & 0.05, $p=1.00$  \\
    \hline     
    \end{tabular}
    \label{tab_histograms_DL}
\end{table*}

Equations \eqref{eq:R-L_DL} and \eqref{eq:xuvflux} may be inverted to give $R-L$ and $L_X-L_{UV}$ luminosity distances to a source in terms of measured quantities, 
\begin{equation}
\log D_{L,R-L} = {1 \over 2 \gamma}\bigg\{\log \tau - \beta - \gamma\big[\log (4 \pi) - 44 + \log F_{3000}\big]\bigg\},
\label{eq_DL_RL}
\end{equation}
and
\begin{align}
& \log D_{L, L_X-L_{UV}} = \notag\\ 
& \ \  {1 \over 2(1 - \gamma)}\bigg[\beta + (\gamma -1)\log(4 \pi) + \gamma (\log F_{UV} -29) - \log F_X\bigg],
\label{eq_DL_LXLUV}
\end{align}
with corresponding errors derived by the first-order Taylor expansion,
\begin{align}
\sigma_{\log D_{L,R-L}} = &  \\
  {1 \over 2}\Bigg\{\sigma^2_{\log F_{3000}} 
  & + {1 \over \gamma^2} \left[\sigma^2_{\log \tau} + {1 \over \gamma^2} (\log \tau - \beta)^2 \sigma^2_{\gamma} + \sigma^2_{\beta}\right]\Bigg\}^{1 \over 2}, \notag
\end{align}
and
\begin{align}
\label{eq:er_DL_Lx_Luv}    
    \sigma_{\log D_{L,L_X-L_{UV}}} = & 
    {1 \over 2(\gamma-1)}\Bigg[\sigma^2_{\log F_{X}} + \gamma^2 \sigma^2_{\log F_{UV}} + \sigma^2_{\beta} \\
    & \ \ + \left(\frac{\log F_{X} - \beta -\log F_{UV} + 29}{1-\gamma}\right)^2\sigma^2_{\gamma}\Bigg]^{1 \over 2}. \notag
\end{align}
These expressions allow us to compute the $R-L$ relation and the $L_X-L_{UV}$ relation luminosity distances of all sources in each of the six cosmological models. For each source, $\log \tau$, $\sigma_{\log \tau}$, $\log F_{3000}$, $\sigma_{\log F_{3000}}$, $\log F_{X}$, $\sigma_{\log F_{X}}$, $\log F_{UV}$, and $\sigma_{\log F_{UV}}$ are given in Table \ref{tab_xray_uv_data}, and for each cosmological model, $\gamma$, $\sigma_\gamma$, $\beta$, and $\sigma_\beta$ are listed in Table \ref{tab:1d_BFP}.

In Fig.~\ref{fig:luminosity_distance} we compare $R-L$ and $L_X-L_{UV}$ luminosity distances for each source. In the left panels we plot weighted mean luminosity distances at the average redhsift of the points in the bin \citep[see, e.g.,][]{Podariuetal2001}, in 13 redshift bins spanning $0.4 \leq z \leq 1.7$ and of width $0.1$. Each bin contains at least a single source. Lower redshift data are too sparse to be binned. In the right panels we plot the two luminosity distances of each source. The left panels, especially, show that $L_X-L_{UV}$ distances are significantly shorter than $R-L$ distances and $\Omega_{m0} = 0.3$ flat $\Lambda$CDM model distances, especially in the $0.8 \lesssim z \lesssim 1.3$ range. This explains why these $L_X-L_{UV}$ data favour higher $\Omega_{m0}$ values than 0.3 and higher than those favoured by these $R-L$ data. These results are similar to those of \cite{Lussoetal2020} and \cite{KhadkaRatra2021a, KhadkaRatra2022}. However the causes in the two cases might not be similar as \cite{KhadkaRatra2021a, KhadkaRatra2022} showed that the \cite{Lussoetal2020} $L_X-L_{UV}$ sources are not standardizable, which is not the case with the $L_X-L_{UV}$ sources we study here. The plots of individual distances (Fig.\ref{fig:luminosity_distance}, right panel) show that some of the $R-L$ and $L_X-L_{UV}$ luminosity distance measurements have large errors and some have large offsets from the overall trends. However, simple selective removal of such sources can introduce a bias in the sample.

\begin{figure*}  
\includegraphics[width=\columnwidth]{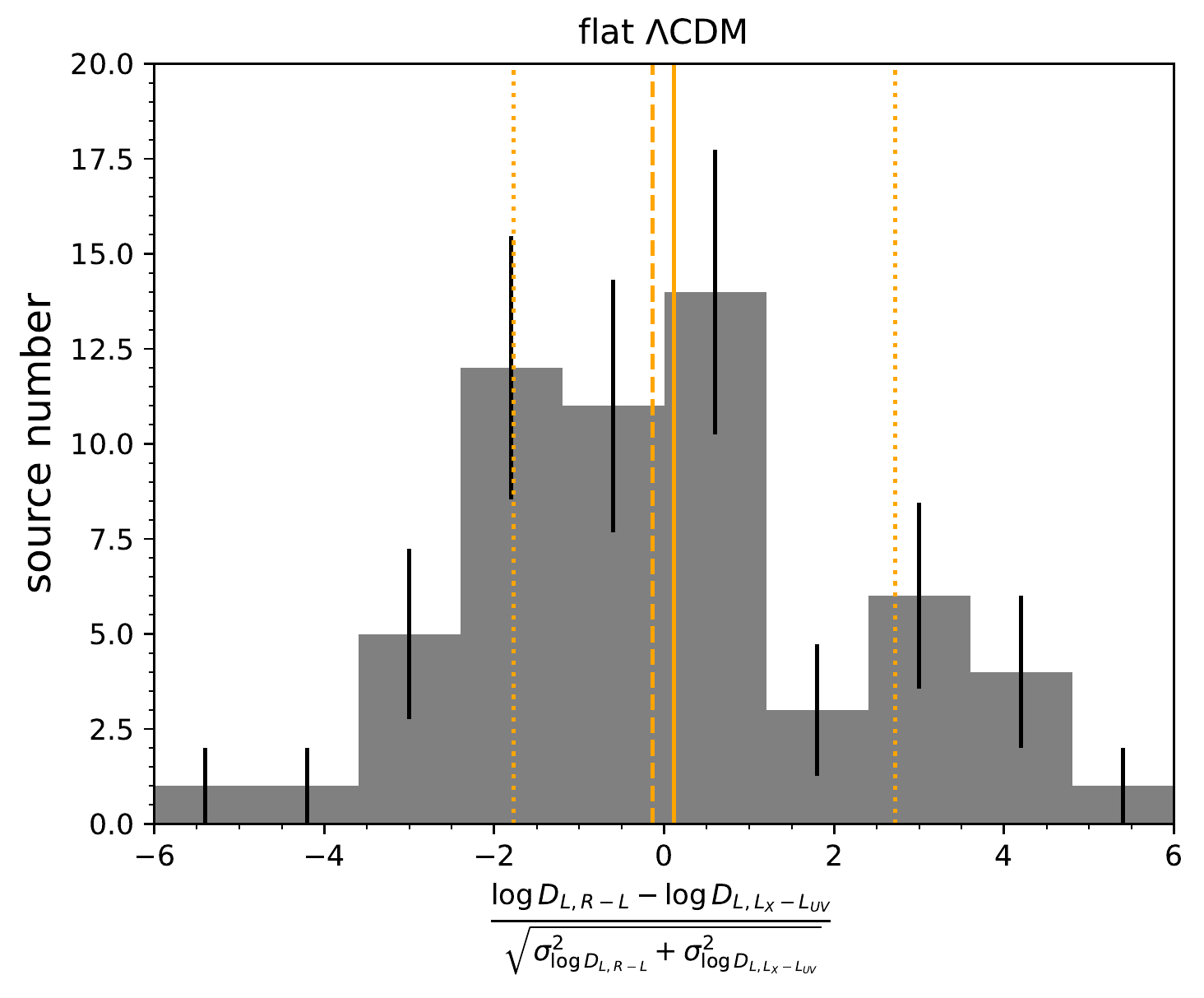}
\includegraphics[width=\columnwidth]{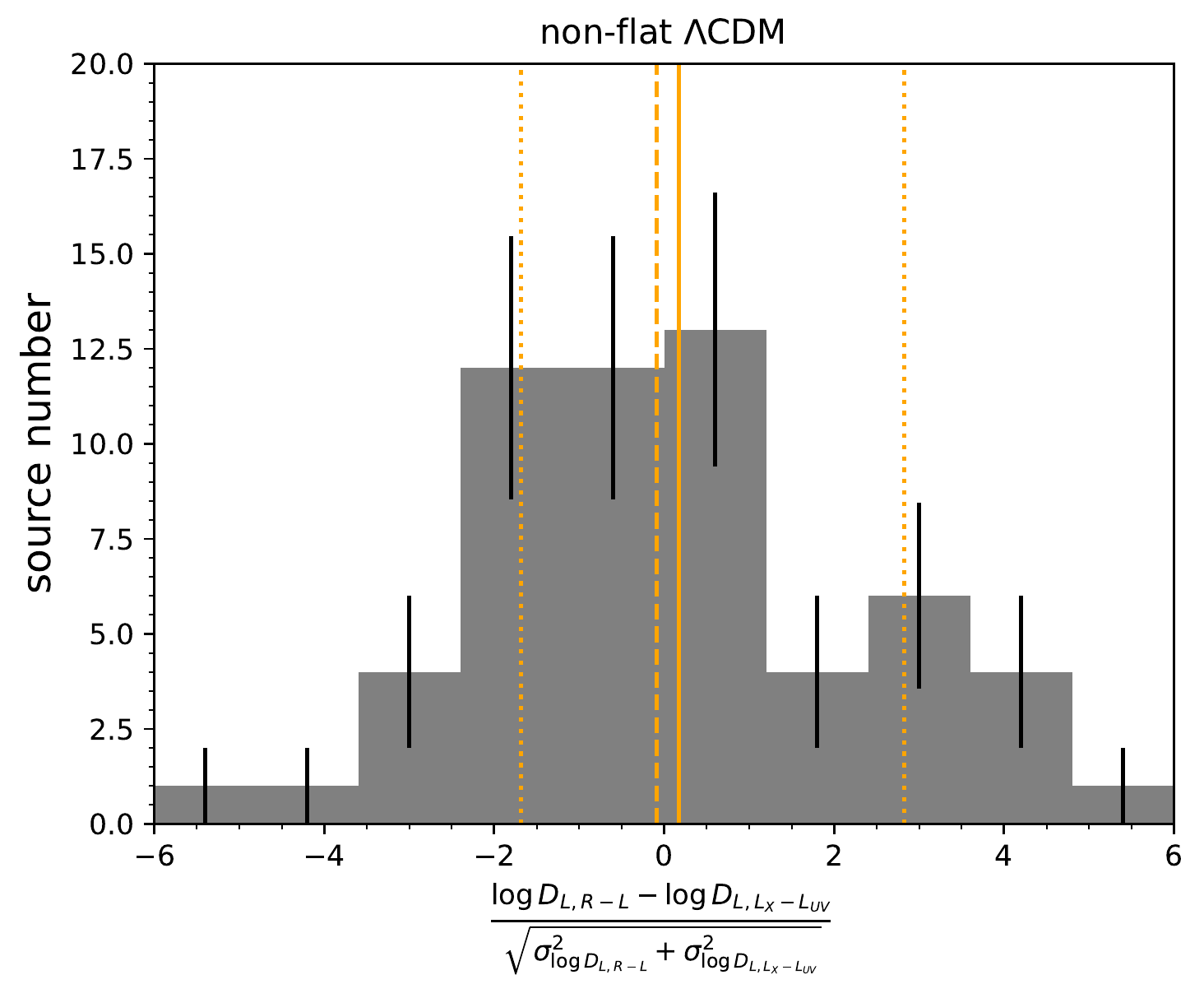}
\includegraphics[width=\columnwidth]{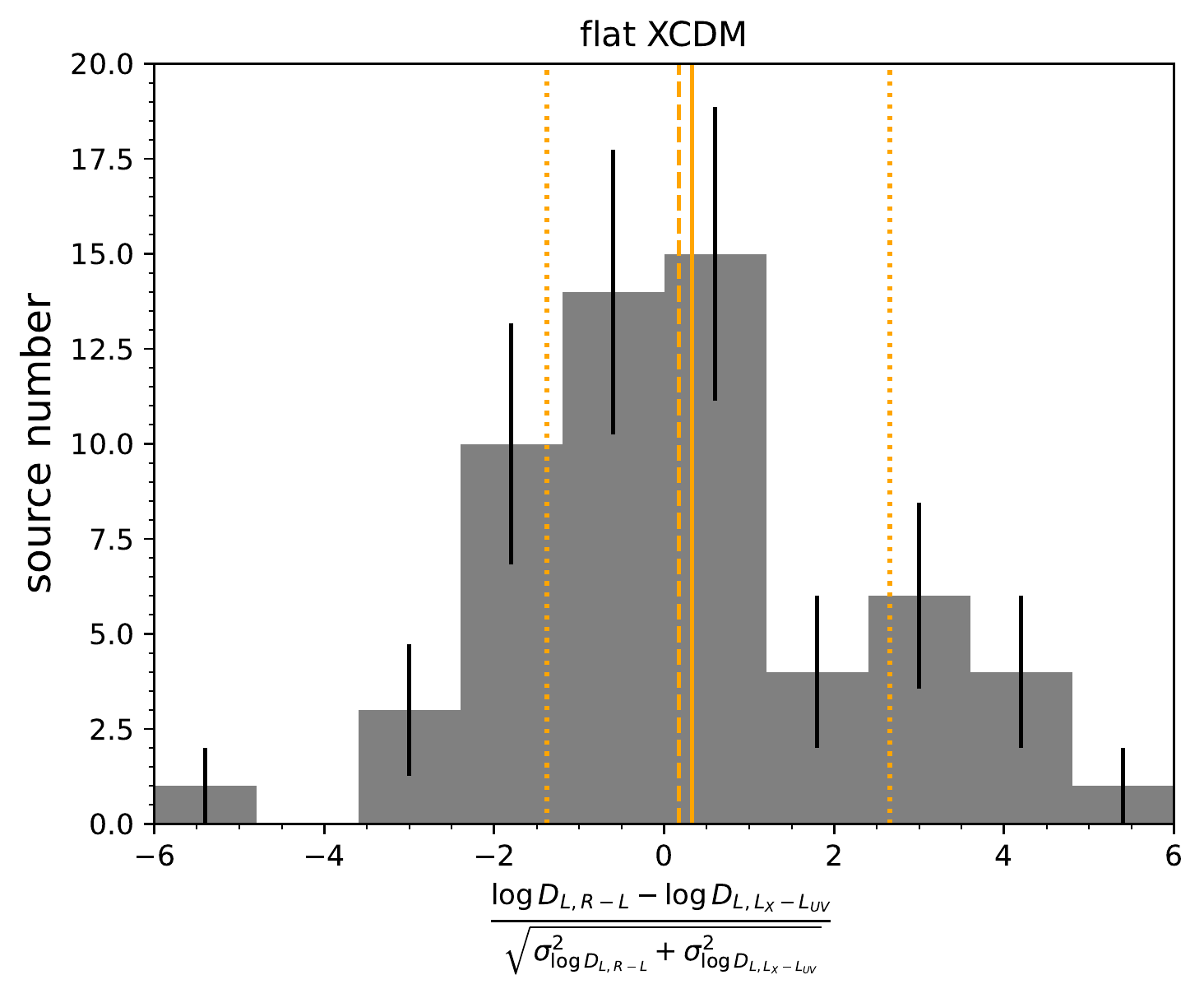}
\includegraphics[width=\columnwidth]{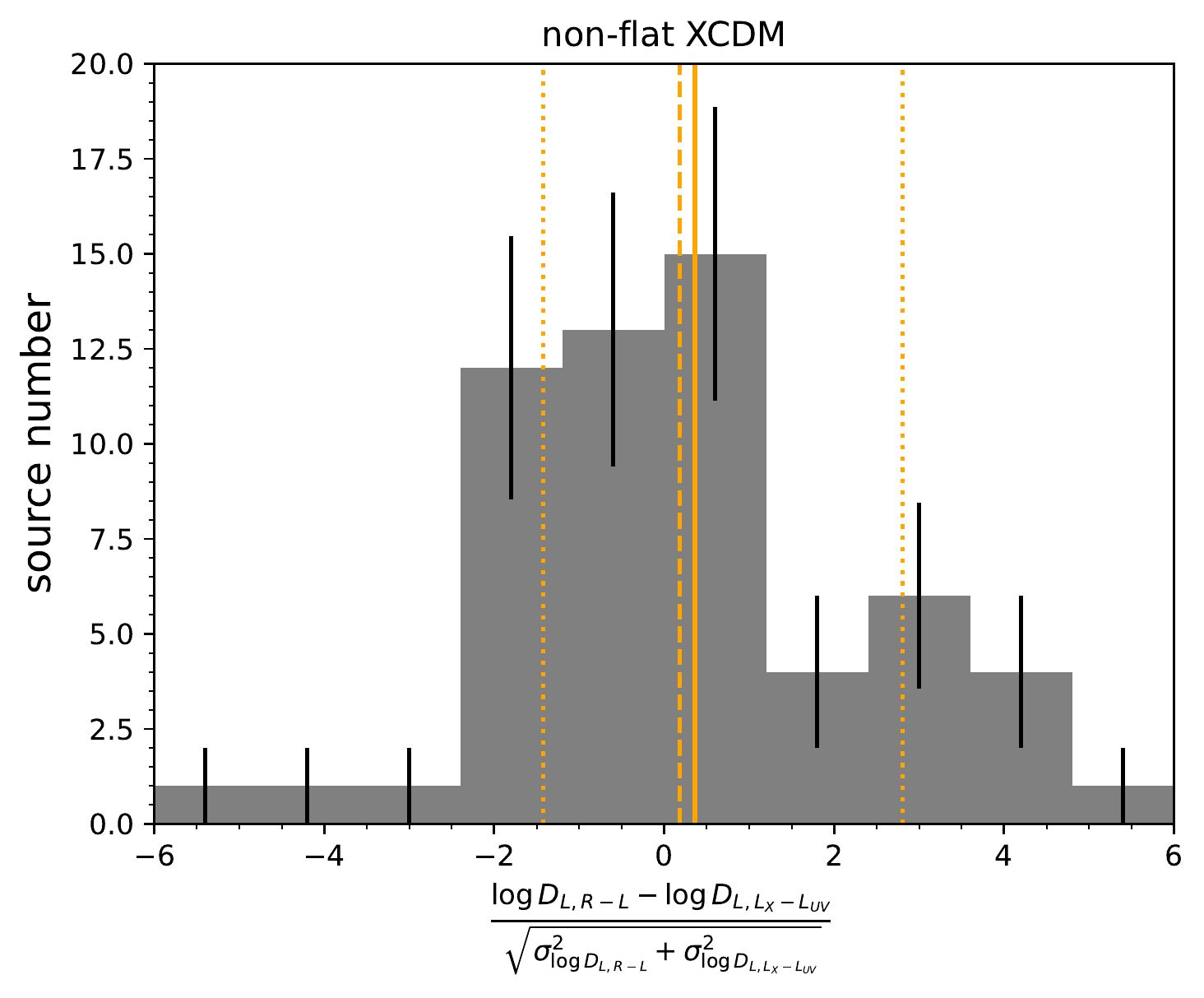}
\includegraphics[width=\columnwidth]{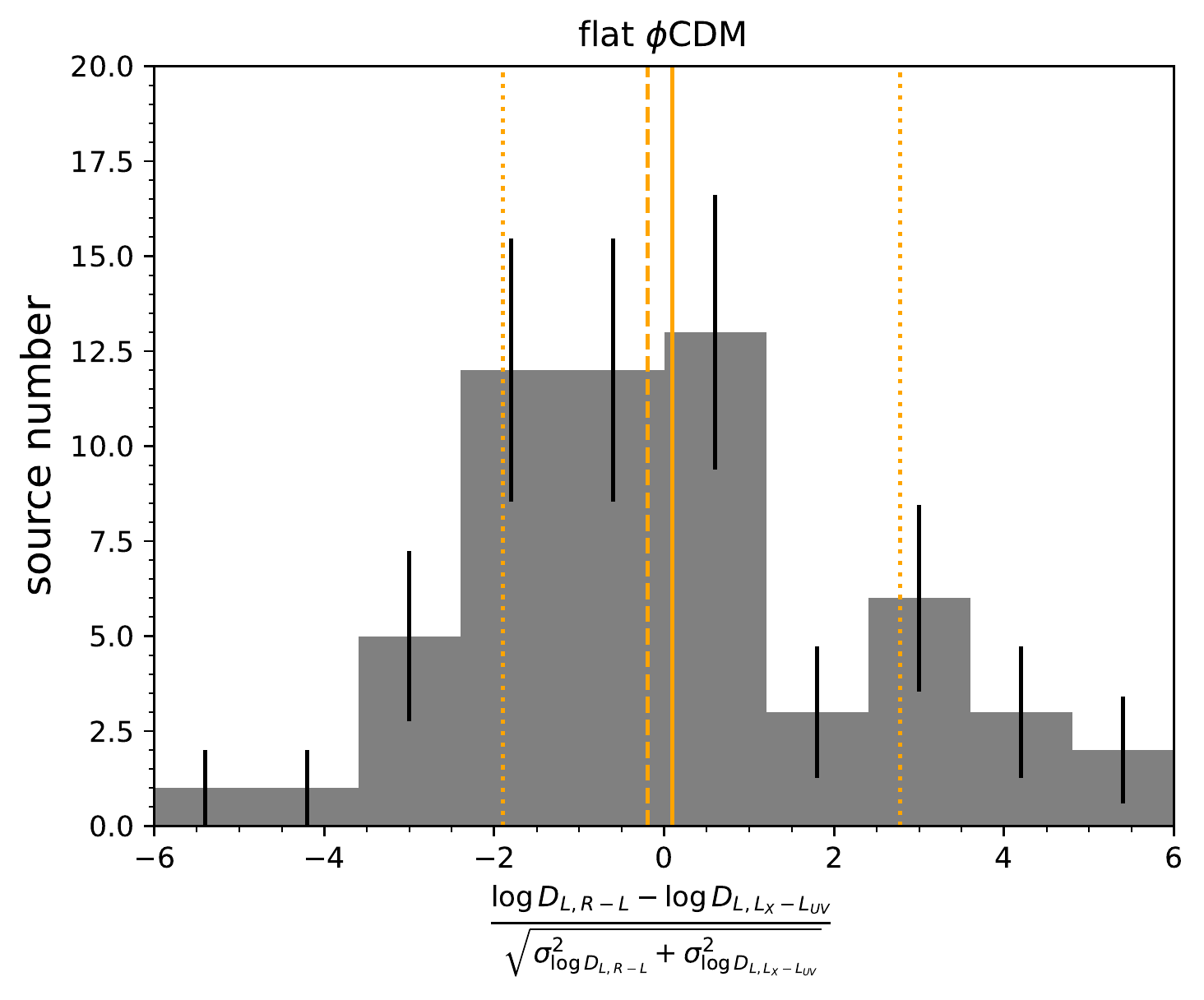}
\includegraphics[width=\columnwidth]{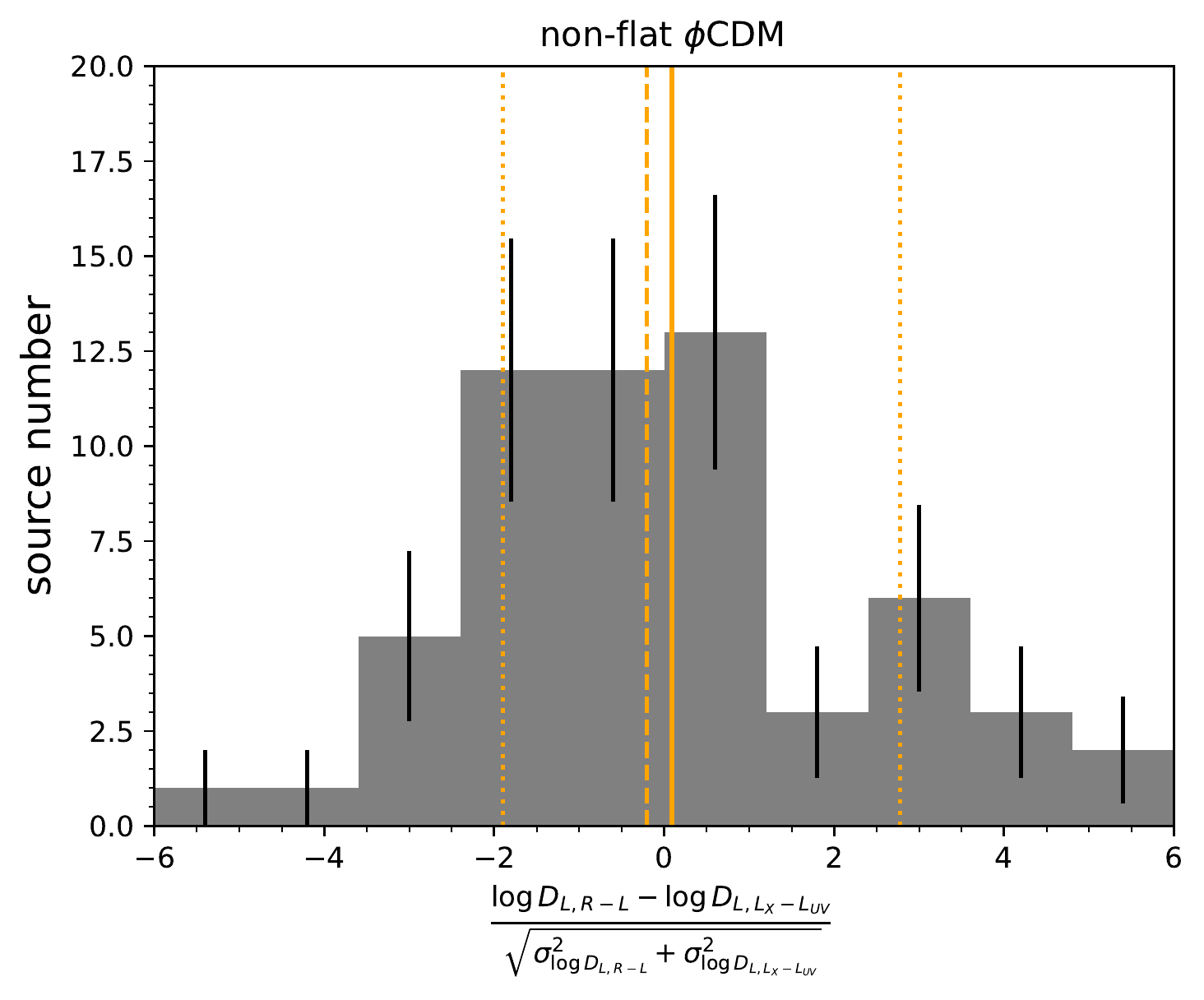}
\caption{Histograms of $(\log{D_{L,R-L}}-\log{D_{L,L_X-L_{UV}}})/(\sigma_{\log{D_{L,R-L}}}^2+\sigma_{\log{D_{L,L_X-L_{UV}}}}^2)^{1/2}$ for 58 sources in the flat and non-flat $\Lambda$CDM, XCDM, and $\phi$CDM cosmological models (from the top to the bottom). The solid vertical line stands for the distribution mean, dashed vertical line for the median, and dotted vertical lines stand for 16\% and 84\% percentiles. The bin size is 1.2 for all the cases (10 bins between $-6$ and 6). The uncertainties for each bin were calculated as $\sigma_{i}=\sqrt{N_i}$ where $N_i$ is the number of sources that belong to the given bin.}
\label{fig_histograms_DL_RL_xuv}
\end{figure*}

To better understand these systematic differences between the two luminosity distances for each source, we study distributions of $(\log{D_{L,R-L}}-\log{D_{L,L_X-L_{UV}}})/(\sigma_{\log{D_{L,R-L}}}^2+\sigma_{\log{D_{L,L_X-L_{UV}}}}^2)^{1/2}$, i.e. histograms of the logarithm of the ratio $D_{L,R-L}/D_{L,L_{X}-L_{UV}}$ normalized by the combined uncertainty of the two luminosity distances for each source, where we discard the second $R-L$ distance measurement for NGC 4151. We study these distributions for the flat and non-flat $\Lambda$CDM, XCDM, and $\phi$CDM models, i.e. a total of six histograms that are shown in Fig.~\ref{fig_histograms_DL_RL_xuv}. The histograms are constructed uniformly for the six cases, with 10 equal-sized bins in the range $(-6,6)$ with bin width 1.2.\footnote{This ensures that differences for all the sources fall in this range for all six cosmological models. The minimum and the maximum normalized differences for flat $\Lambda$CDM, $\phi$CDM, and XCDM are $(-5.11,5.51)$, $(-5.20, 5.66)$, and $(-4.91, 5.30)$, respectively; for non-flat $\Lambda$CDM, $\phi$CDM, and XCDM models we have $(-5.07, 5.59)$, $(-5.20, 5.66)$, and $(-4.87, 5.38)$, respectively.} The basic statistical properties of the 58 point distributions of normalized luminosity-distance differences, including the median and mean values as well as skewness and kurtosis coefficients among other things, are summarized in Table~\ref{tab_histograms_DL}. The minimum and maximum median values are $-0.200$ and 0.183, respectively, corresponding to non-flat $\phi$CDM and non-flat XCDM, respectively. The minimum and maximum 16\% percentile values are $-1.901$ and $-1.379$, respectively, corresponding to non-flat $\phi$CDM and flat XCDM, respectively. The 84\% percentile value is in the range from $2.656$ to $2.822$, which correspond to flat XCDM and non-flat $\Lambda$CDM, respectively. The mean normalized luminosity-distance difference values have a minimum of $0.091$ for the non-flat $\phi$CDM model and a maximum of $0.362$ for the non-flat XCDM model. The standard deviation values are in the range from $2.031$ to $2.286$ corresponding to flat XCDM and flat and non-flat $\phi$CDM, respectively. The skewness coefficient has a minimum of $0.204$ for the flat XCDM model and a maximum of $0.360$ for the non-flat $\phi$CDM model. The minimum and maximum kurtosis values are $-0.243$ and $-0.029$, respectively, which correspond to the non-flat $\phi$CDM model and to the flat XCDM model, respectively. Overall, for all six cosmological models, the distributions have a positive mean value, are positively skewed, and have a negative kurtosis, which indicates that for the current limited sample of 58 sources the $R-L$ relation has a tendency to yield larger luminosity distances in comparison with the $L_{X}-L_{UV}$ relation. The negative kurtosis  indicates that distribution outliers are suppressed in comparison with the normal distribution and the median value is slightly negative for all models except for flat and non-flat XCDM. The normalized luminosity-distance difference distributions for all the six cosmological models are consistent with being drawn from the same distribution, which is shown by the two-sample Kolmogorov-Smirnov (KS) test statistic calculated between a given 58 point distribution and the distribution corresponding to the flat $\Lambda$CDM model, see Table~\ref{tab_histograms_DL} (last column). The KS statistic is in the range 0.05-0.17 (excluding the comparison of the flat $\Lambda$CDM distribution with itself), with the $p$-value in the range $0.36-1.00$, hence the null hypothesis that an underlying distribution of a given normalized $D_{L}$ difference distribution and the underlying distribution corresponding to the flat $\Lambda$CDM case are identical is confirmed for all the models. 

\begin{figure*}  
\includegraphics[width=0.95\linewidth]{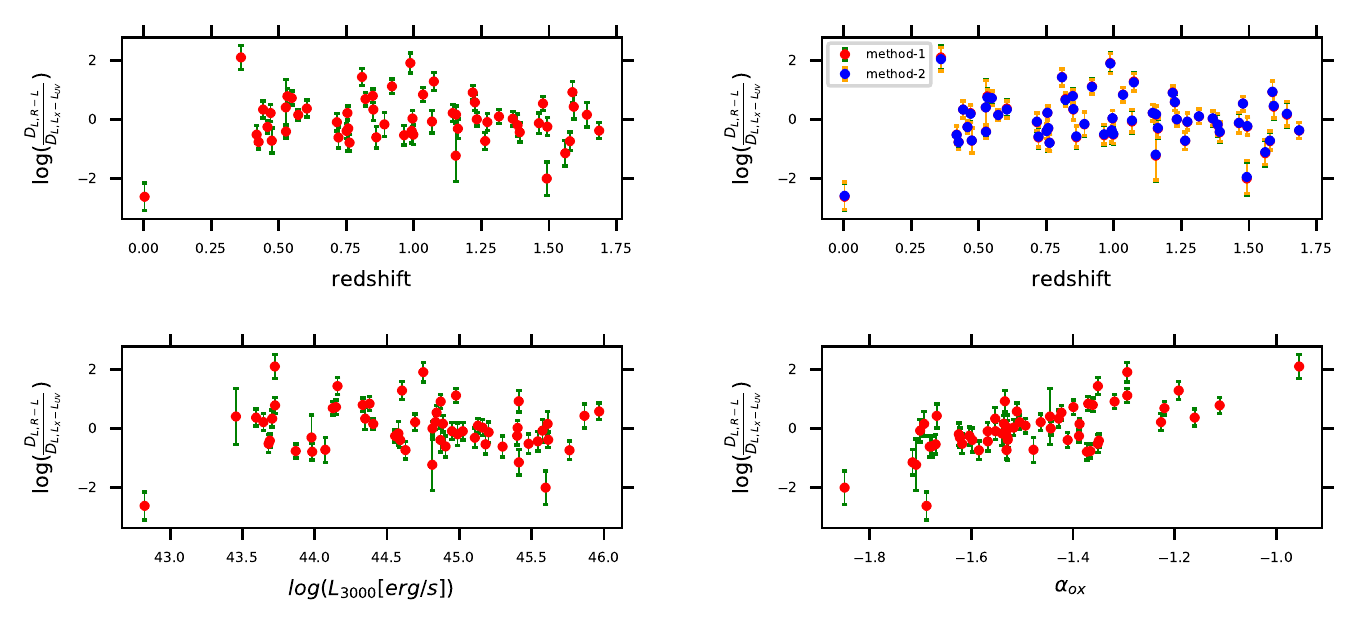}
\includegraphics[width=0.95\linewidth]{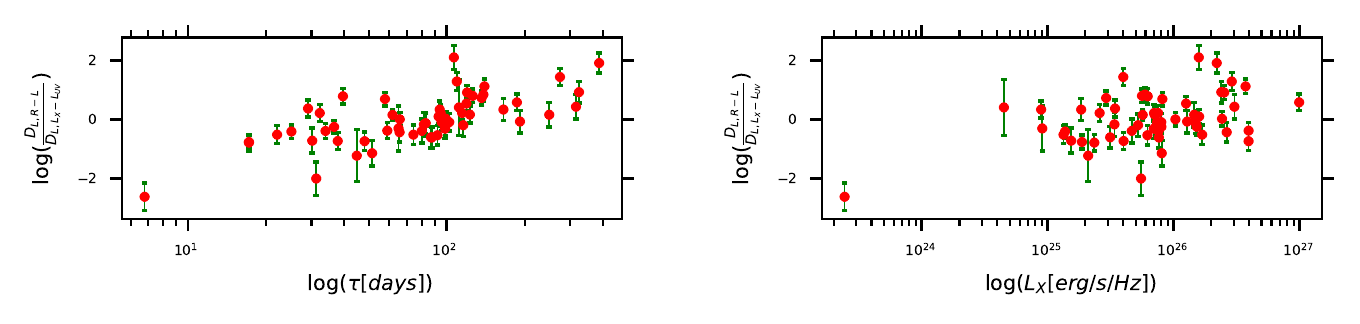}
\caption{Ratio of $R-L$ and $L_{X}-L_{UV}$ source luminosity distances as a function of redshift, with red (blue) points in the upper left (right) panel based on method-1 (method-2). Middle panels show method-1 distance ratios as a function of source absolute luminosity at 3000 \AA\ (middle left panel) and as a 
function of the $\alpha_{OX}$ parameter (middle right panel). Bottom panels show method-1 distance ratios as a function of the time delay (bottom left panel) and as a function of the source X-ray absolute luminosity.}
\label{fig:rat_z_L}
\end{figure*}

Next we analyze differences between the $R-L$ and $L_X-L_{UV}$ luminosity distances for each source in the fixed $\Omega_{m0} = 0.3$ flat $\Lambda$CDM model using the slope and intercept values given in eqs.~\eqref{eq_Lx_Luv} and \eqref{eq_RL}, again discarding the second time-delay distance measurement to NGC 4151. The resulting ratios of the $R-L$ and $L_X-L_{UV}$ luminosity distances, $\log{(D_{L,R-L}/D_{L,L_X-L_{UV}})}$, are plotted as a function of $z$ in Fig.~\ref{fig:rat_z_L} using red points. [We call this technique, based on eqs.~\eqref{eq_Lx_Luv} and \eqref{eq_RL}, "method-1".] The dispersion in the measured ratios is large, necessitating the use of the log scale: the smallest ratio is $2.4 \times 10^{-3}$, for NGC 4151, while the largest ratio is 126, for object No. 160 from \citet{Homayouni2020}. Half of the sources have a ratio lower than 1 (30 out of 58), and the average value of the logarithm of the ratio is $1.14 \times 10^{-3}$, corresponding to a mean factor of 1.002. This indication that $L_X-L_{UV}$ distances on average are shorter than $R-L$ distances is qualitatively consistent with the results from the histogram in the upper left panel of Fig.~\ref{fig_histograms_DL_RL_xuv} that fully accounts for all uncertainties (and so is more accurate than the method-1 technique result). 
 
If instead of the fixed $\Omega_{m0} = 0.3$ flat $\Lambda$CDM model slope and intercept values given in eqs.~\eqref{eq_Lx_Luv} and \eqref{eq_RL} we use the flat $\Lambda$CDM model coefficients from Table~\ref{tab:1d_BFP} for the correlation coefficients, i.e. we allow for different $R-L$ and $L_X-L_{UV}$ cosmological models and also account for the uncertainties in more parameters, we find the results shown in the top right panel of Fig.~\ref{fig:rat_z_L} using blue points. (We call this technique, based on Table~\ref{tab:1d_BFP} values, "method-2".) In this case the same number of sources have a ratio below unity, but the mean value of the logarithm of the ratio is $-0.01$, corresponding to a mean factor of 0.98, i.e. $\sim 2\%$ offset in the opposite direction, with $R-L$ placing sources closer by than $L_X-L_{UV}$. While differing from the result of Table~\ref{tab_histograms_DL}, which indicates that $L_X-L_{UV}$ places sources closer by than does $R-L$, the Table~\ref{tab_histograms_DL} result is the more correct one as it accounts for the error bars of the two luminosity distances for each source, while method-2 does not account for these.

\begin{table*}
    \centering
    \caption{Pearson correlation coefficients (PCC) and Spearman rank-order correlation coefficients (SCC) with the corresponding $p$-values of the distance ratio trends of Fig.~\ref{fig:rat_z_L} . All ratios are calculated for the fixed flat $\Lambda$CDM model with $\Omega_{m0}=0.3$, except for $\log{(D_{L,R-L}/D_{L,L_X-L_{UV}}}, z )$ in the second row which is computed for the flat $\Lambda$CDM model values taken from Table \ref{tab:1d_BFP}.}  
    \begin{tabular}{c|c|c|c|c|c}
    \hline
    \hline
    Correlation & PCC  & $p$-value & SCC & $p$-value &  plot position   \\
    \hline     
     $\log{(D_{L,R-L}/D_{L,L_X-L_{UV}})}$, $z$ & $-0.0126$ &0.925 & $-0.0349$   & 0.795 & upper left panel  
     \\
     $\log{(D_{L,R-L}/D_{L,L_X-L_{UV}})}$, $z$& $6.75\times 10^{-5}$ &$1.000$ & $-0.0235$  & 0.861 & upper right panel
     \\
     $\log{(D_{L,R-L}/D_{L,L_X-L_{UV}})}$, $\log{L_{3000}}$& $-0.0698$ &0.603 & $-0.160$  &  0.232 &middle left panel
     \\
      $\log{(D_{L,R-L}/D_{L,L_X-L_{UV}})}$, $\alpha_{OX}$&0.679& $4.80 \times 10^{-9}$ & 0.593  & $9.15 \times 10^{-7}$  & middle right panel
      \\
       $\log{(D_{L,R-L}/D_{L,L_X-L_{UV}})}$, $\log{\tau}$& 0.686 &$2.75 \times 10^{-9}$& $0.707$  & $5.62 \times 10^{-10}$    &bottom left panel
       \\
       $\log{(D_{L,R-L}/D_{L,L_X-L_{UV}})}$, $\log{L_X}$& 0.444& $4.84 \times 10^{-4}$ &  0.304 &  $0.0204$  &bottom right panel
       \\
     \hline
    \end{tabular}
    \label{tab_PCC}
\end{table*}

The most extreme $D_L$ ratio outliers are due to the unfortunate coincidence of a short distance value from one technique and a long distance value from the other. Most of the $R-L$ and $L_{X}-L_{UV}$ source luminosity distances (107 out of 116 measurements) are within one order of magnitude from the predicted distances in the $\Omega_{m0} = 0.3$ flat $\Lambda$CDM model.

However, there is no simple way of pre-selecting better measured sources to include in our sample. For example, the distance to NGC 4151 is a major problem. According to the $\Omega_{m0} = 0.3$ flat $\Lambda$CDM model, it should be at 14.15 Mpc while the $R-L$ relation and the $L_X-L_{UV}$ relation give the values $1.59 \pm 1.59$ Mpc and $698 \pm 455$ Mpc, respectively (from the complete flat $\Lambda$CDM model analysis). The problem is that NGC 4151 is heavily absorbed at 3000 \AA\ rendering the $R-L$ distance to NGC 4151 unreliable. It is also heavily absorbed in the X-ray band \citep[e.g.][]{2007A&A...463..903D} and in the UV band, so the index $\alpha_{OX}$ seems typical while both measured fluxes (UV and X-ray) do not represent the intrinsic properties of the source as the source does not seem to posses the usual Big Blue Bump \citep{dermer1995}. This is confirmed by the latest studies of the spectrum decomposition for this source by \citet{mahmoud2020}.

We searched the NED database for independent measurements of distances for all the sources in our sample. Unfortunately, except for NGC 4151, all other sources only have the $R-L$ and $L_X-L_{UV}$ luminosity distance measurements that we have used. Only the NGC 4151 distance has been measured using other techniques, such as the Tip of the Red Giant Branch (TRGB) (14.20 $\pm$ 0.88 Mpc, \citealt{tikhonov2021}), and the Cepheid period-luminosity relation (15.8 $\pm$ 0.4 Mpc, \citealt{yuan2020}).  

We also explored the potential of selecting a more reliable subsample of these sources by using only those sources whose $R-L$ and $L_X-L_{UV}$ luminosity distances differ by less than a factor of 3 (there are 30 such sources). For the fixed values of $\beta$ and $\gamma$ in the $R-L$ relation in eq.~(\ref{eq_RL}) we find from these 30 sources the constraint $\Omega_{m0} = 0.4^{+0.4}_{-0.2}$ in the flat $\Lambda$CDM model. (Note that this constraint does not fully account for all the uncertainties.) This might be an interesting option in the future when there are more sources with the required measurements for both distance measurement techniques to be applicable, although it is based on the assumption that both the $R-L$ and $L_X-L_{UV}$ relations hold for the data set being used for this purpose and as of now this is an open question for the $L_X-L_{UV}$ case, for both the data set we consider here as well as for the \cite{Lussoetal2020} data set (see \citealt{KhadkaRatra2021a, KhadkaRatra2022}).

In the bottom four panels of Fig.~\ref{fig:rat_z_L} we also plot method-1 ratios of the $R-L$ and $L_X-L_{UV}$ luminosity distances, $\log{(D_{L,R-L}/D_{L,L_X-L_{UV}})}$, as functions of $L_{3000}$, $\alpha_{OX}$, $\tau$, and $L_X$, with red points. The $L_{3000}$, $\alpha_{OX}$, $\tau$, and $L_X$ error bars are not shown in these panels. Two systematic trends are visible in these plots. In the middle right panel of Fig.~\ref{fig:rat_z_L}, we see a dependence of the distance ratio on the broad-band $\alpha_{OX}$ index. The overall trend reflects the fact that $\alpha_{OX}$ is a function of $L_{UV}$, $\alpha_{OX} \propto 0.384 (1 - \gamma) \log L_{UV}$. However, a significant departure of QSOs from this expected correlation also affects the luminosity distance estimation. In addition, we do not see strong correlation with the X-ray luminosity but the correlation with the time delay is again significant.

We also examine and record the quality of the correlations between the distance ratios and the other quantities shown in Fig.~\ref{fig:rat_z_L}. The Pearson as well as Spearman rank-order correlation coefficients and the corresponding $p$-values for the presented relations are listed in Table~\ref{tab_PCC}. The significant correlations are between the distance ratio and $\alpha_{OX}$ as well as the rest-frame time delay, which are both positive. The mildly significant correlation between the distance ratio and the X-ray luminosity $L_X$ is positive as well. These correlations are partially driven by the fact that the luminosity distance ratio depends on $\log{F_X}$ and $\log{\tau}$, see eqs.~\eqref{eq_DL_RL} and \eqref{eq_DL_LXLUV}, and the $\alpha_{OX}$ parameter depends again on $\log{F_{X}}$, see eq.~\eqref{eq_alpha_OX}, hence there is an intrinsic dependence that enhances the correlation. This can explicitly be inferred by evaluating $\log{(D_{L,R-L}/D_{L,L_{X}-L_{UV}})}$ using Eqs.~\eqref{eq_DL_RL} and \eqref{eq_DL_LXLUV},
\begin{align}
   \log & {\left(\frac{D_{L,R-L}}{D_{L,L_{X}-L_{UV}}}\right)} \\
        & \simeq \frac{\log{\tau}}{2\gamma}-\frac{\log{F_{UV}}}{2(1-\gamma')}+\frac{\log{F_{X}}}{2(1-\gamma')}+C(\gamma, \beta, \gamma', \beta') \notag
   \label{eq_DL_ratio}
\end{align}
where $\gamma$ and $\beta$ are the slope and the intercept of the $R-L$ relation and $\gamma'$ and $\beta'$ are the slope and the intercept of the $L_X-L_{UV}$ relation. Hence $C(\gamma, \beta, \gamma',\beta')$ denotes the function of these parameters. In Eq.~\eqref{eq_DL_ratio} we considered $F_{3000}\approx F_{\rm UV}$ for simplicity. Using the approximate values for the slopes, $\gamma\sim 0.3$ and $\gamma'\sim 0.6$, using Eq.~\eqref{eq_alpha_OX} for $\alpha_{OX}$, and using the $R-L$ relation, we can numerically evaluate Eq.~\eqref{eq_DL_ratio},
\begin{align}
    \log{\left(\frac{D_{L,R-L}}{D_{L,L_{X}-L_{UV}}}\right)}&\approx 1.67\log{\tau}-1.25\log{F_{UV}}+1.25\log{F_{X}}+C\,\notag\\
    &\approx 1.67\log\tau + 3.256\alpha_{OX}+C\,\notag\\
    &\approx -0.75 \log{F_{\rm UV}}+1.25\log{F_{\rm X}}+C'\,,
\end{align}
which implies strong dependence of the luminosity distance ratio on $\tau$ and $\alpha_{OX}$ (and proportionally on $L_{X}\propto F_{\rm X}$) while at the same time there is a weaker and negative correlation with the UV flux/luminosity. This is in accordance with the correlation coefficients listed in Table~\ref{tab_PCC}. Taking this into account, the redshift as such is explicitly not present in the luminosity distance ratio, hence the missing correlation implies that there is no significant systematic effect with redshift for our current sample. The correlations among all the quantities should be further evaluated when the sample size increases since then some of the systematic effects may be more apparent on top of the expected intrinsically driven correlations.

\section{Conclusions}
\label{sec:conclusions}

Given that the previous large \cite{Lussoetal2020} $L_X-L_{UV}$ QSO data set that includes 2036 (better) QSOs that span $0.009 \leq z \leq 7.5413$ is not standardizable (see \citealt{KhadkaRatra2021a, KhadkaRatra2022}), our hope here was to compile a new set of such QSOs to investigate the prospects and potential of using $L_X-L_{UV}$ QSO data to constrain cosmological parameters

To this end, we have compiled a small set of 58 QSOs, that span a smaller redshift range $0.0041 \leq z \leq 1.686$, but with both $(\tau, F_{3000})$ and $(F_X, F_{UV})$ data, that allow us to test both the $R-L$ and $L_X-L_{UV}$ relations, to compare these relations, and use these relations to jointly constrain both $R-L$ or $L_X-L_{UV}$ relation parameters and cosmological model parameters.

We have shown that the $R-L$ relation and $L_X-L_{UV}$ relation $\gamma$ and $\beta$ values are almost completely independent of cosmological model, indicating that these $R-L$ relation and $L_X-L_{UV}$ relation QSOs are standardizable.

While the slope, $\gamma$, and intercept, $\beta$, values are significantly better determined for the $L_X-L_{UV}$ data set, the slightly larger value of $\sigma_{\rm ext}$ makes the $L_X-L_{UV}$ data set slightly less reliable than the $R-L$ data set. However, both these data sets are small and both provide very weak constraints on cosmological parameters, with the $R-L$ relation cosmological constraints being slightly more restrictive (see Table \ref{tab:1d_BFP} and Figs.\ \ref{fig:LCDM_triangle}--\ref{fig:phiCDM_triangle}), possibly because of the smaller value of $\sigma_{\rm ext}$.

More importantly, if we look at the trend in the $\Omega_{m0}$ posterior distributions, the $R-L$ relation ones favour lower values of $\Omega_{m0}$ compared to the $L_X-L_{UV}$ relation posterior distributions.\footnote{This is similar to the findings of \citet{Lussoetal2020} and \citet{KhadkaRatra2021a, KhadkaRatra2022} for the larger but non-standardizable \citet{Lussoetal2020} QSO compilation.} This is supported by the results shown in Fig.~\ref{fig:luminosity_distance} where we compare $R-L$ and $L_X-L_{UV}$ luminosity distances for each source (also see Fig.~\ref{fig_histograms_DL_RL_xuv} and the related discussion), which show that $L_X-L_{UV}$ relation luminosity distances are significantly shorter than $R-L$ distances and $\Omega_{m0} = 0.3$ flat $\Lambda$CDM model distances, especially in the $0.8 \lesssim z \lesssim 1.3$ range. This explains why $L_X-L_{UV}$ data favour higher $\Omega_{m0}$ values than 0.3 and higher $\Omega_{m0}$ values than those favoured by $R-L$ data. While there are no independent distance measurements for any of the sources besides NGC 4151 (which is not a good $R-L$ or $L_X-L_{UV}$ source), \Mgii\ and \Civ\ $R-L$ sources are standardizable and provide cosmological constraints consistent with those from better-established data \citep{khadka2021, Khadkaetal2022, Caoetal2022, CaoRatra2022, CaoRatra2023, Czernyetal2022}. Consequently, more work is needed to determine whether the $L_X-L_{UV}$ relation can be used to standardize QSOs.

\section*{Acknowledgements}

This research was supported in part by Dr.\ Richard Jelsma (a Bellarmine University donor), US DOE grant DE-SC0011840, by the Polish Funding Agency National Science Centre, project 2017/26/A/ST9/00756 (Maestro 9), by GAČR EXPRO grant 21-13491X, by Millenium Nucleus NCN$19\_058$ (TITANs), and by the Conselho Nacional de Desenvolvimento Científico e Tecnológico (CNPq) Fellowships (164753/2020-6 and 313497/2022-2). BC and MZ  acknowledge the Czech-Polish mobility program (M\v{S}MT 8J20PL037 and PPN/BCZ/2019/1/00069). Part of the computation for this project was performed on the Beocat Research Cluster at Kansas State University.

\section*{Data Availability}
Data used in this paper are listed in App.\ \ref{sec:appendix_golden}.



\bibliographystyle{mnras}
\bibliography{mybibfile} 

\begin{thebibliography}{}
\makeatletter
\relax
\def\mn@urlcharsother{\let\do\@makeother \do\$\do\&\do\#\do\^\do\_\do\%\do\~}
\def\mn@doi{\begingroup\mn@urlcharsother \@ifnextchar [ {\mn@doi@}
  {\mn@doi@[]}}
\def\mn@doi@[#1]#2{\def\@tempa{#1}\ifx\@tempa\@empty \href
  {http://dx.doi.org/#2} {doi:#2}\else \href {http://dx.doi.org/#2} {#1}\fi
  \endgroup}
\def\mn@eprint#1#2{\mn@eprint@#1:#2::\@nil}
\def\mn@eprint@arXiv#1{\href {http://arxiv.org/abs/#1} {{\tt arXiv:#1}}}
\def\mn@eprint@dblp#1{\href {http://dblp.uni-trier.de/rec/bibtex/#1.xml}
  {dblp:#1}}
\def\mn@eprint@#1:#2:#3:#4\@nil{\def\@tempa {#1}\def\@tempb {#2}\def\@tempc
  {#3}\ifx \@tempc \@empty \let \@tempc \@tempb \let \@tempb \@tempa \fi \ifx
  \@tempb \@empty \def\@tempb {arXiv}\fi \@ifundefined
  {mn@eprint@\@tempb}{\@tempb:\@tempc}{\expandafter \expandafter \csname
  mn@eprint@\@tempb\endcsname \expandafter{\@tempc}}}

\bibitem[\protect\citeauthoryear{{Abdalla} et~al.,}{{Abdalla}
  et~al.}{2022}]{Abdallaetal2022}
{Abdalla} E.,  et~al., 2022, \mn@doi [Journal of High Energy Astrophysics]
  {10.1016/j.jheap.2022.04.002}, \href
  {https://ui.adsabs.harvard.edu/abs/2022JHEAp..34...49A} {34, 49}

\bibitem[\protect\citeauthoryear{{Adil}, {Albrecht}  \& {Knox}}{{Adil}
  et~al.}{2022}]{Adiletal2022}
{Adil} A.,  {Albrecht} A.,   {Knox} L.,  2022, arXiv e-prints, \href
  {https://ui.adsabs.harvard.edu/abs/2022arXiv220710235A} {p. arXiv:2207.10235}

\bibitem[\protect\citeauthoryear{{Arjona} \& {Nesseris}}{{Arjona} \&
  {Nesseris}}{2021}]{ArjonaNesseris2021}
{Arjona} R.,  {Nesseris} S.,  2021, \mn@doi [\prd]
  {10.1103/PhysRevD.103.103539}, \href
  {https://ui.adsabs.harvard.edu/abs/2021PhRvD.103j3539A} {103, 103539}

\bibitem[\protect\citeauthoryear{{Avni} \& {Tananbaum}}{{Avni} \&
  {Tananbaum}}{1986}]{AvniTananbaum1986}
{Avni} Y.,  {Tananbaum} H.,  1986, \mn@doi [\apj] {10.1086/164230}, \href
  {https://ui.adsabs.harvard.edu/abs/1986ApJ...305...83A} {305, 83}

\bibitem[\protect\citeauthoryear{{Banerjee}, {{\'O} Colg{\'a}in}, {Sasaki},
  {Sheikh-Jabbari}  \& {Yang}}{{Banerjee} et~al.}{2021}]{Banerjeeetal2021}
{Banerjee} A.,  {{\'O} Colg{\'a}in} E.,  {Sasaki} M.,  {Sheikh-Jabbari} M.~M.,
   {Yang} T.,  2021, \mn@doi [Physics Letters B]
  {10.1016/j.physletb.2021.136366}, \href
  {https://ui.adsabs.harvard.edu/abs/2021PhLB..81836366B} {818, 136366}

\bibitem[\protect\citeauthoryear{{Bechtold} et~al.,}{{Bechtold}
  et~al.}{1994}]{1994AJ....108..374B}
{Bechtold} J.,  et~al., 1994, \mn@doi [\aj] {10.1086/117076}, \href
  {https://ui.adsabs.harvard.edu/abs/1994AJ....108..374B} {108, 374}

\bibitem[\protect\citeauthoryear{{Bentz} et~al.,}{{Bentz}
  et~al.}{2013}]{2013ApJ...767..149B}
{Bentz} M.~C.,  et~al., 2013, \mn@doi [\apj] {10.1088/0004-637X/767/2/149},
  \href {https://ui.adsabs.harvard.edu/abs/2013ApJ...767..149B} {767, 149}

\bibitem[\protect\citeauthoryear{{Brinckmann} \& {Lesgourgues}}{{Brinckmann} \&
  {Lesgourgues}}{2019}]{Brinckmann2019}
{Brinckmann} T.,  {Lesgourgues} J.,  2019, \mn@doi [Physics of the Dark
  Universe] {10.1016/j.dark.2018.100260}, \href
  {https://ui.adsabs.harvard.edu/abs/2019PDU....24..260B} {24, 100260}

\bibitem[\protect\citeauthoryear{{Cao} \& {Ratra}}{{Cao} \&
  {Ratra}}{2022}]{CaoRatra2022}
{Cao} S.,  {Ratra} B.,  2022, \mn@doi [\mnras] {10.1093/mnras/stac1184}, \href
  {https://ui.adsabs.harvard.edu/abs/2022MNRAS.513.5686C} {513, 5686}

\bibitem[\protect\citeauthoryear{{Cao} \& {Ratra}}{{Cao} \&
  {Ratra}}{2023}]{CaoRatra2023}
{Cao} S.,  {Ratra} B.,  2023, \mn@doi [arXiv e-prints]
  {10.48550/arXiv.2302.14203}, \href
  {https://ui.adsabs.harvard.edu/abs/2023arXiv230214203C} {p. arXiv:2302.14203}

\bibitem[\protect\citeauthoryear{{Cao}, {Zheng}, {Biesiada}, {Qi}, {Chen}  \&
  {Zhu}}{{Cao} et~al.}{2017}]{Caoetal2017}
{Cao} S.,  {Zheng} X.,  {Biesiada} M.,  {Qi} J.,  {Chen} Y.,   {Zhu} Z.-H.,
  2017, \mn@doi [\aap] {10.1051/0004-6361/201730551}, \href
  {https://ui.adsabs.harvard.edu/abs/2017A&A...606A..15C} {606, A15}

\bibitem[\protect\citeauthoryear{{Cao}, {Ryan}  \& {Ratra}}{{Cao}
  et~al.}{2020}]{Caoetal2020}
{Cao} S.,  {Ryan} J.,   {Ratra} B.,  2020, \mn@doi [\mnras]
  {10.1093/mnras/staa2190}, \href
  {https://ui.adsabs.harvard.edu/abs/2020MNRAS.497.3191C} {497, 3191}

\bibitem[\protect\citeauthoryear{{Cao}, {Ryan}, {Khadka}  \& {Ratra}}{{Cao}
  et~al.}{2021a}]{Caoetal2021a}
{Cao} S.,  {Ryan} J.,  {Khadka} N.,   {Ratra} B.,  2021a, \mn@doi [\mnras]
  {10.1093/mnras/staa3748}, \href
  {https://ui.adsabs.harvard.edu/abs/2021MNRAS.501.1520C} {501, 1520}

\bibitem[\protect\citeauthoryear{{Cao}, {Ryan}  \& {Ratra}}{{Cao}
  et~al.}{2021b}]{Caoetal2021b}
{Cao} S.,  {Ryan} J.,   {Ratra} B.,  2021b, \mn@doi [\mnras]
  {10.1093/mnras/stab942}, \href
  {https://ui.adsabs.harvard.edu/abs/2021MNRAS.504..300C} {504, 300}

\bibitem[\protect\citeauthoryear{{Cao}, {Ryan}  \& {Ratra}}{{Cao}
  et~al.}{2022a}]{Caoetal_2021c}
{Cao} S.,  {Ryan} J.,   {Ratra} B.,  2022a, \mn@doi [\mnras]
  {10.1093/mnras/stab3304}, \href
  {https://ui.adsabs.harvard.edu/abs/2022MNRAS.509.4745C} {509, 4745}

\bibitem[\protect\citeauthoryear{{Cao}, {Khadka}  \& {Ratra}}{{Cao}
  et~al.}{2022b}]{CaoKhadkaRatra2022}
{Cao} S.,  {Khadka} N.,   {Ratra} B.,  2022b, \mn@doi [\mnras]
  {10.1093/mnras/stab3559}, \href
  {https://ui.adsabs.harvard.edu/abs/2022MNRAS.510.2928C} {510, 2928}

\bibitem[\protect\citeauthoryear{{Cao}, {Dainotti}  \& {Ratra}}{{Cao}
  et~al.}{2022c}]{CaoDainottiRatra2022a}
{Cao} S.,  {Dainotti} M.,   {Ratra} B.,  2022c, \mn@doi [\mnras]
  {10.1093/mnras/stac517}, \href
  {https://ui.adsabs.harvard.edu/abs/2022MNRAS.512..439C} {512, 439}

\bibitem[\protect\citeauthoryear{{Cao}, {Dainotti}  \& {Ratra}}{{Cao}
  et~al.}{2022d}]{CaoDainottiRatra2022b}
{Cao} S.,  {Dainotti} M.,   {Ratra} B.,  2022d, \mn@doi [\mnras]
  {10.1093/mnras/stac2170}, \href
  {https://ui.adsabs.harvard.edu/abs/2022MNRAS.516.1386C} {516, 1386}

\bibitem[\protect\citeauthoryear{{Cao}, {Zaja{\v{c}}ek}, {Panda},
  {Mart{\'\i}nez-Aldama}, {Czerny}  \& {Ratra}}{{Cao}
  et~al.}{2022e}]{Caoetal2022}
{Cao} S.,  {Zaja{\v{c}}ek} M.,  {Panda} S.,  {Mart{\'\i}nez-Aldama} M.~L.,
  {Czerny} B.,   {Ratra} B.,  2022e, \mn@doi [\mnras] {10.1093/mnras/stac2325},
  \href {https://ui.adsabs.harvard.edu/abs/2022MNRAS.516.1721C} {516, 1721}

\bibitem[\protect\citeauthoryear{{Cardelli}, {Clayton}  \& {Mathis}}{{Cardelli}
  et~al.}{1989}]{cardelli1989}
{Cardelli} J.~A.,  {Clayton} G.~C.,   {Mathis} J.~S.,  1989, \mn@doi [\apj]
  {10.1086/167900}, \href
  {https://ui.adsabs.harvard.edu/abs/1989ApJ...345..245C} {345, 245}

\bibitem[\protect\citeauthoryear{{Ch{\'a}vez}, {Terlevich}, {Terlevich},
  {Bresolin}, {Melnick}, {Plionis}  \& {Basilakos}}{{Ch{\'a}vez}
  et~al.}{2014}]{Chavezetal2014}
{Ch{\'a}vez} R.,  {Terlevich} R.,  {Terlevich} E.,  {Bresolin} F.,  {Melnick}
  J.,  {Plionis} M.,   {Basilakos} S.,  2014, \mn@doi [\mnras]
  {10.1093/mnras/stu987}, \href
  {https://ui.adsabs.harvard.edu/abs/2014MNRAS.442.3565C} {442, 3565}

\bibitem[\protect\citeauthoryear{{Colg{\'a}in}, {Sheikh-Jabbari}, {Solomon},
  {Bargiacchi}, {Capozziello}, {Dainotti}  \& {Stojkovic}}{{Colg{\'a}in}
  et~al.}{2022}]{Colgainetal2022}
{Colg{\'a}in} E.~{\'O}.,  {Sheikh-Jabbari} M.~M.,  {Solomon} R.,  {Bargiacchi}
  G.,  {Capozziello} S.,  {Dainotti} M.~G.,   {Stojkovic} D.,  2022, arXiv
  e-prints, \href {https://ui.adsabs.harvard.edu/abs/2022arXiv220310558C} {p.
  arXiv:2203.10558}

\bibitem[\protect\citeauthoryear{{Czerny}, {Hryniewicz}, {Maity},
  {Schwarzenberg-Czerny}, {{\.Z}ycki}  \& {Bilicki}}{{Czerny}
  et~al.}{2013}]{czerny2013}
{Czerny} B.,  {Hryniewicz} K.,  {Maity} I.,  {Schwarzenberg-Czerny} A.,
  {{\.Z}ycki} P.~T.,   {Bilicki} M.,  2013, \mn@doi [\aap]
  {10.1051/0004-6361/201220832}, \href
  {https://ui.adsabs.harvard.edu/abs/2013A&A...556A..97C} {556, A97}

\bibitem[\protect\citeauthoryear{{Czerny} et~al.,}{{Czerny}
  et~al.}{2019}]{2019ApJ...880...46C}
{Czerny} B.,  et~al., 2019, \mn@doi [\apj] {10.3847/1538-4357/ab2913}, \href
  {https://ui.adsabs.harvard.edu/abs/2019ApJ...880...46C} {880, 46}

\bibitem[\protect\citeauthoryear{{Czerny} et~al.,}{{Czerny}
  et~al.}{2021}]{2021AcPPA.139..389C}
{Czerny} B.,  et~al., 2021, \mn@doi [Acta Physica Polonica A]
  {10.12693/APhysPolA.139.389}, \href
  {https://ui.adsabs.harvard.edu/abs/2021AcPPA.139..389C} {139, 389}

\bibitem[\protect\citeauthoryear{{Czerny} et~al.,}{{Czerny}
  et~al.}{2022}]{Czernyetal2022}
{Czerny} B.,  et~al., 2022, arXiv e-prints, \href
  {https://ui.adsabs.harvard.edu/abs/2022arXiv220906563C} {p. arXiv:2209.06563}

\bibitem[\protect\citeauthoryear{{Czerny} et~al.,}{{Czerny}
  et~al.}{2023}]{2023arXiv230108975C}
{Czerny} B.,  et~al., 2023, \mn@doi [arXiv e-prints]
  {10.48550/arXiv.2301.08975}, \href
  {https://ui.adsabs.harvard.edu/abs/2023arXiv230108975C} {p. arXiv:2301.08975}

\bibitem[\protect\citeauthoryear{{D'Agostini}}{{D'Agostini}}{2005}]{Dago2005}
{D'Agostini} G.,  2005, arXiv e-prints, \href
  {https://ui.adsabs.harvard.edu/abs/2005physics..11182D} {p. physics/0511182}

\bibitem[\protect\citeauthoryear{{DES Collaboration}}{{DES
  Collaboration}}{2019}]{DESCollaboration2019}
{DES Collaboration} 2019, \mn@doi [\prd] {10.1103/PhysRevD.99.123505}, \href
  {https://ui.adsabs.harvard.edu/abs/2019PhRvD..99l3505A} {99, 123505}

\bibitem[\protect\citeauthoryear{{Dahiya} \& {Jain}}{{Dahiya} \&
  {Jain}}{2022}]{DahiyaJain2022}
{Dahiya} D.,  {Jain} D.,  2022, arXiv e-prints, \href
  {https://ui.adsabs.harvard.edu/abs/2022arXiv221204751D} {p. arXiv:2212.04751}

\bibitem[\protect\citeauthoryear{{Dainotti}, {Nielson}, {Sarracino}, {Rinaldi},
  {Nagataki}, {Capozziello}, {Gnedin}  \& {Bargiacchi}}{{Dainotti}
  et~al.}{2022a}]{Dainottietal2022b}
{Dainotti} M.~G.,  {Nielson} V.,  {Sarracino} G.,  {Rinaldi} E.,  {Nagataki}
  S.,  {Capozziello} S.,  {Gnedin} O.~Y.,   {Bargiacchi} G.,  2022a, \mn@doi
  [\mnras] {10.1093/mnras/stac1141}, \href
  {https://ui.adsabs.harvard.edu/abs/2022MNRAS.514.1828D} {514, 1828}

\bibitem[\protect\citeauthoryear{{Dainotti}, {Bargiacchi}, {Lenart},
  {Capozziello}, {{\'O} Colg{\'a}in}, {Solomon}, {Stojkovic}  \&
  {Sheikh-Jabbari}}{{Dainotti} et~al.}{2022b}]{dainotti2022}
{Dainotti} M.~G.,  {Bargiacchi} G.,  {Lenart} A.~{\L}.,  {Capozziello} S.,
  {{\'O} Colg{\'a}in} E.,  {Solomon} R.,  {Stojkovic} D.,   {Sheikh-Jabbari}
  M.~M.,  2022b, \mn@doi [\apj] {10.3847/1538-4357/ac6593}, \href
  {https://ui.adsabs.harvard.edu/abs/2022ApJ...931..106D} {931, 106}

\bibitem[\protect\citeauthoryear{{\MakeLowercase{D}e Cruz Perez}, {Sola
  Peracaula}, {Gomez-Valent}  \& {Moreno-Pulido}}{{\MakeLowercase{D}e Cruz
  Perez} et~al.}{2021}]{deCruzetal2021}
{\MakeLowercase{D}e Cruz Perez} J.,  {Sola Peracaula} J.,  {Gomez-Valent} A.,
  {Moreno-Pulido} C.,  2021, preprint, \href
  {https://ui.adsabs.harvard.edu/abs/2021arXiv211007569D} {} (\mn@eprint {}
  {2110.07569})

\bibitem[\protect\citeauthoryear{{\MakeLowercase{D}e Cruz P{\'e}rez}, {Park}
  \& {Ratra}}{{\MakeLowercase{D}e Cruz P{\'e}rez}
  et~al.}{2022}]{deCruzPerez2022}
{\MakeLowercase{D}e Cruz P{\'e}rez} J.,  {Park} C.-G.,   {Ratra} B.,  2022,
  arXiv e-prints, \href {https://ui.adsabs.harvard.edu/abs/2022arXiv221104268D}
  {p. arXiv:2211.04268}

\bibitem[\protect\citeauthoryear{{\MakeLowercase{D}e~Rosa}, {Piro}, {Perola},
  {Capalbi}, {Cappi}, {Grandi}, {Maraschi}  \&
  {Petrucci}}{{\MakeLowercase{D}e~Rosa} et~al.}{2007}]{2007A&A...463..903D}
{\MakeLowercase{D}e~Rosa} A.,  {Piro} L.,  {Perola} G.~C.,  {Capalbi} M.,
  {Cappi} M.,  {Grandi} P.,  {Maraschi} L.,   {Petrucci} P.~O.,  2007, \mn@doi
  [\aap] {10.1051/0004-6361:20065028}, \href
  {https://ui.adsabs.harvard.edu/abs/2007A&A...463..903D} {463, 903}

\bibitem[\protect\citeauthoryear{Demianski, Piedipalumbo, Sawant  \&
  Amati}{Demianski et~al.}{2021}]{Demianskietal_2021}
Demianski M.,  Piedipalumbo E.,  Sawant D.,   Amati L.,  2021, \mn@doi [\mnras]
  {10.1093/mnras/stab1669}, 506, 903

\bibitem[\protect\citeauthoryear{{Dermer} \& {Gehrels}}{{Dermer} \&
  {Gehrels}}{1995}]{dermer1995}
{Dermer} C.~D.,  {Gehrels} N.,  1995, \mn@doi [\apj] {10.1086/175859}, \href
  {https://ui.adsabs.harvard.edu/abs/1995ApJ...447..103D} {447, 103}

\bibitem[\protect\citeauthoryear{{Dhawan}, {Alsing}  \& {Vagnozzi}}{{Dhawan}
  et~al.}{2021}]{Dhawanetal2021}
{Dhawan} S.,  {Alsing} J.,   {Vagnozzi} S.,  2021, \mn@doi [\mnras]
  {10.1093/mnrasl/slab058}, \href
  {https://ui.adsabs.harvard.edu/abs/2021MNRAS.506L...1D} {506, L1}

\bibitem[\protect\citeauthoryear{{Di Valentino} et~al.,}{{Di Valentino}
  et~al.}{2021a}]{DiValentinoetal2021b}
{Di Valentino} E.,  et~al., 2021a, \mn@doi [Classical and Quantum Gravity]
  {10.1088/1361-6382/ac086d}, \href
  {https://ui.adsabs.harvard.edu/abs/2021CQGra..38o3001D} {38, 153001}

\bibitem[\protect\citeauthoryear{{Di Valentino}, {Melchiorri}  \& {Silk}}{{Di
  Valentino} et~al.}{2021b}]{DiValentinoetal2021a}
{Di Valentino} E.,  {Melchiorri} A.,   {Silk} J.,  2021b, \mn@doi [\apjl]
  {10.3847/2041-8213/abe1c4}, \href
  {https://ui.adsabs.harvard.edu/abs/2021ApJ...908L...9D} {908, L9}

\bibitem[\protect\citeauthoryear{{\MakeLowercase{E}BOSS
  Collaboration}}{{\MakeLowercase{E}BOSS
  Collaboration}}{2021}]{eBOSSCollaboration2021}
{\MakeLowercase{E}BOSS Collaboration} 2021, \mn@doi [\prd]
  {10.1103/PhysRevD.103.083533}, \href
  {https://ui.adsabs.harvard.edu/abs/2021PhRvD.103h3533A} {103, 083533}

\bibitem[\protect\citeauthoryear{{Efstathiou} \& {Gratton}}{{Efstathiou} \&
  {Gratton}}{2020}]{EfstathiouGratton2020}
{Efstathiou} G.,  {Gratton} S.,  2020, \mn@doi [\mnras]
  {10.1093/mnrasl/slaa093}, \href
  {https://ui.adsabs.harvard.edu/abs/2020MNRAS.496L..91E} {496, L91}

\bibitem[\protect\citeauthoryear{{Fana Dirirsa} et~al.,}{{Fana Dirirsa}
  et~al.}{2019}]{Dirirsa2019}
{Fana Dirirsa} F.,  et~al., 2019, \mn@doi [\apj] {10.3847/1538-4357/ab4e11},
  \href {https://ui.adsabs.harvard.edu/abs/2019ApJ...887...13F} {887, 13}

\bibitem[\protect\citeauthoryear{{Geng}, {Hsu}  \& {Lu}}{{Geng}
  et~al.}{2022}]{Gengetal2022}
{Geng} C.-Q.,  {Hsu} Y.-T.,   {Lu} J.-R.,  2022, \mn@doi [\apj]
  {10.3847/1538-4357/ac4495}, \href
  {https://ui.adsabs.harvard.edu/abs/2022ApJ...926...74G} {926, 74}

\bibitem[\protect\citeauthoryear{{Glanville}, {Howlett}  \&
  {Davis}}{{Glanville} et~al.}{2022}]{Glanvilleetal2022}
{Glanville} A.,  {Howlett} C.,   {Davis} T.,  2022, \mn@doi [\mnras]
  {10.1093/mnras/stac2891}, \href
  {https://ui.adsabs.harvard.edu/abs/2022MNRAS.517.3087G} {517, 3087}

\bibitem[\protect\citeauthoryear{{Gonz{\'a}lez-Mor{\'a}n}
  et~al.,}{{Gonz{\'a}lez-Mor{\'a}n} et~al.}{2021}]{GonzalezMoranetal2021}
{Gonz{\'a}lez-Mor{\'a}n} A.~L.,  et~al., 2021, \mn@doi [\mnras]
  {10.1093/mnras/stab1385}, \href
  {https://ui.adsabs.harvard.edu/abs/2021MNRAS.tmp.1358G} {}

\bibitem[\protect\citeauthoryear{{Green} et~al.,}{{Green}
  et~al.}{2009}]{Greenetal2009}
{Green} P.~J.,  et~al., 2009, \mn@doi [\apj] {10.1088/0004-637X/690/1/644},
  \href {https://ui.adsabs.harvard.edu/abs/2009ApJ...690..644G} {690, 644}

\bibitem[\protect\citeauthoryear{{Grupe}, {Komossa}, {Leighly}  \&
  {Page}}{{Grupe} et~al.}{2010}]{Grupeetal2010}
{Grupe} D.,  {Komossa} S.,  {Leighly} K.~M.,   {Page} K.~L.,  2010, \mn@doi
  [\apjs] {10.1088/0067-0049/187/1/64}, \href
  {https://ui.adsabs.harvard.edu/abs/2010ApJS..187...64G} {187, 64}

\bibitem[\protect\citeauthoryear{{Haas}, {Chini}, {Ramolla}, {Pozo Nu{\~n}ez},
  {Westhues}, {Watermann}, {Hoffmeister}  \& {Murphy}}{{Haas}
  et~al.}{2011}]{2011A&A...535A..73H}
{Haas} M.,  {Chini} R.,  {Ramolla} M.,  {Pozo Nu{\~n}ez} F.,  {Westhues} C.,
  {Watermann} R.,  {Hoffmeister} V.,   {Murphy} M.,  2011, \mn@doi [\aap]
  {10.1051/0004-6361/201117325}, \href
  {https://ui.adsabs.harvard.edu/abs/2011A&A...535A..73H} {535, A73}

\bibitem[\protect\citeauthoryear{{Homayouni} et~al.,}{{Homayouni}
  et~al.}{2020}]{Homayouni2020}
{Homayouni} Y.,  et~al., 2020, \mn@doi [\apj] {10.3847/1538-4357/ababa9}, \href
  {https://ui.adsabs.harvard.edu/abs/2020ApJ...901...55H} {901, 55}

\bibitem[\protect\citeauthoryear{{Hu} \& {Wang}}{{Hu} \&
  {Wang}}{2022}]{HuWang2022}
{Hu} J.~P.,  {Wang} F.~Y.,  2022, \mn@doi [\aap] {10.1051/0004-6361/202142162},
  \href {https://ui.adsabs.harvard.edu/abs/2022A&A...661A..71H} {661, A71}

\bibitem[\protect\citeauthoryear{{Hu}, {Wang}  \& {Dai}}{{Hu}
  et~al.}{2021}]{Huetal2021}
{Hu} J.~P.,  {Wang} F.~Y.,   {Dai} Z.~G.,  2021, \mn@doi [\mnras]
  {10.1093/mnras/stab2180}, \href
  {https://ui.adsabs.harvard.edu/abs/2021MNRAS.507..730H} {507, 730}

\bibitem[\protect\citeauthoryear{{Jesus}, {Valentim}, {Escobal}, {Pereira}  \&
  {Benndorf}}{{Jesus} et~al.}{2022}]{Jesusetal2021}
{Jesus} J.~F.,  {Valentim} R.,  {Escobal} A.~A.,  {Pereira} S.~H.,   {Benndorf}
  D.,  2022, \mn@doi [\jcap] {10.1088/1475-7516/2022/11/037}, \href
  {https://ui.adsabs.harvard.edu/abs/2022JCAP...11..037J} {2022, 037}

\bibitem[\protect\citeauthoryear{{Jia}, {Hu}, {Yang}, {Zhang}  \& {Wang}}{{Jia}
  et~al.}{2022}]{Jiaetal2022}
{Jia} X.~D.,  {Hu} J.~P.,  {Yang} J.,  {Zhang} B.~B.,   {Wang} F.~Y.,  2022,
  \mn@doi [\mnras] {10.1093/mnras/stac2356}, \href
  {https://ui.adsabs.harvard.edu/abs/2022MNRAS.516.2575J} {516, 2575}

\bibitem[\protect\citeauthoryear{{Johnson}, {Sangwan}  \&
  {Shankaranarayanan}}{{Johnson} et~al.}{2022}]{Johnsonetal2022}
{Johnson} J.~P.,  {Sangwan} A.,   {Shankaranarayanan} S.,  2022, \mn@doi
  [\jcap] {10.1088/1475-7516/2022/01/024}, \href
  {https://ui.adsabs.harvard.edu/abs/2022JCAP...01..024J} {2022, 024}

\bibitem[\protect\citeauthoryear{{Just}, {Brandt}, {Shemmer}, {Steffen},
  {Schneider}, {Chartas}  \& {Garmire}}{{Just} et~al.}{2007}]{Justetal2007}
{Just} D.~W.,  {Brandt} W.~N.,  {Shemmer} O.,  {Steffen} A.~T.,  {Schneider}
  D.~P.,  {Chartas} G.,   {Garmire} G.~P.,  2007, \mn@doi [\apj]
  {10.1086/519990}, \href
  {https://ui.adsabs.harvard.edu/abs/2007ApJ...665.1004J} {665, 1004}

\bibitem[\protect\citeauthoryear{{Karas}, {Svoboda}  \&
  {Zaja{\v{c}}ek}}{{Karas} et~al.}{2021}]{2021bhns.confE...1K}
{Karas} V.,  {Svoboda} J.,   {Zaja{\v{c}}ek} M.,  2021, in RAGtime: Workshops
  on black holes and netron stars. p.~E1 (\mn@eprint {arXiv} {1901.06507})

\bibitem[\protect\citeauthoryear{{Khadka} \& {Ratra}}{{Khadka} \&
  {Ratra}}{2020a}]{KhadkaRatra2020a}
{Khadka} N.,  {Ratra} B.,  2020a, \mn@doi [\mnras] {10.1093/mnras/staa101},
  \href {https://ui.adsabs.harvard.edu/abs/2020MNRAS.492.4456K} {492, 4456}

\bibitem[\protect\citeauthoryear{{Khadka} \& {Ratra}}{{Khadka} \&
  {Ratra}}{2020b}]{KhadkaRatra2020b}
{Khadka} N.,  {Ratra} B.,  2020b, \mn@doi [\mnras] {10.1093/mnras/staa1855},
  \href {https://ui.adsabs.harvard.edu/abs/2020MNRAS.497..263K} {497, 263}

\bibitem[\protect\citeauthoryear{{Khadka} \& {Ratra}}{{Khadka} \&
  {Ratra}}{2020c}]{KhadkaRatra2020c}
{Khadka} N.,  {Ratra} B.,  2020c, \mn@doi [\mnras] {10.1093/mnras/staa2779},
  \href {https://ui.adsabs.harvard.edu/abs/2020MNRAS.499..391K} {499, 391}

\bibitem[\protect\citeauthoryear{{Khadka} \& {Ratra}}{{Khadka} \&
  {Ratra}}{2021}]{KhadkaRatra2021a}
{Khadka} N.,  {Ratra} B.,  2021, \mn@doi [\mnras] {10.1093/mnras/stab486},
  \href {https://ui.adsabs.harvard.edu/abs/2021MNRAS.502.6140K} {502, 6140}

\bibitem[\protect\citeauthoryear{{Khadka} \& {Ratra}}{{Khadka} \&
  {Ratra}}{2022}]{KhadkaRatra2022}
{Khadka} N.,  {Ratra} B.,  2022, \mn@doi [\mnras] {10.1093/mnras/stab3678},
  \href {https://ui.adsabs.harvard.edu/abs/2022MNRAS.510.2753K} {510, 2753}

\bibitem[\protect\citeauthoryear{{Khadka}, {Yu}, {Zaja{\v{c}}ek},
  {Martinez-Aldama}, {Czerny}  \& {Ratra}}{{Khadka} et~al.}{2021a}]{khadka2021}
{Khadka} N.,  {Yu} Z.,  {Zaja{\v{c}}ek} M.,  {Martinez-Aldama} M.~L.,  {Czerny}
  B.,   {Ratra} B.,  2021a, \mn@doi [\mnras] {10.1093/mnras/stab2807}, \href
  {https://ui.adsabs.harvard.edu/abs/2021MNRAS.508.4722K} {508, 4722}

\bibitem[\protect\citeauthoryear{{Khadka}, {Luongo}, {Muccino}  \&
  {Ratra}}{{Khadka} et~al.}{2021b}]{Khadkaetal2021a}
{Khadka} N.,  {Luongo} O.,  {Muccino} M.,   {Ratra} B.,  2021b, \mn@doi [\jcap]
  {10.1088/1475-7516/2021/09/042}, \href
  {https://ui.adsabs.harvard.edu/abs/2021JCAP...09..042K} {2021, 042}

\bibitem[\protect\citeauthoryear{{Khadka}, {Mart{\'\i}nez-Aldama},
  {Zaja{\v{c}}ek}, {Czerny}  \& {Ratra}}{{Khadka}
  et~al.}{2022a}]{2022MNRAS.513.1985K}
{Khadka} N.,  {Mart{\'\i}nez-Aldama} M.~L.,  {Zaja{\v{c}}ek} M.,  {Czerny} B.,
   {Ratra} B.,  2022a, \mn@doi [\mnras] {10.1093/mnras/stac914}, \href
  {https://ui.adsabs.harvard.edu/abs/2022MNRAS.513.1985K} {513, 1985}

\bibitem[\protect\citeauthoryear{{Khadka}, {Zaja{\v{c}}ek}, {Panda},
  {Mart{\'\i}nez-Aldama}  \& {Ratra}}{{Khadka} et~al.}{2022b}]{Khadkaetal2022}
{Khadka} N.,  {Zaja{\v{c}}ek} M.,  {Panda} S.,  {Mart{\'\i}nez-Aldama} M.~L.,
  {Ratra} B.,  2022b, \mn@doi [\mnras] {10.1093/mnras/stac1940}, \href
  {https://ui.adsabs.harvard.edu/abs/2022MNRAS.515.3729K} {515, 3729}

\bibitem[\protect\citeauthoryear{{KiDS Collaboration}}{{KiDS
  Collaboration}}{2021}]{KiDSCollaboration2021}
{KiDS Collaboration} 2021, \mn@doi [\aap] {10.1051/0004-6361/202039805}, \href
  {https://ui.adsabs.harvard.edu/abs/2021A&A...649A..88T} {649, A88}

\bibitem[\protect\citeauthoryear{{Kumar}, {Rani}, {Jain}, {Mahajan}  \&
  {Mukherjee}}{{Kumar} et~al.}{2022}]{Kumaretal2022}
{Kumar} D.,  {Rani} N.,  {Jain} D.,  {Mahajan} S.,   {Mukherjee} A.,  2022,
  arXiv e-prints, \href {https://ui.adsabs.harvard.edu/abs/2022arXiv221205731K}
  {p. arXiv:2212.05731}

\bibitem[\protect\citeauthoryear{{Lewis}}{{Lewis}}{2019}]{Lewis_2019}
{Lewis} A.,  2019, preprint, \href
  {https://ui.adsabs.harvard.edu/abs/2019arXiv191013970L} {} (\mn@eprint
  {arXiv} {1910.13970})

\bibitem[\protect\citeauthoryear{{Li}, {Huang}  \& {Wang}}{{Li}
  et~al.}{2022}]{2022MNRAS.517.1901L}
{Li} Z.,  {Huang} L.,   {Wang} J.,  2022, \mn@doi [\mnras]
  {10.1093/mnras/stac2735}, \href
  {https://ui.adsabs.harvard.edu/abs/2022MNRAS.517.1901L} {517, 1901}

\bibitem[\protect\citeauthoryear{{Lian}, {Cao}, {Biesiada}, {Chen}, {Zhang}  \&
  {Guo}}{{Lian} et~al.}{2021}]{Lianetal2021}
{Lian} Y.,  {Cao} S.,  {Biesiada} M.,  {Chen} Y.,  {Zhang} Y.,   {Guo} W.,
  2021, \mn@doi [\mnras] {10.1093/mnras/stab1373}, \href
  {https://ui.adsabs.harvard.edu/abs/2021MNRAS.tmp.1338L} {505, 2111–2123}

\bibitem[\protect\citeauthoryear{{Liang}, {Li}, {Xie}  \& {Wu}}{{Liang}
  et~al.}{2022}]{Liangetal2022}
{Liang} N.,  {Li} Z.,  {Xie} X.,   {Wu} P.,  2022, \mn@doi [\apj]
  {10.3847/1538-4357/aca08a}, \href
  {https://ui.adsabs.harvard.edu/abs/2022ApJ...941...84L} {941, 84}

\bibitem[\protect\citeauthoryear{{Liu}, {Liang}, {Xie}, {Yuan}, {Yu}  \&
  {Wu}}{{Liu} et~al.}{2022}]{Liuetal2022}
{Liu} Y.,  {Liang} N.,  {Xie} X.,  {Yuan} Z.,  {Yu} H.,   {Wu} P.,  2022,
  \mn@doi [\apj] {10.3847/1538-4357/ac7de5}, \href
  {https://ui.adsabs.harvard.edu/abs/2022ApJ...935....7L} {935, 7}

\bibitem[\protect\citeauthoryear{{{\L}ukasz Lenart}, {Bargiacchi}, {Dainotti},
  {Nagataki}  \& {Capozziello}}{{{\L}ukasz Lenart}
  et~al.}{2022}]{2022arXiv221110785L}
{{\L}ukasz Lenart} A.,  {Bargiacchi} G.,  {Dainotti} M.~G.,  {Nagataki} S.,
  {Capozziello} S.,  2022, arXiv e-prints, \href
  {https://ui.adsabs.harvard.edu/abs/2022arXiv221110785L} {p. arXiv:2211.10785}

\bibitem[\protect\citeauthoryear{Luongo \& Muccino}{Luongo \&
  Muccino}{2021}]{OrlandoMarco2021}
Luongo O.,  Muccino M.,  2021, \mn@doi [Galaxies] {10.3390/galaxies9040077}, 9

\bibitem[\protect\citeauthoryear{{Luongo}, {Muccino}, {Colg{\'a}in},
  {Sheikh-Jabbari}  \& {Yin}}{{Luongo} et~al.}{2022}]{Luongoetal2021}
{Luongo} O.,  {Muccino} M.,  {Colg{\'a}in} E.~{\'O}.,  {Sheikh-Jabbari} M.~M.,
   {Yin} L.,  2022, \mn@doi [\prd] {10.1103/PhysRevD.105.103510}, \href
  {https://ui.adsabs.harvard.edu/abs/2022PhRvD.105j3510L} {105, 103510}

\bibitem[\protect\citeauthoryear{{Lusso} \& {Risaliti}}{{Lusso} \&
  {Risaliti}}{2016}]{2016ApJ...819..154L}
{Lusso} E.,  {Risaliti} G.,  2016, \mn@doi [\apj]
  {10.3847/0004-637X/819/2/154}, \href
  {https://ui.adsabs.harvard.edu/abs/2016ApJ...819..154L} {819, 154}

\bibitem[\protect\citeauthoryear{{Lusso} et~al.,}{{Lusso}
  et~al.}{2010}]{Lussoetal2010}
{Lusso} E.,  et~al., 2010, \mn@doi [\aap] {10.1051/0004-6361/200913298}, \href
  {https://ui.adsabs.harvard.edu/abs/2010A&A...512A..34L} {512, A34}

\bibitem[\protect\citeauthoryear{{Lusso} et~al.,}{{Lusso}
  et~al.}{2020}]{Lussoetal2020}
{Lusso} E.,  et~al., 2020, \mn@doi [\aap] {10.1051/0004-6361/202038899}, \href
  {https://ui.adsabs.harvard.edu/abs/2020A&A...642A.150L} {642, A150}

\bibitem[\protect\citeauthoryear{{Mahmoud} \& {Done}}{{Mahmoud} \&
  {Done}}{2020}]{mahmoud2020}
{Mahmoud} R.~D.,  {Done} C.,  2020, \mn@doi [\mnras] {10.1093/mnras/stz3196},
  \href {https://ui.adsabs.harvard.edu/abs/2020MNRAS.491.5126M} {491, 5126}

\bibitem[\protect\citeauthoryear{{Mania} \& {Ratra}}{{Mania} \&
  {Ratra}}{2012}]{ManiaRatra2012}
{Mania} D.,  {Ratra} B.,  2012, \mn@doi [Physics Letters B]
  {10.1016/j.physletb.2012.07.011}, \href
  {https://ui.adsabs.harvard.edu/abs/2012PhLB..715....9M} {715, 9}

\bibitem[\protect\citeauthoryear{{Mart{\'\i}nez-Aldama}, {Czerny}, {Kawka},
  {Karas}, {Panda}, {Zaja{\v{c}}ek}  \& {{\.Z}ycki}}{{Mart{\'\i}nez-Aldama}
  et~al.}{2019}]{Mary2019}
{Mart{\'\i}nez-Aldama} M.~L.,  {Czerny} B.,  {Kawka} D.,  {Karas} V.,  {Panda}
  S.,  {Zaja{\v{c}}ek} M.,   {{\.Z}ycki} P.~T.,  2019, \mn@doi [\apj]
  {10.3847/1538-4357/ab3728}, \href
  {https://ui.adsabs.harvard.edu/abs/2019ApJ...883..170M} {883, 170}

\bibitem[\protect\citeauthoryear{{Mart{\'\i}nez-Aldama}, {Zaja{\v{c}}ek},
  {Czerny}  \& {Panda}}{{Mart{\'\i}nez-Aldama} et~al.}{2020}]{Mary2020}
{Mart{\'\i}nez-Aldama} M.~L.,  {Zaja{\v{c}}ek} M.,  {Czerny} B.,   {Panda} S.,
  2020, \mn@doi [\apj] {10.3847/1538-4357/abb6f8}, \href
  {https://ui.adsabs.harvard.edu/abs/2020ApJ...903...86M} {903, 86}

\bibitem[\protect\citeauthoryear{{Mehrabi} et~al.,}{{Mehrabi}
  et~al.}{2022}]{Mehrabietal2022}
{Mehrabi} A.,  et~al., 2022, \mn@doi [\mnras] {10.1093/mnras/stab2915}, \href
  {https://ui.adsabs.harvard.edu/abs/2022MNRAS.509..224M} {509, 224}

\bibitem[\protect\citeauthoryear{{Metzroth}, {Onken}  \& {Peterson}}{{Metzroth}
  et~al.}{2006}]{2006ApJ...647..901M}
{Metzroth} K.~G.,  {Onken} C.~A.,   {Peterson} B.~M.,  2006, \mn@doi [\apj]
  {10.1086/505525}, \href
  {https://ui.adsabs.harvard.edu/abs/2006ApJ...647..901M} {647, 901}

\bibitem[\protect\citeauthoryear{{Mukherjee} \& {Banerjee}}{{Mukherjee} \&
  {Banerjee}}{2022}]{MukherjeeBanerjee2022}
{Mukherjee} P.,  {Banerjee} N.,  2022, \mn@doi [\prd]
  {10.1103/PhysRevD.105.063516}, \href
  {https://ui.adsabs.harvard.edu/abs/2022PhRvD.105f3516M} {105, 063516}

\bibitem[\protect\citeauthoryear{{Ooba}, {Ratra}  \& {Sugiyama}}{{Ooba}
  et~al.}{2018a}]{Oobaetal2018a}
{Ooba} J.,  {Ratra} B.,   {Sugiyama} N.,  2018a, \mn@doi [\apj]
  {10.3847/1538-4357/aad633}, \href
  {https://ui.adsabs.harvard.edu/abs/2018ApJ...864...80O} {864, 80}

\bibitem[\protect\citeauthoryear{{Ooba}, {Ratra}  \& {Sugiyama}}{{Ooba}
  et~al.}{2018b}]{Oobaetal2018c}
{Ooba} J.,  {Ratra} B.,   {Sugiyama} N.,  2018b, \mn@doi [\apj]
  {10.3847/1538-4357/aadcf3}, \href
  {https://ui.adsabs.harvard.edu/abs/2018ApJ...866...68O} {866, 68}

\bibitem[\protect\citeauthoryear{{Ooba}, {Ratra}  \& {Sugiyama}}{{Ooba}
  et~al.}{2018c}]{Oobaetal2018b}
{Ooba} J.,  {Ratra} B.,   {Sugiyama} N.,  2018c, \mn@doi [\apj]
  {10.3847/1538-4357/aaec6f}, \href
  {https://ui.adsabs.harvard.edu/abs/2018ApJ...869...34O} {869, 34}

\bibitem[\protect\citeauthoryear{{Ooba}, {Ratra}  \& {Sugiyama}}{{Ooba}
  et~al.}{2019}]{Oobaetal2019}
{Ooba} J.,  {Ratra} B.,   {Sugiyama} N.,  2019, \mn@doi [\apss]
  {10.1007/s10509-019-3663-4}, \href
  {https://ui.adsabs.harvard.edu/abs/2019Ap&SS.364..176O} {364, 176}

\bibitem[\protect\citeauthoryear{{Panda}}{{Panda}}{2022}]{Panda_2022FrASS...950409P}
{Panda} S.,  2022, \mn@doi [Frontiers in Astronomy and Space Sciences]
  {10.3389/fspas.2022.850409}, \href
  {https://ui.adsabs.harvard.edu/abs/2022FrASS...950409P} {9, 850409}

\bibitem[\protect\citeauthoryear{{Panda} \& {Marziani}}{{Panda} \&
  {Marziani}}{2022}]{2022arXiv221015041P}
{Panda} S.,  {Marziani} P.,  2022, arXiv e-prints, \href
  {https://ui.adsabs.harvard.edu/abs/2022arXiv221015041P} {p. arXiv:2210.15041}

\bibitem[\protect\citeauthoryear{{Park} \& {Ratra}}{{Park} \&
  {Ratra}}{2018}]{ParkRatra2018}
{Park} C.-G.,  {Ratra} B.,  2018, \mn@doi [\apj] {10.3847/1538-4357/aae82d},
  \href {https://ui.adsabs.harvard.edu/abs/2018ApJ...868...83P} {868, 83}

\bibitem[\protect\citeauthoryear{{Park} \& {Ratra}}{{Park} \&
  {Ratra}}{2019a}]{ParkRatra2019b}
{Park} C.-G.,  {Ratra} B.,  2019a, \mn@doi [\apss] {10.1007/s10509-019-3567-3},
  \href {https://ui.adsabs.harvard.edu/abs/2019Ap&SS.364...82P} {364, 82}

\bibitem[\protect\citeauthoryear{{Park} \& {Ratra}}{{Park} \&
  {Ratra}}{2019b}]{ParkRatra2019c}
{Park} C.-G.,  {Ratra} B.,  2019b, \mn@doi [\apss] {10.1007/s10509-019-3627-8},
  \href {https://ui.adsabs.harvard.edu/abs/2019Ap&SS.364..134P} {364, 134}

\bibitem[\protect\citeauthoryear{{Park} \& {Ratra}}{{Park} \&
  {Ratra}}{2019c}]{ParkRatra2019a}
{Park} C.-G.,  {Ratra} B.,  2019c, \mn@doi [\apj] {10.3847/1538-4357/ab3641},
  \href {https://ui.adsabs.harvard.edu/abs/2019ApJ...882..158P} {882, 158}

\bibitem[\protect\citeauthoryear{{Park} \& {Ratra}}{{Park} \&
  {Ratra}}{2020}]{ParkRatra2020}
{Park} C.-G.,  {Ratra} B.,  2020, \mn@doi [\prd] {10.1103/PhysRevD.101.083508},
  \href {https://ui.adsabs.harvard.edu/abs/2020PhRvD.101h3508P} {101, 083508}

\bibitem[\protect\citeauthoryear{{Pavlov}, {Westmoreland}, {Saaidi}  \&
  {Ratra}}{{Pavlov} et~al.}{2013}]{Pavlovetal2013}
{Pavlov} A.,  {Westmoreland} S.,  {Saaidi} K.,   {Ratra} B.,  2013, \mn@doi
  [\prd] {10.1103/PhysRevD.88.123513}, \href
  {https://ui.adsabs.harvard.edu/abs/2013PhRvD..88l3513P} {88, 123513}

\bibitem[\protect\citeauthoryear{{Peebles}}{{Peebles}}{1984}]{Peebles1984}
{Peebles} P.~J.~E.,  1984, \mn@doi [\apj] {10.1086/162425}, \href
  {https://ui.adsabs.harvard.edu/abs/1984ApJ...284..439P} {284, 439}

\bibitem[\protect\citeauthoryear{{Peebles} \& {Ratra}}{{Peebles} \&
  {Ratra}}{1988}]{PeeblesRatra1988}
{Peebles} P.~J.~E.,  {Ratra} B.,  1988, \mn@doi [\apjl] {10.1086/185100}, \href
  {https://ui.adsabs.harvard.edu/abs/1988ApJ...325L..17P} {325, L17}

\bibitem[\protect\citeauthoryear{{Perivolaropoulos} \&
  {Skara}}{{Perivolaropoulos} \& {Skara}}{2022}]{PerivolaropoulosSkara2021}
{Perivolaropoulos} L.,  {Skara} F.,  2022, \mn@doi [\nar]
  {10.1016/j.newar.2022.101659}, \href
  {https://ui.adsabs.harvard.edu/abs/2022NewAR..9501659P} {95, 101659}

\bibitem[\protect\citeauthoryear{{Petrosian}, {Singal}  \&
  {Mutchnick}}{{Petrosian} et~al.}{2022}]{Petrosianetal2022}
{Petrosian} V.,  {Singal} J.,   {Mutchnick} S.,  2022, \mn@doi [\apjl]
  {10.3847/2041-8213/ac85ac}, \href
  {https://ui.adsabs.harvard.edu/abs/2022ApJ...935L..19P} {935, L19}

\bibitem[\protect\citeauthoryear{{Planck Collaboration}}{{Planck
  Collaboration}}{2020}]{PlanckCollaboration2020}
{Planck Collaboration} 2020, \mn@doi [\aap] {10.1051/0004-6361/201833910},
  \href {https://ui.adsabs.harvard.edu/abs/2020A&A...641A...6P} {641, A6}

\bibitem[\protect\citeauthoryear{{Podariu}, {Souradeep}, {Gott}, {Ratra}  \&
  {Vogeley}}{{Podariu} et~al.}{2001}]{Podariuetal2001}
{Podariu} S.,  {Souradeep} T.,  {Gott} J.~Richard I.,  {Ratra} B.,   {Vogeley}
  M.~S.,  2001, \mn@doi [\apj] {10.1086/322409}, \href
  {https://ui.adsabs.harvard.edu/abs/2001ApJ...559....9P} {559, 9}

\bibitem[\protect\citeauthoryear{{Pourojaghi}, {Zabihi}  \&
  {Malekjani}}{{Pourojaghi} et~al.}{2022}]{Pourojaghietal2022}
{Pourojaghi} S.,  {Zabihi} N.~F.,   {Malekjani} M.,  2022, arXiv e-prints,
  \href {https://ui.adsabs.harvard.edu/abs/2022arXiv221204118P} {p.
  arXiv:2212.04118}

\bibitem[\protect\citeauthoryear{{Prince} et~al.,}{{Prince}
  et~al.}{2022}]{2022Prince}
{Prince} R.,  et~al., 2022, \mn@doi [\aap] {10.1051/0004-6361/202243194}, \href
  {https://ui.adsabs.harvard.edu/abs/2022A&A...667A..42P} {667, A42}

\bibitem[\protect\citeauthoryear{{Rana}, {Jain}, {Mahajan}  \&
  {Mukherjee}}{{Rana} et~al.}{2017}]{Ranaetal2017}
{Rana} A.,  {Jain} D.,  {Mahajan} S.,   {Mukherjee} A.,  2017, \mn@doi [\jcap]
  {10.1088/1475-7516/2017/03/028}, \href
  {https://ui.adsabs.harvard.edu/abs/2017JCAP...03..028R} {2017, 028}

\bibitem[\protect\citeauthoryear{{Ratra} \& {Peebles}}{{Ratra} \&
  {Peebles}}{1988}]{RatraPeebles1988}
{Ratra} B.,  {Peebles} P.~J.~E.,  1988, \mn@doi [\prd]
  {10.1103/PhysRevD.37.3406}, \href
  {https://ui.adsabs.harvard.edu/abs/1988PhRvD..37.3406R} {37, 3406}

\bibitem[\protect\citeauthoryear{{Renzi}, {Hogg}  \& {Giar{\`e}}}{{Renzi}
  et~al.}{2022}]{Renzietal2021}
{Renzi} F.,  {Hogg} N.~B.,   {Giar{\`e}} W.,  2022, \mn@doi [\mnras]
  {10.1093/mnras/stac1030}, \href
  {https://ui.adsabs.harvard.edu/abs/2022MNRAS.513.4004R} {513, 4004}

\bibitem[\protect\citeauthoryear{{Rezaei}, {Sol{\`a} Peracaula}  \&
  {Malekjani}}{{Rezaei} et~al.}{2022}]{Rezaeietal2022}
{Rezaei} M.,  {Sol{\`a} Peracaula} J.,   {Malekjani} M.,  2022, \mn@doi
  [\mnras] {10.1093/mnras/stab3117}, \href
  {https://ui.adsabs.harvard.edu/abs/2022MNRAS.509.2593R} {509, 2593}

\bibitem[\protect\citeauthoryear{{Risaliti} \& {Lusso}}{{Risaliti} \&
  {Lusso}}{2015}]{RisalitiLusso2015}
{Risaliti} G.,  {Lusso} E.,  2015, \mn@doi [\apj] {10.1088/0004-637X/815/1/33},
  \href {https://ui.adsabs.harvard.edu/abs/2015ApJ...815...33R} {815, 33}

\bibitem[\protect\citeauthoryear{{Risaliti} \& {Lusso}}{{Risaliti} \&
  {Lusso}}{2019}]{RisalitiLusso2019}
{Risaliti} G.,  {Lusso} E.,  2019, \mn@doi [Nature Astronomy]
  {10.1038/s41550-018-0657-z}, \href
  {https://ui.adsabs.harvard.edu/abs/2019NatAs...3..272R} {3, 272}

\bibitem[\protect\citeauthoryear{{Ryan}, {Doshi}  \& {Ratra}}{{Ryan}
  et~al.}{2018}]{Ryanetal2018}
{Ryan} J.,  {Doshi} S.,   {Ratra} B.,  2018, \mn@doi [\mnras]
  {10.1093/mnras/sty1922}, \href
  {https://ui.adsabs.harvard.edu/abs/2018MNRAS.480..759R} {480, 759}

\bibitem[\protect\citeauthoryear{{Ryan}, {Chen}  \& {Ratra}}{{Ryan}
  et~al.}{2019}]{Ryanetal2019}
{Ryan} J.,  {Chen} Y.,   {Ratra} B.,  2019, \mn@doi [\mnras]
  {10.1093/mnras/stz1966}, \href
  {https://ui.adsabs.harvard.edu/abs/2019MNRAS.488.3844R} {488, 3844}

\bibitem[\protect\citeauthoryear{{Sacchi} et~al.,}{{Sacchi}
  et~al.}{2022}]{sacchi2022}
{Sacchi} A.,  et~al., 2022, \mn@doi [\aap] {10.1051/0004-6361/202243411}, \href
  {https://ui.adsabs.harvard.edu/abs/2022A&A...663L...7S} {663, L7}

\bibitem[\protect\citeauthoryear{{Scolnic} et~al.,}{{Scolnic}
  et~al.}{2018}]{Scolnicetal2018}
{Scolnic} D.~M.,  et~al., 2018, \mn@doi [\apj] {10.3847/1538-4357/aab9bb},
  \href {https://ui.adsabs.harvard.edu/abs/2018ApJ...859..101S} {859, 101}

\bibitem[\protect\citeauthoryear{{Shen} et~al.,}{{Shen}
  et~al.}{2016}]{2016ApJ...818...30S}
{Shen} Y.,  et~al., 2016, \mn@doi [\apj] {10.3847/0004-637X/818/1/30}, \href
  {https://ui.adsabs.harvard.edu/abs/2016ApJ...818...30S} {818, 30}

\bibitem[\protect\citeauthoryear{{Shen} et~al.,}{{Shen}
  et~al.}{2019}]{shen_2019_sample_characterization}
{Shen} Y.,  et~al., 2019, \mn@doi [\apjs] {10.3847/1538-4365/ab074f}, \href
  {https://ui.adsabs.harvard.edu/abs/2019ApJS..241...34S} {241, 34}

\bibitem[\protect\citeauthoryear{{Shull}, {Stevans}  \& {Danforth}}{{Shull}
  et~al.}{2012}]{shull_2012}
{Shull} J.~M.,  {Stevans} M.,   {Danforth} C.~W.,  2012, \mn@doi [\apj]
  {10.1088/0004-637X/752/2/162}, \href
  {https://ui.adsabs.harvard.edu/abs/2012ApJ...752..162S} {752, 162}

\bibitem[\protect\citeauthoryear{{Singh}, {Sangwan}  \& {Jassal}}{{Singh}
  et~al.}{2019}]{Singhetal2019}
{Singh} A.,  {Sangwan} A.,   {Jassal} H.~K.,  2019, \mn@doi [\jcap]
  {10.1088/1475-7516/2019/04/047}, \href
  {https://ui.adsabs.harvard.edu/abs/2019JCAP...04..047S} {2019, 047}

\bibitem[\protect\citeauthoryear{{Sinha} \& {Banerjee}}{{Sinha} \&
  {Banerjee}}{2021}]{SinhaBanerjee2021}
{Sinha} S.,  {Banerjee} N.,  2021, \mn@doi [\jcap]
  {10.1088/1475-7516/2021/04/060}, \href
  {https://ui.adsabs.harvard.edu/abs/2021JCAP...04..060S} {2021, 060}

\bibitem[\protect\citeauthoryear{{Sol{\`a} Peracaula}, {G{\'o}mez-Valent}  \&
  {de Cruz P{\'e}rez}}{{Sol{\`a} Peracaula}
  et~al.}{2019}]{SolaPercaulaetal2019}
{Sol{\`a} Peracaula} J.,  {G{\'o}mez-Valent} A.,   {de Cruz P{\'e}rez} J.,
  2019, \mn@doi [Physics of the Dark Universe] {10.1016/j.dark.2019.100311},
  \href {https://ui.adsabs.harvard.edu/abs/2019PDU....25..311S} {25, 100311}

\bibitem[\protect\citeauthoryear{{Steffen}, {Strateva}, {Brandt}, {Alexander},
  {Koekemoer}, {Lehmer}, {Schneider}  \& {Vignali}}{{Steffen}
  et~al.}{2006}]{Steffenetal2006}
{Steffen} A.~T.,  {Strateva} I.,  {Brandt} W.~N.,  {Alexander} D.~M.,
  {Koekemoer} A.~M.,  {Lehmer} B.~D.,  {Schneider} D.~P.,   {Vignali} C.,
  2006, \mn@doi [\aj] {10.1086/503627}, \href
  {https://ui.adsabs.harvard.edu/abs/2006AJ....131.2826S} {131, 2826}

\bibitem[\protect\citeauthoryear{{Tananbaum} et~al.,}{{Tananbaum}
  et~al.}{1979}]{Tananbaumetal1979}
{Tananbaum} H.,  et~al., 1979, \mn@doi [\apjl] {10.1086/183100}, \href
  {https://ui.adsabs.harvard.edu/abs/1979ApJ...234L...9T} {234, L9}

\bibitem[\protect\citeauthoryear{{Tikhonov} \& {Galazutdinova}}{{Tikhonov} \&
  {Galazutdinova}}{2021}]{tikhonov2021}
{Tikhonov} N.~A.,  {Galazutdinova} O.~A.,  2021, \mn@doi [Astrophysical
  Bulletin] {10.1134/S199034132103010X}, \href
  {https://ui.adsabs.harvard.edu/abs/2021AstBu..76..255T} {76, 255}

\bibitem[\protect\citeauthoryear{{Ure{\~n}a-L{\'o}pez} \&
  {Roy}}{{Ure{\~n}a-L{\'o}pez} \& {Roy}}{2020}]{UrenaLopezRoy2020}
{Ure{\~n}a-L{\'o}pez} L.~A.,  {Roy} N.,  2020, \mn@doi [\prd]
  {10.1103/PhysRevD.102.063510}, \href
  {https://ui.adsabs.harvard.edu/abs/2020PhRvD.102f3510U} {102, 063510}

\bibitem[\protect\citeauthoryear{{Vagnetti}, {Turriziani}, {Trevese}  \&
  {Antonucci}}{{Vagnetti} et~al.}{2010}]{Vagnettietal2010}
{Vagnetti} F.,  {Turriziani} S.,  {Trevese} D.,   {Antonucci} M.,  2010,
  \mn@doi [\aap] {10.1051/0004-6361/201014320}, \href
  {https://ui.adsabs.harvard.edu/abs/2010A&A...519A..17V} {519, A17}

\bibitem[\protect\citeauthoryear{{Vanden Berk} et~al.,}{{Vanden Berk}
  et~al.}{2001}]{2001AJ....122..549V}
{Vanden Berk} D.~E.,  et~al., 2001, \mn@doi [\aj] {10.1086/321167}, \href
  {https://ui.adsabs.harvard.edu/abs/2001AJ....122..549V} {122, 549}

\bibitem[\protect\citeauthoryear{{Vanden Berk}, {Wesolowski}, {Yeckley},
  {Marcinik}, {Quashnock}, {Machia}  \& {Wu}}{{Vanden Berk}
  et~al.}{2020}]{VandenBerk_etal_2020}
{Vanden Berk} D.~E.,  {Wesolowski} S.~C.,  {Yeckley} M.~J.,  {Marcinik} J.~M.,
  {Quashnock} J.~M.,  {Machia} L.~M.,   {Wu} J.,  2020, \mn@doi [\mnras]
  {10.1093/mnras/staa411}, \href
  {https://ui.adsabs.harvard.edu/abs/2020MNRAS.493.2745V} {493, 2745}

\bibitem[\protect\citeauthoryear{{Wang}, {Wang}, {Cheng}  \& {Dai}}{{Wang}
  et~al.}{2016}]{Wang_2016}
{Wang} J.~S.,  {Wang} F.~Y.,  {Cheng} K.~S.,   {Dai} Z.~G.,  2016, \mn@doi
  [\aap] {10.1051/0004-6361/201526485}, \href
  {https://ui.adsabs.harvard.edu/abs/2016A&A...585A..68W} {585, A68}

\bibitem[\protect\citeauthoryear{{Wang} et~al.,}{{Wang}
  et~al.}{2021}]{2021ApJ...908...53W}
{Wang} F.,  et~al., 2021, \mn@doi [\apj] {10.3847/1538-4357/abcc5e}, \href
  {https://ui.adsabs.harvard.edu/abs/2021ApJ...908...53W} {908, 53}

\bibitem[\protect\citeauthoryear{{Wang}, {Hu}, {Zhang}  \& {Dai}}{{Wang}
  et~al.}{2022a}]{Wangetal2022}
{Wang} F.~Y.,  {Hu} J.~P.,  {Zhang} G.~Q.,   {Dai} Z.~G.,  2022a, \mn@doi
  [\apj] {10.3847/1538-4357/ac3755}, \href
  {https://ui.adsabs.harvard.edu/abs/2022ApJ...924...97W} {924, 97}

\bibitem[\protect\citeauthoryear{{Wang}, {Liu}, {Yuan}, {Liang}, {Yu}  \&
  {Wu}}{{Wang} et~al.}{2022b}]{2022arXiv221014432W}
{Wang} B.,  {Liu} Y.,  {Yuan} Z.,  {Liang} N.,  {Yu} H.,   {Wu} P.,  2022b,
  \mn@doi [\apj] {10.3847/1538-4357/ac9df8}, \href
  {https://ui.adsabs.harvard.edu/abs/2022ApJ...940..174W} {940, 174}

\bibitem[\protect\citeauthoryear{{Watson}, {Denney}, {Vestergaard}  \&
  {Davis}}{{Watson} et~al.}{2011}]{2011ApJ...740L..49W}
{Watson} D.,  {Denney} K.~D.,  {Vestergaard} M.,   {Davis} T.~M.,  2011,
  \mn@doi [\apjl] {10.1088/2041-8205/740/2/L49}, \href
  {https://ui.adsabs.harvard.edu/abs/2011ApJ...740L..49W} {740, L49}

\bibitem[\protect\citeauthoryear{{Wei} \& {Melia}}{{Wei} \&
  {Melia}}{2022}]{WeiMelia2022}
{Wei} J.-J.,  {Melia} F.,  2022, \mn@doi [\apj] {10.3847/1538-4357/ac562c},
  \href {https://ui.adsabs.harvard.edu/abs/2022ApJ...928..165W} {928, 165}

\bibitem[\protect\citeauthoryear{{Wu}, {Qi}  \& {Zhang}}{{Wu}
  et~al.}{2022}]{Wuetal2022}
{Wu} P.-J.,  {Qi} J.-Z.,   {Zhang} X.,  2022, arXiv e-prints, \href
  {https://ui.adsabs.harvard.edu/abs/2022arXiv220908502W} {p. arXiv:2209.08502}

\bibitem[\protect\citeauthoryear{{Xu}, {Chen}, {Xu}  \& {Cao}}{{Xu}
  et~al.}{2022}]{Xuetal2021}
{Xu} T.,  {Chen} Y.,  {Xu} L.,   {Cao} S.,  2022, \mn@doi [Physics of the Dark
  Universe] {10.1016/j.dark.2022.101023}, \href
  {https://ui.adsabs.harvard.edu/abs/2022PDU....3601023X} {36, 101023}

\bibitem[\protect\citeauthoryear{{Young}, {Elvis}  \& {Risaliti}}{{Young}
  et~al.}{2010}]{Youngetal2010}
{Young} M.,  {Elvis} M.,   {Risaliti} G.,  2010, \mn@doi [\apj]
  {10.1088/0004-637X/708/2/1388}, \href
  {https://ui.adsabs.harvard.edu/abs/2010ApJ...708.1388Y} {708, 1388}

\bibitem[\protect\citeauthoryear{{Yu}, {Ratra}  \& {Wang}}{{Yu}
  et~al.}{2018}]{Yuetal2018}
{Yu} H.,  {Ratra} B.,   {Wang} F.-Y.,  2018, \mn@doi [\apj]
  {10.3847/1538-4357/aab0a2}, \href
  {https://ui.adsabs.harvard.edu/abs/2018ApJ...856....3Y} {856, 3}

\bibitem[\protect\citeauthoryear{{Yu} et~al.,}{{Yu} et~al.}{2021}]{Zhefu2021}
{Yu} Z.,  et~al., 2021, \mn@doi [\mnras] {10.1093/mnras/stab2244}, \href
  {https://ui.adsabs.harvard.edu/abs/2021MNRAS.507.3771Y} {507, 3771}

\bibitem[\protect\citeauthoryear{{Yu} et~al.,}{{Yu} et~al.}{2022}]{Yuetal2022}
{Yu} Z.,  et~al., 2022, preprint, \href
  {https://ui.adsabs.harvard.edu/abs/2022arXiv220805491Y} {} (\mn@eprint {}
  {2208.05491})

\bibitem[\protect\citeauthoryear{{Yuan} et~al.,}{{Yuan}
  et~al.}{2020}]{yuan2020}
{Yuan} W.,  et~al., 2020, \mn@doi [\apj] {10.3847/1538-4357/abb377}, \href
  {https://ui.adsabs.harvard.edu/abs/2020ApJ...902...26Y} {902, 26}

\bibitem[\protect\citeauthoryear{{Zaja{\v{c}}ek} et~al.,}{{Zaja{\v{c}}ek}
  et~al.}{2020}]{Michal2020}
{Zaja{\v{c}}ek} M.,  et~al., 2020, \mn@doi [\apj] {10.3847/1538-4357/ab94ae},
  \href {https://ui.adsabs.harvard.edu/abs/2020ApJ...896..146Z} {896, 146}

\bibitem[\protect\citeauthoryear{{Zaja{\v{c}}ek} et~al.,}{{Zaja{\v{c}}ek}
  et~al.}{2021}]{Michal2021}
{Zaja{\v{c}}ek} M.,  et~al., 2021, \mn@doi [\apj] {10.3847/1538-4357/abe9b2},
  \href {https://ui.adsabs.harvard.edu/abs/2021ApJ...912...10Z} {912, 10}

\bibitem[\protect\citeauthoryear{{Zamorani} et~al.,}{{Zamorani}
  et~al.}{1981}]{Zamoranietal1981}
{Zamorani} G.,  et~al., 1981, \mn@doi [\apj] {10.1086/158815}, \href
  {https://ui.adsabs.harvard.edu/abs/1981ApJ...245..357Z} {245, 357}

\bibitem[\protect\citeauthoryear{{Zhai}, {Blanton}, {Slosar}  \&
  {Tinker}}{{Zhai} et~al.}{2017}]{Zhaietal2017}
{Zhai} Z.,  {Blanton} M.,  {Slosar} A.,   {Tinker} J.,  2017, \mn@doi [\apj]
  {10.3847/1538-4357/aa9888}, \href
  {https://ui.adsabs.harvard.edu/abs/2017ApJ...850..183Z} {850, 183}

\bibitem[\protect\citeauthoryear{{Zheng}, {Cao}, {Biesiada}, {Li}, {Liu}  \&
  {Liu}}{{Zheng} et~al.}{2021}]{Zhengetal2021}
{Zheng} X.,  {Cao} S.,  {Biesiada} M.,  {Li} X.,  {Liu} T.,   {Liu} Y.,  2021,
  \mn@doi [Science China Physics, Mechanics, and Astronomy]
  {10.1007/s11433-020-1664-9}, \href
  {https://ui.adsabs.harvard.edu/abs/2021SCPMA..6459511Z} {64, 259511}

\makeatother
\end{thebibliography}




\onecolumn
\begin{appendix}
\section{X-ray detected \Mgii\ QSOs}
\label{sec:appendix_golden}
\addtolength{\tabcolsep}{0pt}
\LTcapwidth=\linewidth
\begin{longtable}{lccccccc}
\caption{Sample of 58 X-ray detected reverberation-mapped \Mgii\ QSOs. From left to right, the columns stand for object ID, redshift, 2 keV X-ray flux density per frequency, 2500 \AA\ UV flux density per frequency, 3000 \AA\ UV flux density, \Mgii\ broad-line rest-frame time delay, the spectral slope between 2 keV and 2500 \AA, (using the convention $F_{\nu}\propto \nu^{+\alpha_{OX}}$), and the original reference. For the \Mgii\ radius-luminosity analysis we consider two \Mgii\ time-delay determinations for NGC 4151, hence the total number of $R-L$ measurements is 59. Reference notations stand for: (a) \citet{Homayouni2020} (object IDs stand for RMIDs from the original catalogue), (b) \citet{2016ApJ...818...30S}, (c) \citet{shen_2019_sample_characterization}, (d) \citet{2006ApJ...647..901M}, and (e) \citet{Zhefu2021}.}
\label{tab_xray_uv_data}\\
\hline\hline
Object &  $z$ &  $\log \left(\frac{F_X}{{\rm erg\,s^{-1}\,cm^{-2}\,Hz^{-1}}} \right)$  & $\log \left(\frac{F_{UV}}{{\rm erg\,s^{-1}\,cm^{-2}\,Hz^{-1}}} \right)$ & $\log \left(\frac{F_{\rm 3000}}{{\rm erg\,s^{-1}\,cm^{-2}}} \right)$ & $\frac{\tau}{{\rm days}}$  & $\alpha_{\rm OX}$ & Ref.\\
\hline
\endfirsthead
\endhead
\hline
\endfoot
        18  &   0.8480 &  $-31.6897 \pm 0.0764$ &  $-28.1437 \pm 0.0012$  & $-13.1079 \pm  0.0009$  &    $125.9^{+6.8}_{-7.0}$ &     $-1.36 \pm 0.03$ & (a) \\
        28  &   1.3920 & $-31.5542 \pm 0.0961$ & $-27.4696 \pm 0.0009$   & $-12.4342 \pm  0.0004$  &     $65.7^{+24.8}_{-14.2}$ &     $-1.57\pm 0.04$ & (a) \\
        44  &   1.2330 &  $-31.8278 \pm 0.0707$ &  $-28.0663 \pm 0.0015$ &  $-13.0287 \pm  0.0013$  &     $65.8^{+18.8}_{-4.8}$ &     $-1.44\pm  0.03$ & (a)\\
       102  &   0.8610 &  $-31.9586 \pm 0.1101$ & $-27.5864 \pm 0.0009$ &  $-12.5513 \pm 0.0005$  &     $86.9^{+16.2}_{-13.3}$ &   $-1.68 \pm 0.04$ &  (a)\\
       114  &   1.2260 &  $-30.8388 \pm 0.0438$ &  $-26.9031 \pm 0.0008$ &  $-11.8697  \pm 0.0003$  &    $186.6^{+20.3}_{-15.4}$ &   $-1.51\pm 0.02$ &  (a)\\
       118  &   0.7150 & $-31.3581 \pm 0.0508$ & $-27.2730 \pm 0.0010$ &  $-12.2373  \pm   0.0006$  &    $102.2^{+27.0}_{-19.5}$ &     $-1.57 \pm  0.02$ &  (a)\\
       123  &   0.8910 & $-31.9610 \pm 0.1903$ & $-27.9488 \pm  0.0012$  &  $-12.9136 \pm  0.0009$  &     $81.6^{+28.0}_{-26.6}$ &      $-1.54 \pm 0.07$ &  (a)\\
       135  &   1.3150 &  $-31.7104 \pm  0.0859$ & $-27.8191 \pm 0.0009$ &   $-12.7852 \pm 0.0005$ &      $93.0^{+9.6}_{-9.8}$ &    $-1.49 \pm 0.03$  & (a)\\
       158  &   1.4780 &  $-31.9433 \pm  0.0986$ & $-28.2366 \pm 0.0014$ &  $-13.2007 \pm  0.0012$ &     $119.1^{+4.0}_{-11.8}$ &     $-1.42 \pm 0.04$  &  (a)\\
       159  &   1.5870 &  $-31.7435 \pm 0.0961$ & $-27.7467 \pm  0.0010$  &  $-12.7122 \pm 0.0006$ &     $324.2^{+25.3}_{-19.4}$ &      $-1.53 \pm 0.04$ &  (a)\\
       160  &   0.3600 &  $-30.3706 \pm  0.0146$ &  $-27.8833 \pm 0.0015$  &  $-12.8477 \pm 0.0013$ &     $106.5^{+18.2}_{-16.6}$ &      $-0.95 \pm  0.01$ &  (a)\\
       170  &   1.1630 &   $-31.9355 \pm 0.1140$ & $-27.7081 \pm 0.0009$   &   $-12.6716 \pm     0.0005$ &       $98.5^{+6.7}_{-17.7}$  &    $-1.62 \pm  0.04$  & (a)\\
       185  &   0.9870 & $-31.2580 \pm 0.0179$ & $-27.8888  \pm 0.0094$  &  $-12.8508  \pm 0.0093$  &    $387.9^{+3.3}_{-3.0}$ &      $-1.29 \pm  0.01$  &  (a)\\
       191  &   0.4420 & $-31.8241 \pm 0.0899$ & $-28.1033 \pm  0.0014$  &  $-13.0675 \pm    0.0012$ &      $93.9^{+24.3}_{-29.1}$ &     $-1.43 \pm  0.03$  &  (a)\\
       228  &   1.2640 &  $-32.2662 \pm  0.0542$ & $-28.2778 \pm 0.0014$  &  $-13.2426 \pm 0.0011$ &      $37.9^{+14.4}_{-9.1}$ &     $-1.53 \pm  0.02$ &  (a)\\
       232  &   0.8080 &  $-31.7874 \pm  0.0308$ & $-28.2676 \pm 0.0016$   &  $-13.2314 \pm 0.0014$ &     $273.8^{+5.1}_{-4.1}$ &      $-1.35\pm  0.01$ & (a)\\
       240  &   0.7620 &  $-31.9561 \pm 0.1068$ &  $-28.3802 \pm 0.0022$  &  $-13.3449 \pm  0.0021$ &       $17.2^{+3.5}_{-2.8}$ &      $-1.37\pm  0.04$ & (a)\\
       260  &   0.9950 & $-31.4365 \pm  0.0640$ & $-27.4848 \pm  0.0009$  &   $-12.4498 \pm  0.0004$ &       $94.9^{+18.7}_{-17.2}$ &      $-1.52 \pm 0.02$ & (a)\\
       280  &   1.3660 & $-31.5705 \pm  0.0752$ & $-27.5878 \pm  0.0009$   &  $-12.5528 \pm  0.0003$  &     $99.1^{+3.3}_{-9.5}$ &      $-1.53 \pm 0.03$ & (a)\\
       285  &   1.0340 &  $-31.8776 \pm 0.0473$ & $-28.3083 \pm 0.0022$  &  $-13.2725 \pm    0.0020$  &     $138.5^{+15.2}_{-21.1}$ &      $-1.37 \pm       0.02$ & (a)\\
       291  &   0.5320 &  $-31.1633 \pm 0.0174$ &  $-28.2683 \pm 0.0018$  &  $-13.2328 \pm   0.0016$  &      $39.7^{+4.2}_{-2.6}$ &      $-1.11 \pm       0.01$ &  (a)\\
       301  &   0.5480 &   $-31.5231 \pm 0.0643$ & $-27.8778 \pm 0.0013$  &  $-12.8416 \pm    0.0011$  &    $136.3^{+17.0}_{-16.9}$ &      $-1.40 \pm 0.02$ &  (a)\\
       303  &   0.8210 &  $-31.4942 \pm 0.0340$ & $-28.3158 \pm  0.0016$  &   $-13.2807  \pm  0.0014$ &      $57.7^{+10.5}_{-8.3}$ &    $-1.22 \pm 0.01$  &  (a)\\
       329  &   0.7210 &  $-31.3856 \pm 0.0767$ & $-27.0036 \pm  0.0011$  &  $-11.9706 \pm  0.0007$ &      $87.5^{+23.8}_{-14.0}$ &     $-1.68 \pm 0.03$  &   (a)\\
       338  &   0.4180 &  $-31.5911 \pm  0.0768$ & $-28.0706 \pm 0.0015$ &  $-13.0372 \pm 0.0013$ &      $22.1^{+8.8}_{-6.2}$ &     $-1.35\pm  0.03$  &  (a)\\
       419  &   1.2720 &  $-32.0069 \pm 0.0993$ & $-27.9619 \pm 0.0014$  &  $-12.9281 \pm 0.0011$ &      $95.5^{+15.2}_{-15.5}$ &      $-1.55 \pm 0.04$ &  (a)\\
       422  &   1.0740 &  $-31.2296 \pm  0.0440$ & $-28.1245 \pm 0.0014$   &  $-13.0888 \pm 0.0011$  &    $109.3^{+25.4}_{-29.6}$ &       $-1.19 \pm 0.02$ &  (a)\\
       440  &   0.7540 & $-31.4390 \pm  0.0669$ &  $-27.5181 \pm 0.0009$  &  $-12.4815 \pm 0.0004$  &    $114.6^{+7.4}_{-10.8}$ &      $-1.50 \pm 0.03$  &  (a)\\
       449  &   1.2180 & $-31.4268 \pm 0.0489$ & $-27.9928 \pm 0.0015$   &  $-12.9586 \pm 0.0013$ &     $119.8^{+14.7}_{-24.4}$ &      $-1.32 \pm 0.02$  &  (a)\\
       459  &   1.1560 &  $-31.9226 \pm 0.1129$ & $-27.9208 \pm  0.0013$ &  $-12.8861  \pm    0.0011$ &     $122.8^{+5.1}_{-5.7}$ &      $-1.54 \pm 0.04$ &  (a)\\
       492  &   0.9640 & $-31.7830 \pm 0.0715$ & $-27.4308 \pm  0.0009$ &  $-12.3947 \pm 0.0004$ &      $92.0^{+16.3}_{-12.7}$ &      $-1.67 \pm 0.03$ &  (a)\\
       493  &   1.5920 & $-31.6431 \pm  0.0721$ & $-27.2967 \pm 0.0009$   &  $-12.2612 \pm 0.0004$ &     $315.6^{+30.7}_{-35.7}$ &      $-1.67 \pm 0.03$ &  (a)\\
       501  &   1.1550 & $-32.4490 \pm  0.5935$ & $-27.9964 \pm  0.0012$  &   $-12.9586 \pm  0.0009$ &       $44.9^{+11.7}_{-10.4}$ &      $-1.71 \pm 0.15$ &  (a)\\
       505  &   1.1440 &  $-31.9151 \pm 0.1309$ &  $-28.1019 \pm 0.0014$   &  $-13.0665  \pm 0.0011$ &      $94.7^{+10.8}_{-16.7}$  &      $-1.46  \pm  0.05$ &  (a)\\
       522  &   1.3840 &  $-32.2515 \pm 0.1738$ & $-28.0185 \pm 0.0010$  &  $-12.9830 \pm  0.0006$ &     $115.8^{+11.3}_{-16.0}$  &      $-1.62 \pm  0.06$ &  (a)\\
       556  &   1.4940 &  $-31.8709 \pm  0.0928$ & $-27.6918 \pm 0.0009$  &  $-12.6556 \pm  0.0005$  &     $98.7^{+13.9}_{-10.8}$ &     $-1.60 \pm   0.04 $ &  (a)\\
       588  &   0.9980 & $-31.3856 \pm 0.0604$ & $-27.1696 \pm 0.0008$  &  $-12.1343 \pm 0.0002$  &     $74.3^{+23.0}_{-18.2}$ &     $-1.62 \pm 0.02$ &  (a)\\
       593  &   0.9920 & $-31.9378 \pm 0.1941$ &  $-27.7738 \pm 0.0010$   &  $-12.7375 \pm  0.0006$  &     $80.1^{+21.4}_{-20.8}$ &     $-1.60 \pm  0.07$ &  (a)\\
       622  &   0.5720 & $-31.2762 \pm 0.0372$ & $-27.6625 \pm 0.0009$   &  $-12.6271 \pm  0.0005$  &     $61.7^{+6.0}_{-4.3}$ &     $-1.39  \pm  0.01$ &  (a)\\
       645  &   0.4740 & $-31.6551 \pm 0.0970$ & $-27.8050 \pm 0.0012$   &  $-12.7696  \pm   0.0009$  &     $30.2^{+26.8}_{-8.9}$ &     $-1.48 \pm  0.04$ &  (a)\\
       649  &   0.8500 & $-32.1753 \pm 0.1679$ & $-28.1288 \pm 0.0015$  &  $-13.0931 \pm  0.0013$  &    $165.5^{+22.2}_{-25.1}$ &     $-1.55  \pm  0.06$ &  (a)\\
       675  &   0.9190 &  $-30.9529 \pm 0.0389$ & $-27.5836 \pm  0.0009$ &  $-12.5482 \pm  0.0005$  &    $139.8^{+12.0}_{-22.6}$ &     $-1.29 \pm 0.01$ &  (a)\\
       678  &   1.4630 &  $-31.8552 \pm 0.1940$ & $-27.8643 \pm  0.0010$ &  $-12.8297 \pm 0.0007$  &     $82.9^{+11.9}_{-10.2}$ &     $-1.53  \pm  0.07$  & (a)\\
       771  &   1.4920 & $-32.3116 \pm 0.2587$ & $-27.4926 \pm 0.0009$  &  $-12.4572 \pm 0.0004$  &      $31.3^{+8.1}_{-4.6}$ &       $-1.85 \pm  0.09$ &  (a)\\
       774  &   1.6860 &  $-31.5946 \pm 0.0877$ & $-27.6123 \pm 0.0009$ &  $-12.5768 \pm 0.0004$  &     $58.9^{+13.7}_{-10.1}$  &      $-1.53 \pm  0.03$ &  (a)\\
       792  &   0.5260 & $-32.2935 \pm  0.6754$ & $-28.5277 \pm  0.0030$   &  $-13.4921 \pm  0.0030$  &    $111.4^{+29.5}_{-20.0}$ &     $-1.45  \pm  0.15$ & (a)\\
       848  &   0.7570 & $-32.3623 \pm  0.5161$ & $-28.3785 \pm  0.0017$    &  $-13.3429  \pm  0.0015$  &     $65.1^{+29.4}_{-16.3}$ &     $-1.53  \pm  0.20$ & (a)\\
    J141214  &   0.4581 &  $-30.9049 \pm 0.0278$ & $-27.2882 \pm 0.0009$   & $-12.2526 \pm    0.0004$  &     $36.7^{+10.4}_{-4.8}$ &      $-1.39 \pm 0.01$ & (b), (c) \\
    J141018  &   0.4696 & $-31.4197 \pm 0.0797$ & $-28.2237 \pm 0.0052$  &  $-13.1884 \pm 0.0051$  &     $32.3^{+12.9}_{-5.3}$ &     $-1.23  \pm  0.03$ & (b), (c)\\
    J141417  &   0.6037 &  $-31.5509 \pm 0.0656$ & $-28.5277 \pm 0.0030$ &  $-13.4921 \pm 0.0029$  &     $29.1^{+3.6}_{-8.8}$ &     $-1.16 \pm  0.03$ & (b), (c)\\
    J142049  &   0.7510 & $-31.4316 \pm 0.0665$ & $-27.7570 \pm  0.0012$    &  $-12.7212 \pm 0.0009$  &     $34.0^{+6.7}_{-12.0}$ &  $-1.41 \pm  0.03$ & (b), (c)\\
    J141650  &   0.5266 & $-31.8094 \pm 0.1051$ & $-28.2946 \pm 0.0022$   &  $-13.2588 \pm  0.0020$  &       $25.1^{+2.0}_{-2.6}$ &      $-1.35 \pm  0.04$  & (b), (c)\\
    J141644  &   0.4253 & $-31.4618 \pm 0.0554$ & $-27.9031 \pm 0.0013$   &  $-12.8665 \pm  0.0010$ &      $17.2^{+2.7}_{-2.7}$ &     $-1.37 \pm  0.02$  & (b), (c)\\
      NGC4151  &   0.0041 & $-28.9908 \pm 0.0004$ &  $-24.5920 \pm 0.2015$  &   $-9.5560  \pm 0.2006$  &      $6.8^{+1.7}_{-2.1}$ &     $-1.69 \pm 0.07$ & (d) \\
      NGC4151  &   0.0041 & \ldots & \ldots  &  $-9.5560 \pm 0.2006$  &      $5.3^{+1.9}_{-1.8}$ &    $-1.69 \pm 0.07$ & (d)\\
     J021612  &   1.5604 & $-32.1973 \pm 0.1710$ &  $-27.7270 \pm 0.0410$ &  $-12.6925 \pm    0.0412$  &     $51.5^{+14.4}_{-8.9}$ &     $-1.72 \pm 0.06$ & (e)\\
     J033553 &   1.5777 & $-31.5201 \pm  0.0310$ & $-27.3899 \pm 0.0427$    &  $-12.3546 \pm  0.0427$  &     $48.1^{+22.1}_{-8.8}$    &  $-1.59 \pm  0.02$  & (e)\\
     J003710  &   1.0670 & $-31.5773 \pm 0.0412$ & $-27.1457 \pm 0.0444$  &  $-12.1101 \pm 0.0444$  &    $191.8^{+27.6}_{-18.5}$  &      $-1.70  \pm 0.02$ & (e)\\
     J003234 &   1.6406 &  $-31.9961 \pm 0.1090$ & $-27.5850 \pm 0.0303$  &  $-12.5498 \pm 0.0302$  &    $248.8^{+18.1}_{-11.6}$    &  $-1.69 \pm  0.04$ & (e)\\
\hline
\end{longtable}

\end{appendix}


\bsp	
\label{lastpage}
\end{document}